\documentclass[8pt]{article}
\usepackage{mathpazo}
\usepackage[utf8]{inputenc}
\usepackage{fullpage, charter}
\usepackage{amsmath}
\usepackage{amssymb}
\usepackage{mathtools}
\usepackage{amsthm}
\usepackage{graphicx}
\usepackage{adjustbox}
\usepackage{textcomp, gensymb}
\usepackage{enumerate}
\usepackage{titling}
\usepackage{booktabs}
\usepackage[colorlinks,citecolor=red,urlcolor=blue,bookmarks=true,hypertexnames=true]{hyperref}
\usepackage{subfigure}
\usepackage{float}
\usepackage[toc,page]{appendix}
\usepackage{makecell}
\usepackage[parfill]{parskip}
\usepackage{xcolor}
\usepackage{rotating}
\usepackage{tablefootnote}
\usepackage[detect-all]{siunitx}
\usepackage{pdflscape}
\usepackage{soul}
\usepackage{color}
\usepackage[most]{tcolorbox}
\usepackage{soul}

\usepackage{tikz}
\usetikzlibrary{positioning}
\usetikzlibrary { decorations.pathmorphing, decorations.pathreplacing, decorations.shapes}
\usetikzlibrary{patterns}
\usetikzlibrary{bayesnet}
\usetikzlibrary{positioning}

\usepackage{algpseudocode}
\usepackage{algorithm}
\usepackage{algorithmic}

\usepackage{longtable}

\usepackage[authordate,bibencoding=auto,strict,backend=bibtex,natbib]{biblatex-chicago}

\allowdisplaybreaks
\title{Probabilistic Modelling of Multiple Long-Term Condition Onset Times}
\author{Kieran Richards$^1$, Kelly Fleetwood$^2$, Regina Prigge$^2$, Paolo Missier$^3$\\Michael Barnes$^4$, Nick J. Reynolds$^{5,6}$, Bruce Guthrie$^2$, Sohan Seth$^1$}
\date{%
    $^1$ School of Informatics, University of Edinburgh\\%
    $^2$ Usher Institute, University of Edinburgh\\%
    $^3$ School of Computer Science, University of Birmingham\\%
    $^4$ Centre for Translational Bioinformatics, Queen Mary University of London\\%
    $^5$ Institute of Translational and Clinical Medicine, Faculty of Medical Sciences, Newcastle University\\%
    $^6$ NIHR Newcastle Biomedical Research Centre \& Department of Dermatology, Royal Victoria Infirmary, Newcastle Hospitals NHS Foundation Trust\\%
    \vspace{3mm}
    \today
}

\renewcommand{\d}{\textrm{d}} 
 
\newcommand{\E}{\mathbb{E}} 
\newcommand{\g}{\,|\,} 

\newcommand{\distas}{\quad\sim\quad}

\newcommand{\obs}{{\textrm{obs}}}

\newcommand{\bfone}{\mathbf{1}}

\newcommand{\bfgamma}{{\boldsymbol{\gamma}}}

\newcommand{\bfeta}{{\boldsymbol{\eta}}}
\newcommand{\bftheta}{{\boldsymbol{\theta}}}

\newcommand{\bfkappa}{{\boldsymbol{\kappa}}}

\newcommand{\bfpi}{{\boldsymbol{\pi}}}

\newcommand{\bfchi}{{\boldsymbol{\chi}}}

\newcommand{\bfPhi}{\boldsymbol{\Phi}}

\newcommand{\bfd}{\mathbf{d}}

\newcommand{\bfr}{\mathbf{r}}

\newcommand{\bft}{\mathbf{t}}

\newcommand{\bfz}{\mathbf{z}}

\newcommand{\bfE}{\mathbf{E}}

\newcommand{\bfG}{\mathbf{G}}

\newcommand{\bfK}{\mathbf{K}}

\newcommand{\bfP}{\mathbf{P}}

\newcommand{\bfT}{\mathbf{T}}

\newcommand{\bfZ}{\mathbf{Z}}

\newcommand{\primebfd}{\bfd^{\prime}}
\newcommand{\primebft}{\bft^{\prime}}

\newcommand{\calP}{\mathcal{P}}

\newcommand{\bernoulli}{\textrm{Bern}}
\newcommand{\betarand}{\mathrm{Beta}}

\newcommand{\multinomial}{\mathrm{Mult}}

\newcommand{\Dirichlet}{\mathrm{Dir}}

\newcommand{\normal}{\mathcal{N}}

\newcommand{\uniform}{\mathcal{U}}
\newcommand{\expfam}{\mathrm{ExpFam}}
\newcommand{\expfamc}{\mathrm{ExpFamConj}}
\newcommand{\expfamt}{\mathrm{ExpFamTrunc}}
\newcommand{\normalIG}{\mathcal{NIG}}

\bibliography{LTCtrajectoryclustering}

\definecolor{reg}{RGB}{77, 175, 74}
\definecolor{death}{RGB}{228, 26, 28}
\definecolor{eof}{RGB}{255, 127, 0}
\definecolor{heart}{RGB}{55, 126, 184}
\definecolor{lungs}{RGB}{152, 78, 163}

\begin{document}
\maketitle

\newcommand{\promote}{\texttt{ProMOTe}}
\allowdisplaybreaks

\begin{abstract}
The co-occurrence of multiple long-term conditions (MLTC), or multimorbidity, in an individual can reduce their lifespan and severely impact their quality of life. Exploring the longitudinal patterns, e.g. clusters, of disease accrual can help better understand the genetic and environmental drivers of multimorbidity, and potentially identify individuals who may benefit from early targeted intervention. We introduce \emph{probabilistic modelling of onset times}, or {\promote}, for  clustering and forecasting MLTC trajectories. {\promote} seamlessly learns from incomplete and unreliable disease trajectories that is commonplace in Electronic Health Records but often ignored in existing longitudinal clustering methods. We analyse data from 150,000 individuals in the UK Biobank and identify 50 clusters showing patterns of disease accrual that have also been reported by some recent studies. 
We further discuss the forecasting capabilities of the model given the history of disease accrual.
\end{abstract}

\section{Introduction}

Multimorbidity, or the presence of Multiple Long-Term Conditions (MLTC), is a widespread public health concern with around one third of adults globally estimated to have multimorbidity \citep{hajat_global_2018}. Multimorbidity is associated with poorer quality of life \citep{makovski_multimorbidity_2019} and higher mortality \citep{nunes_multimorbidity_2016}. Early prediction of multimorbidity can, therefore, facilitate intervention and improve quality of life in vulnerable individuals, while grouping together similar trajectories can facilitate exploring the underlying genetic and sociodemographic drivers of these trajectories \citep{zhang_topic_2023}. Electronic Health Records (EHR) provide unprecedented information on the presence or absence of long-term conditions (LTCs) in individuals, which together with their respective date of diagnosis can be used to identify groups of individuals with similar accrual characteristics \citep{cezard_studying_2021}, and predict their future disease accrual.

Most attempts at modelling MLTC groups have used a cross-sectional approach without temporal information \citep{busija_replicable_2019, cornell_multimorbidity_2008, garin_global_2016}. Alternative longitudinal approaches are relatively recent, and offer insight into the \emph{onset} of MLTC over life course \citep{cezard_studying_2021, jiang_age-dependent_2023}. Several longitudinal approaches model only the count of MLTC affecting an individual over time, ignoring the specific conditions themselves \citep{dekhtyar_association_2019, perez_glutathione_2020}. Few studies have considered specific conditions and their diagnosis dates in clustering \citep{cond-NMF2019,cond-dtw2020}.
A more recent study introduced a topic modelling approach, i.e. Age-dependent Topic Modelling (ATM), that allows disease topic loadings to vary with age \citep{jiang_age-dependent_2023}.

None of the studies, to the best of our knowledge, address the unreliability and incompleteness of the longitudinal EHR data in the respective analysis while these features are ubiquitous. For example, chronic conditions which developed prior to registration often do not have their proper diagnosis date reported, i.e. they are \emph{unreliable}, while any information of disease accrual is censored at the current age, i.e. one does not have knowledge of the conditions an individual will develop in the future or the respective trajectories are \emph{incomplete}.
The existing approaches are also not fully generative, i.e. they do not allow, given history, forecasting future disease accrual alongside their respective onset times and they do not capture the associated uncertainty. 

\begin{figure*}[t!]
\centering
{\includegraphics[width=1\textwidth]{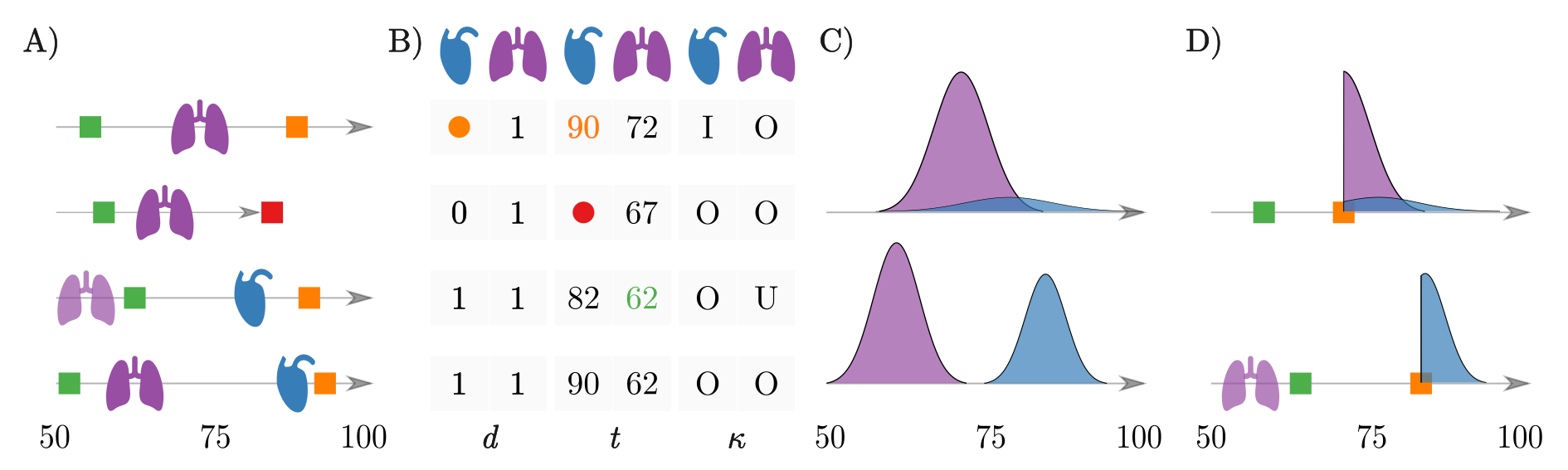}}
\caption{ \textbf{A)} $N=4$ individuals with different trajectories $\Phi^{(n)}$ with $M=2$ possible conditions affecting the \textcolor{heart}{heart} and \textcolor{lungs}{lungs} respectively. Here,  \textcolor{reg}{green} denotes baseline, \textcolor{death}{red} denotes death, and \textcolor{eof}{orange} denotes data extraction time. 
\textbf{B)} Data representation of each of the $N=4$ trajectories showing the respective $\bfd,\bft$ and $\bfkappa$.
\textbf{C)} Two potential clusters, i.e. (top) $k=1$ and (bottom) $k=2$ with respective parameters $(\bar{\pi}_{\textcolor{heart}{1},k},\bar{\bfeta}_{\textcolor{heart}{1},k})$ and $(\bar{\pi}_{\textcolor{lungs}{2},k},\bar{\bfeta}_{\textcolor{lungs}{2},k})$ for each condition,
\textbf{D)}  (top) $\Phi^\prime_1$: an individual without any condition and (bottom) $\Phi^\prime_2$: an individual with the \textcolor{lungs}{lung} condition recorded prior to baseline. Their respective predictive distribution of disease accrual beyond age at data extraction \textcolor{eof}{orange} is shown.}
\end{figure*}

We propose a probabilistic framework, referred to as \emph{Probabilistic Modelling of Onset-Time} or \textbf{{\promote}}, for longitudinal clustering of multimorbidity accrual, that considers both the LTCs present in an individual and their respective time of onset (using respective age at diagnosis as proxy). Our approach, inspired by Latent Class Analysis (LCA) \citep{hagenaars2002applied}, models the age at diagnosis of each LTC alongside their presence or absence in an individual. Unlike existing methods for clustering multimorbidity trajectories, {\promote} (1) learns from incomplete and unreliable disease trajectories seamlessly, (2) allows predicting future LTC accrual based on the current, \emph{possibly unreliable or partially observed}, disease trajectory, and (3) estimates uncertainty at a population (i.e. cluster) and at an individual level (i.e. personalized disease trajectory), both of which are desirable for clinical AI applications  \citep{banerji2023clinical}.   We assess {\promote}'s clustering and predictive capabilities experimentally, using both synthetic and UK Biobank data. 

\newcommand{\alive}{\textsc{\textcolor{cyan}{a}}}
\newcommand{\deceased}{\textsc{\textcolor{red}{d}}}
\newcommand{\LTC}{\textsc{\textcolor{black}{ltc}}}
\section{Probabilistic Trajectory Model}

Consider that we have EHR of $N$ individuals with $\iota^{(n)} \in \{\alive,\deceased\}$ denoting if they are alive or dead at the end of follow up,  $\rho^{(n)}$ denoting their \emph{baseline} age and
$\tau^{(n)}$ denoting their age at the end of follow up (if $\iota^{(n)} = \alive$) or their time of death (if $\iota^{(n)} = \deceased$). We assume that there are $M$ LTCs in the dataset, and we represent the health trajectory of an individual as 
\renewcommand{\obs}{ \textcolor{orange}{ \textsc{o}}}
\renewcommand{\part}{  \textcolor{orange}{ \textsc{u}}}
\newcommand{\unobs}{  \textcolor{orange}{ \textsc{i}}}
\newcommand{\obstar}{  \textcolor{orange}{\ast}}
 \[\Phi^{(n)} = \left\{
 \bfd^{(n)},\bft^{(n)},\bfkappa^{(n)}
 \right\} 
 \] 
using vectors $\bfd^{(n)},\bft^{(n)},$ and $\bfkappa^{(n)}$ of length $M$ each. Here $\bfd^{(n)}$ is an indicator vector with $d^{(n)}_m = 1$ if the $m$-th LTC is present with respective age at diagnosis $t^{(n)}_m$, and $d^{(n)}_m = 0$ if the respective LTC is absent with respective $t^{(n)}_m$ irrelevant.
We consider these data in $4$ different cases. (1) Typically we can assume that diagnosis dates are recorded accurately beyond the baseline date. For these conditions we have $d^{(n)}_m = 1$, $\rho^{(n)} < t^{(n)}_m \leq \tau^{(n)}$ and denote these data as observed, $\kappa^{(n)}_m = {\obs}$. (2) If an individual is recorded as not having a condition and is dead then we can also consider the data to be observed, i.e. $\kappa^{(n)}_m = {\obs}$ but we have, $d^{(n)}_m = 0$, $t^{(n)}_m = \bullet$. (3) Diagnosis dates recorded before baseline can be subject to the recollection of the individual or to incorrect recording of historical dates by the clinician \citep{lewis2005relationship}. In such situations, we use $d^{(n)}_m = 1$ and $t_{m}^{(n)}=\rho^{(n)}$ implying that the respective diagnosis dates are unreliable, denoted as $\kappa^{(n)}_m  = {\part}$. (4) For any individual who is still alive, conditions developed beyond the current age are not observed yet. In these situations, we use $d^{(n)}_m = \bullet$ and $t_{m}^{(n)}=\tau^{(n)}$ implying that the respective diagnosis dates are incomplete, denoted as $\kappa^{(n)}_m = {\unobs}$. We denote by $M_{\obs,\unobs,\part}^{(n)} = |\{m : \kappa_{\obs,\unobs,\part}^{(n)}\}|$ the number of observed, incomplete and unreliable events for an individual. We denote the complete dataset as $\bfPhi = \{\Phi^{(n)}\}_{n=1}^N$. A table of all the notation used here and throughout the rest of the paper is available in Appendix \ref{tableofnotation}.

\newcommand{\suffstat}{\bfT}

\subsection{The Model}

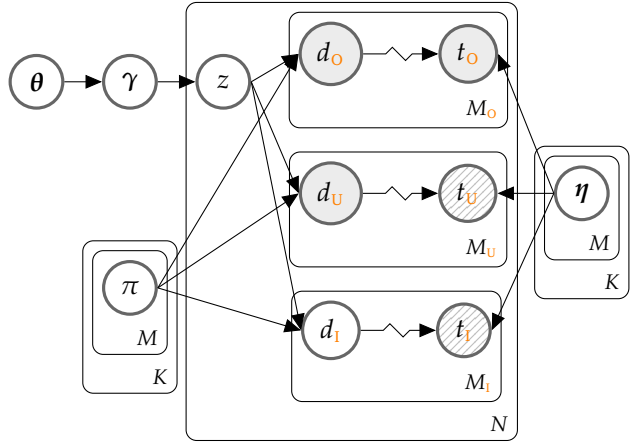
\begin{figure}[ht]
\begin{centering}
\begin{tikzpicture}[
block/.style =
    {
        rectangle
      , draw
      , fill=white!20
      , text width=5.0em
      , text centered
      , node distance=2.5cm
      , rounded corners
      , minimum height=1em
      },
roundnode/.style={circle, draw=black!60, fill=white!5, very thick, minimum size=7mm},
squarednode/.style={rectangle, draw=black!60, fill=black!5, very thick, minimum size=7mm},
]
\node[roundnode]      (nu)  {$\bftheta$};
\node[roundnode]      (gamma)  [right =0.5cm of nu]{$\bfgamma$};
\node[roundnode]      (z)  [right=0.5cm of gamma]{$z$};
\node[roundnode]      (dobs)  [right=1cm of z.north, fill=gray!15]{$d_{\obs}$};
\node[roundnode]      (dpart)  [below=1cm of dobs, fill=gray!15]{$d_{\part}$};
\node[roundnode]      (tpart)  [right =1cm of dpart, pattern=north east lines, pattern color = gray!50]{$t_{\part}$};
\node[roundnode]      (dcen)  [below=1cm of dpart]{$d_{\unobs}$};
\node[roundnode]      (tobs)  [right =1cm of dobs, fill=gray!15]{$t_{\obs}$};
\node[roundnode]      (tcen)  [right =1cm of dcen, pattern=north east lines, pattern color = gray!50]{$t_{\unobs}$};
\node[roundnode]      (mu)  [right =0.75cm  of tpart]{$\bfeta$};
\node[roundnode]      (pi)  [below=2cm of gamma]{$\pi$};
\plate {cen} {(dcen)(tcen)} {$M_{\unobs}$};
\plate {part} {(dpart)(tpart)} {$M_{\part}$};
\plate  {obs} {(dobs)(tobs)} {$M_{\obs}$};
\plate {} {(obs)(cen)(part)(z)(dobs)(tobs)(dcen)(tcen) (dpart) (tpart)} {$N$};
\plate {muK} {(mu)} {$M$};
\plate {} {(muK)(mu)} {$K$};
\plate {piK} {(pi)} {$M$};
\plate {} {(piK)(pi)} {$K$};

\draw[->] (nu.east) -- (gamma.west);
\draw[->] (gamma.east) -- (z.west);
\draw[->, decorate, decoration = {zigzag, pre length = 0.3cm, post length = 0.3cm}] (dobs.east) -- (tobs.west);
\draw[->, decorate, decoration = {zigzag, pre length = 0.3cm, post length = 0.3cm}] (dcen.east) -- (tcen.west);
\draw[->, decorate, decoration = {zigzag, pre length = 0.3cm, post length = 0.3cm}] (dpart.east) -- (tpart.west);
\draw[->] (z.east) -- (dcen.west);
\draw[->] (z.east) -- (dobs.west);
\draw[->] (z.east) -- (dpart.west);
\draw[->] (pi.east) -- (dobs.west);
\draw[->] (pi.east) to (dpart.west);
\draw[->] (pi.east) -- (dcen.west);
\draw[->] (mu.west) -- (tobs.east);
\draw[->] (mu.west) -- (tcen.east);
\draw[->] (mu.west) to (tpart.east);

\end{tikzpicture}

\caption{The Plate diagram illustrates the different types of censoring that can occur within a trajectory. The fully shaded nodes are observed. The hatched nodes are unreliable. The jagged line indicates conditional dependency.}
\label{fig:diffways}
\end{centering}
\end{figure}

{\promote} aims to assign each individual to one of $K$ clusters. Then, given the cluster, it aims to model the presence and diagnosis ages of each of the pre-defined LTCs. We consider the following generative framework (shown in Figure~\ref{fig:diffways}) for each individual:
\begin{enumerate}
    \item Choose a cluster \[\bfz^{\left(n\right)}\distas\multinomial(1, \bfgamma)\] where $z_{k}^{\left(n\right)} = 1$ if the $n$-th individual belongs to cluster $k$ and it is zero otherwise, and ${\bfgamma}$ is a $K$ dimensional stochastic vector of cluster probabilities.
    \item Choose if the individual has a specific condition 
    \[ d^{(n)}_{m} \distas \bernoulli(\pi_{m,k})\]
    where $\pi_{m,k}$ denotes the probability of having $\LTC_m$ if one belongs to cluster $k$.
    \item Choose the age at diagnosis as
    \[ t^{(n)}_{m} \g d^{(n)}_{m} = 1 \distas \expfam({\bfeta_{m,k}}),\] if the condition is present, where $\expfam(\bfeta)$ denotes the exponential family distribution with natural parameters $\bfeta$, and natural statistics $\suffstat$ such as, $f_\expfam(t \g \bfeta) =   g\left(\bfeta\right)  h\left(t\right)\exp\left(\bfeta^\top \suffstat\right)$
\end{enumerate}

Under this construction we can write the full likelihood as:
\begin{align}
p\left(\bfPhi,\{\bfz^{(n)}\} \g \{\pi_{m,k}\}, \{\bfeta_{m,k}\},\bfgamma\right) \propto& \prod_{n=1}^{N}\prod_{k=1}^{K}\left[\gamma_{k}\prod_{m=1}^{M}\left(1-\pi_{m,k}\right)^{\left(1-d_{m}^{\left(n\right)}\right)}\left(\pi_{m,k} f_\expfam\left(t^{(n)}_{m}\g\bfeta_{m,k}\right)\right)^{d_{m}^{\left(n\right)}}\right]^{z_{k}^{\left(n\right)}}\label{eq:likelihoodGaussian}
\end{align}
We complete the Bayesian framework by assigning independent priors to the model parameters as $\bfgamma \sim\Dirichlet\left(\bftheta\right)$; $\pi_{m,k} \sim\betarand\left(a_{m,k},b_{m,k}\right)$; and $\bfeta_{m,k} \sim \expfamc\left(\nu_{m,k},\bfchi_{m,k}\right)$
where $\bftheta,a_{m,k},b_{m,k},\nu_{m,k},\text{ and } \bfchi_{m,k}$ are hyperparameters and $f_\expfamc\left(\bfeta \g \nu,\bfchi\right) =     g\left(\bfeta\right)^{\nu} f\left(\nu,\bfchi\right)\exp\left(\bfeta^\top\bfchi\right)$
denotes the conjugate prior over exponential family distributions.

\subsection{Variational Bayes}

We propose a mean-field Variational Bayes (VB) approach to approximate the posterior distribution. VB scales much more efficiently to large EHR data than other methods \citep{blei2017variational}. We first describe the approach in the straightforward case of complete data i.e. assuming $\kappa^{(n)}_m = \obs\; \forall n,m$. The full derivation is available in Appendix \ref{DeriveVB}.  Later we will describe how the method can be extended to handle both unreliable and incomplete data.
We factorize the variational approximation as
\begin{equation*}
q\left(\bfgamma\right)\prod_{m=1}^M\prod_{k=1}^Kq\left(\pi_{m,k}\right)\prod_{m=1}^M\prod_{k=1}^Kq\left(\bfeta_{m,k}\right)\prod_{n=1}^Nq\left(\bfz^{(n)}\right)
\end{equation*}
and assume the following variational distributions: $\bfgamma \sim \Dirichlet\left(\bftheta^{*}\right)$, $\pi_{m,k} \sim \betarand\left(a_{m,k}^{*}, b_{m,k}^{*}\right)$,\linebreak $\bfeta_{m,k} \sim \expfamc\left(\nu_{m,k}^{*},\bfchi_{m,k}^{*}\right)$, and $\bfz^{\left(n\right)} \sim \multinomial\left(1, \bfr^{(n)\ast}\right)$.

Due to conjugacy, the respective parameters have closed form update equations as follows (see Appendix \ref{DeriveVB}):
\begin{alignat}{1}
\label{eq:thetastar}\bftheta^\ast &= \bftheta + \bar{\bfz}\\
\label{eq:abstar}a_{m,k}^{\ast} &= a_{m,k}+\bar{n}_{m,k}; \quad b_{m,k}^{\ast} = b_{m,k}+\bar{z}_{k}-\bar{n}_{m,k} \\
\label{eq:nuchistar}\nu^\ast_{m,k} &=\nu_{m,k}+\bar{n}_{m,k}; \quad \bfchi^\ast_{m,k} =\bfchi_{m,k}+\bar{\bfT}_{m,k} \\
\label{eq:gammastar}\bfgamma_{k}^{(n)\ast} & \propto \exp\left(\zeta_{k} + \sum_{m=1}^{M} d_{m}^{(n)}\left(\lambda_{m,k} + u_{\obs,m,k}^{(n)}\right)\right)
\end{alignat}
where expressions for $\bar{\bfz}$, $\bar{n}_{m,k}$, $\bar{\bfT}_{m,k}$, $\zeta_{k}$, $\lambda_{m,k}$, and $u_{\obs,m,k}^{(n)}$ can all be found in Appendix \ref{DeriveVB}.
The quantities $\bar{z}_k, \bar{n}_{m,k}$ and $\bar{\bfT}_{m,k}/\bar{n}_{m,k}$ can be intuitively understood as the expected number of individuals in each cluster, the expected number of individuals with the $m$-th disease in cluster $k$ and the expected (sufficient statistics of) diagnosis age of the $m$-th disease in cluster $k$ respectively. The quantities $\zeta_{k}$ and $\lambda_{m,k}$ depend on the variational parameters $\bftheta^\ast$, $a_{m,k}^\ast$, and $b_{m,k}^\ast$  while $u_{\obs,m,k}^{(n)}$ depends on $\nu_{m,k}^\ast$ and $\bfchi_{m,k}^\ast$.

\subsection{Incomplete Data\texorpdfstring{, i.e. $\kappa^{(n)}_m = \unobs$}{}}

To adapt the VB approach to handle right-censoring, or incomplete trajectories, we consider the respective $d_m^{(n)}$ and $t_m^{(n)}$ to be latent variables with $t_m^{(n)}$ to be partially known, i.e. $t_m^{(n)}> \tau^{(n)}$, and approximate their posterior as $q(t^{(n)}_m,d^{(n)}_m)$ where 
\begin{align}
    d_{m}^{(n)} & \distas \bernoulli \left(\pi^{(n)\ast}_m\right) \\
    \label{eq:rightcens}t_{m}^{(n)} & \distas \expfamt_{l}(\bfeta_{m}^{(n)*})
\end{align}
where $\expfamt_{l}$ denotes a left truncated exponential family distribution, and unless otherwise stated left truncation is always done at $\tau^{(n)}$. The respective update equations are given as
\begin{align}
\label{eq:pistar}\pi_{m}^{(n)\ast} &=\frac{\exp\left(\sum_{k=1}^{K}\bfgamma_{k}^{(n)\ast}\left(\lambda_{m,k} + u_{\unobs,m,k}^{(n)}\right)\right)}{1 + \exp\left(\sum_{k=1}^{K}\bfgamma_{k}^{(n)\ast}\left(\lambda_{m,k} + u_{\unobs,m,k}^{(n)}\right)\right)}\\ \label{eq:etastar}
\bfeta_{m}^{(n)\ast} &= \sum_{k=1}^{K}\bfgamma_{k}^{(n)\ast}\E\left(\bfeta_{m,k}\right).
\end{align}
and an expression for $u_{\unobs,m,k}^{(n)}$ can be found in Appendix \ref{DeriveVB}.

\subsection{Unreliable Data\texorpdfstring{, i.e. $\kappa^{(n)}_m = \part$}{}}

To adapt the VB approach to handle left-censoring, or unreliable information prior to the beginning of the study, 
we consider $d_m^{(n)} = 1$ to be observed (i.e. the respective condition to be present) but $t_m^{(n)}$ to be a latent variable that is partially known, i.e. $t_m^{(n)} < \rho^{(n)}$, and
\begin{align}
    \label{eq:leftcens}
    t_{m}^{(n)} & \distas \expfamt_{r}(\bfeta_{m}^{(n)*}) 
\end{align}
where $\expfamt_{r}$ denotes a right truncated exponential family distribution, and unless otherwise stated right truncation is always done at $\rho^{(n)}$. The respective update equation is given as in \eqref{eq:etastar}. An expression for $u_{\part,m,k}^{(n)}$ can be found in Appendix \ref{DeriveVB}.

\subsection{Variational Bayes for Full Model}

For the full model with observed, unreliable and incomplete data, 
the update rules in \eqref{eq:thetastar}, \eqref{eq:abstar}, \eqref{eq:nuchistar}, \eqref{eq:pistar}, and \eqref{eq:etastar} remain the same whilst the updating rule for $\gamma_k^{(n)\ast}$ becomes: 
\begin{equation}
\bfgamma_{k}^{(n)\ast} \propto \exp\left(\zeta_{k} + \sum_{m : \kappa = {\obs}}  d_{m}^{(n)}\left(\lambda_{m,k}^{(n)} + u_{\obs, m,k}^{(n)}\right)\right. + \left.\sum_{m : \kappa = {\part}} \left(\lambda_{m,k}^{(n)} + u_{\part,m,k}^{(n)}\right)+ \sum_{m : \kappa = {\unobs}} \pi_{m}^{(n)\ast}\left(\lambda_{m,k}^{(n)} + u_{\unobs,m,k}^{(n)}\right)\right) 
\end{equation}
where we write $\kappa$ for $\kappa_m^{(n)}$ to preserve space. Expressions for $\bar{n}_{m,k}$ and $\bar{T}_{m,k}$ are given as:
\begin{alignat}{1}
    \bar{n}_{m,k} & = \sum_{n:\kappa_{m}^{(n)}={\obs}, {\part}} d_{m}^{\left(n\right)}\bfgamma_{k}^{(n)\ast} + \sum_{n:\kappa_{m}^{(n)}={\unobs}} \pi_{m}^{(n)\ast}\bfgamma_{k}^{(n)\ast}\\
    \bar{\bfT}_{m,k} & = \sum_{n:\kappa_{m}^{(n)}={\obs}} d_{m}^{\left(n\right)}\bfgamma_{k}^{(n)\ast}\suffstat_m^{(n)} + \sum_{n:\kappa_{m}^{(n)}={\part}} 
    \bfgamma_{k}^{(n)\ast}\E_{r}\left(\suffstat_m^{(n)}\right) + \sum_{n:\kappa_{m}^{(n)}={\unobs}} \pi_{m}^{(n)\ast}\bfgamma_{k}^{(n)\ast}\E_{l}\left(\suffstat_m^{(n)}\right)
\end{alignat}
where $\E_{l}$ and $\E_{r}$ denote the expectation under left and right truncation respectively. The full algorithm is provided in Algorithm \ref{Algorithm:VBcen} in Appendix \ref{cenalg}.

\subsection{Predictive Posterior}

We now consider a new individual where the trajectory can be partitioned into observed, unreliable and incomplete sets, i.e. $\Phi_{\obstar}^{\prime} = \left(\bfd_{\obstar}^{\prime},\bft_{\obstar}^{\prime},\bfkappa_{\obstar}^{\prime}\right) =  \left(d^{\prime}_{m},t^{\prime}_{m},\kappa^{\prime}_{m}\right)_{m:\kappa_{m}^{\prime}={\obstar}}$ where $\obstar = \obs, \part$ or $\unobs$. We are interested in finding the missing values $\bfd_{\unobs}^{\prime}$ and $\bft_{\unobs}^{\prime}$ or the probability of observing the presence and absence of LTCs that have not been observed by time $\tau^\prime$ and their respective age at diagnosis after $\tau^\prime$, i.e. (with a slight abuse of notation)
\begin{equation}
p\left(\bfd_{\unobs}^{\prime},\bft_{\unobs}^{\prime} \right. \g \left. \Phi_{\obs}^{\prime}, \Phi_{\part}^{\prime}, \bfkappa_{\unobs}^{\prime}, \bfPhi,\rho^{\prime}, \tau^{\prime}\right)= 
\sum_{k=1}^K
\left[p(z_k\g\Phi^\prime_{\obs},  \Phi_{\part}^{\prime},\bfkappa_{\unobs}^{\prime},\bfPhi, \rho^{\prime}, \tau^{\prime}) p(\bfd_{\unobs}^{\prime},\bft_{\unobs}^{\prime} \g z_k,\bfkappa_{\unobs}^{\prime}, \bfPhi,\tau^{\prime})\right]\label{eq:predpost}
\end{equation}
Here the first term is the posterior probability of assigning the observed trajectory to a cluster, i.e. 
\begin{align} 
&p(z_k\g\Phi^\prime_{\obs},  \bfd_{\part}^{\prime},\bfkappa_{\part}^{\prime},\bfkappa_{\unobs}^{\prime},\bfPhi)\nonumber\\
&\propto p(z_k \g \bfPhi) p(\Phi^\prime_{\obs} \g z_k, \bfPhi)p(\bfd_{\part}^{\prime},\bfkappa_{\part}^{\prime} \g z_k, \bfPhi, \rho^{\prime})p(\bfkappa_{\unobs}^{\prime} \g z_k, \bfPhi, \tau^{\prime}) \nonumber\\
 &\propto \bar{\theta}_k \prod_{m:\kappa^\prime=\obs} \bar{\pi}_{m,k} f_{\mathcal{P}}(t^\prime_m \g z_k)\prod_{m:\kappa^\prime=\part} \bar{\pi}_{m,k}F_{\mathcal{P}}\left(\rho^\prime\g z_k\right)\prod_{m:\kappa^\prime=\unobs} \left(1-\bar{\pi}_{m,k}F_{\mathcal{P}}\left(\tau^\prime\g z_k\right)\right),
\end{align}
where $\bar{\theta}$, $\bar{\pi}_{m,k}$ and $\bar{\bfeta}$ are posterior estimates of the model parameters. The posterior predictive distribution for $t^\prime$ for an individual who is zero years old and has no conditions:
\begin{equation}
t^{\prime}_m \g z_k, \bfPhi \distas \mathcal{P}\left(\nu^\ast_{m,k},\bfchi^\ast_{m,k}\right)
\end{equation}
with probability density function, $f_{\mathcal{P}}(t^\prime \g \nu,\bfchi)$, and respective cumulative density and survival functions, $F_{\mathcal{P}}(\rho^{\prime})$ and $S_{\mathcal{P}}(\tau^{\prime})$.

The second term in \eqref{eq:predpost} is the probability density over the trajectory beyond $\tau^\prime$ given the cluster, i.e.
\begin{equation}
p(\bfd_{\unobs}^{\prime},\bft_{\unobs}^{\prime} \g z_k,\bfkappa_{\unobs}^{\prime}, \bfPhi,\tau^\prime)=  \prod_{m:\kappa^\prime_m=\unobs} (1-\tilde{\pi}_{m,k})^{(1-d^\prime_m)}(\tilde{\pi}_{m,k}f_{\mathcal{P}}(t_{m}^{\prime} \g t_{m}^{\prime} > \tau^{\prime}))^{d^\prime_m}
\end{equation}
where $f_\calP(\cdot\g\cdot>\tau)$ denotes truncated distribution and
\begin{align*}
\tilde{\pi}_{m,k} =& p(d_m\g z_k, \kappa_{m}^{\prime} = {\unobs},\bfPhi, \tau^{\prime}) \propto \bar{\pi}_{m,k}S_{\mathcal{P}}(\tau^{\prime} \g \nu_{m,k}^{*},\bfchi_{m,k}^{*})
\end{align*}
is the posterior probability of the $m$-th LTC being active in cluster $k$ conditional on the event that it has not happened before $\tau^\prime$.
\section{A Simulated Example}\label{sec:simex}

To evaluate {\promote} empirically, we apply it to simulated data with $200,000$ individuals, $M=80$ conditions, and $K=10$ known clusters with each cluster containing between $3\%$ and $15\%$ of the trajectories. Within each cluster the presence of each of the $80$ conditions was generated independently with probabilities $\pi_{m,k}$ generated from a  $\betarand\left(0.07, 0.49\right)$ distribution, resulting in an average of $10$  conditions per trajectory. For each condition that was present in the cluster, we simulated the respective diagnosis ages using Gaussian random variables whose means and variances were generated from a  $\normalIG\left(50, 0.3, 5, 300\right)$ distribution, resulting in diagnosis ages with an average mean of $50$ and an average standard deviation of $8.66$. For each individual, we simulated baseline age as $\rho^{(n)}\sim\uniform\left(20, 60\right)$ and data before these ages was left-censored. The current ages were set to be thirty years later, i.e. $\tau^{(n)} = \rho^{(n)} + 30$. Finally we simulated i.i.d. Bernoulli random variables where $p\left(\iota^{(n)} = \deceased\right) = 0.8$ to determine whether each trajectory ended at $\tau^{(n)}$ or was right-censored. We split the simulated data such that $160,000$ trajectories formed a training set used to fit the model and the remaining $40,000$ trajectories were used as a test set to evaluate the model. 

Given the simulated data we fit {\promote} assuming a Gaussian distribution for diagnosis ages $t^{(n)}_{m}\g d^{(n)}_{m} = 1 \sim \normal(\mu_{m,k}, \sigma_{m,k})$
with the following prior distributions, $\bfgamma \sim \Dirichlet\left(\bfone\right)$, $\pi_{m,k} \sim \betarand\left(1,1\right)$, and $\mu_{m,k}, \sigma_{m,k}^2 \sim \normalIG\left(50, 0.3, 5, 750\right)$.
Derivations of the VB updating equations for the Gaussian case are available in Appendix \ref{gaussianVBder}. Here we choose hyperparameters that construct weakly informative priors. The hyperparameters can be adjusted to incorporate stronger prior information. 

To assess the predictive performance we simulate current ages uniformly between $50$ and $90$. We then predict forward from the test data using data from before these ages to predict the data afterwards. We compare these predictions to the test data using both accuracy and area under receiver operator curve (AUROC) \citep{grau2015prroc} as metrics. We also consider the mean absolute error (MAE) between the predicted diagnosis ages when the condition was present in the test data. We stress that the following results present a best case scenario for {\promote}, since the data generating process for the synthetic data is same as the model. 

We first assess the model's ability to recover the true clusters from the synthetic data. We observe that {\promote} performs well, recovering the true clusters in most cases from the test data with the most probable cluster matching the true cluster in $92\%$ of trajectories. The inaccuracies largely belonged to cluster 8 which was conflated with cluster 1 and cluster 5 which contain similar trajectories but are much larger. The model still assigned high probability to that cluster for those individuals and we observe an AUROC of $0.75$ for that cluster, compared to AUROC close to $1$ for the other clusters. Plots comparing the estimated risk profiles of all clusters to the true risk profiles are available in Appendix \ref{simrisks}.

Next we assess the model's ability to predict which diseases will occur in the test trajectories. {\promote} achieves very high accuracy when predicting disease presence with an overall accuracy of $89\%$ and AUROC of $0.99$. This reaffirms the ability of {\promote} to cluster probabilistically, since even in test cases where the true cluster is not recovered with certainty, it still assigns high  probability to the true cluster to predict disease outcomes.
We further assess the diagnosis ages predicted by {\promote} for diseases which occur in the test data. 
Over the test data the MAE of the observed and predicted age at diagnosis is $8.2$ years.

\section{UK Biobank Analysis}

The model was applied to data from the UK Biobank. The UK Biobank is a large prospective population-based cohort study of more than $500,000$ participants who were recruited between 2006 and 2010 aged between 40 and 70 years and provided written consent. Participants consented to linkage to EHRs, including GP, hospital, cancer registry and death records. The study was designed to investigate the effects of established disorders, lifestyle, environmental and genetic determinants on adulthood diseases \citep{sudlow2015uk}. Baseline assessments were performed at 22 assessment centres across the UK. The study involves longitudinal follow-up combined with general practice (GP) EHRs, hospital EHRs, and other national registries and databases. 

This study uses a subset of the UK Biobank data regarding individuals for whom GP data were available for at least one year before baseline and at least one day after baseline. Participants whose only primary care registration was with practices in England that use the Vision practice management system were further excluded because this dataset excluded individuals who died before data extraction. Furthermore, we only included data regarding individuals who have at least one LTC by the end of follow up. At the time of data extraction this subset of the data contained $159,145$ individuals. 
This data includes the presence and ages at diagnosis of $80$ conditions. These conditions were defined from a subset of 308 conditions\citep{kuan2019chronological} determined by clinicians to be LTCs relevant to middle-aged adults \citep{prigge4863974robustly}. Additional conditions were selected based on a recent Delphi consensus study defining conditions to include in multimorbidity studies \citep{ho2022measuring}. Conditions at baseline were identified based on a combination of GP, hospital, and cancer registry records combined with self-reported conditions. Left censoring was imposed on these conditions identified at baseline. By the end of follow-up the average individual had approximately four LTCs and $11,718$ participants had died.  

\begin{figure}[t]
\begin{minipage}{0.48\textwidth}
\centering
\includegraphics[width=\linewidth]{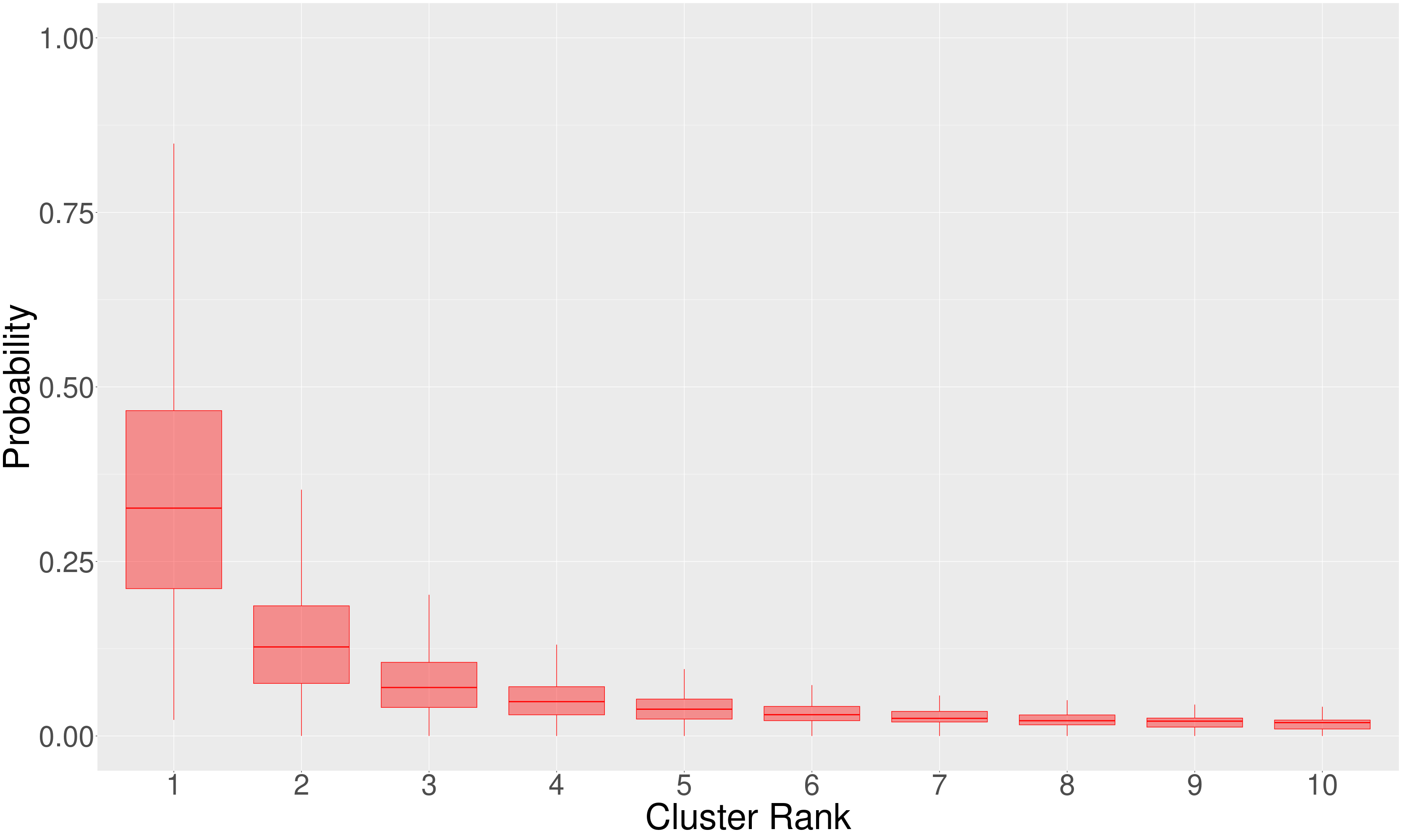}
\caption{The figure illustrates the posterior cluster probabilities for the test data across the $10$ most likely clusters.
}
\label{fig:boxUKB}
\end{minipage}\qquad
\begin{minipage}{0.48\textwidth}
\centering
\includegraphics[width=\linewidth]{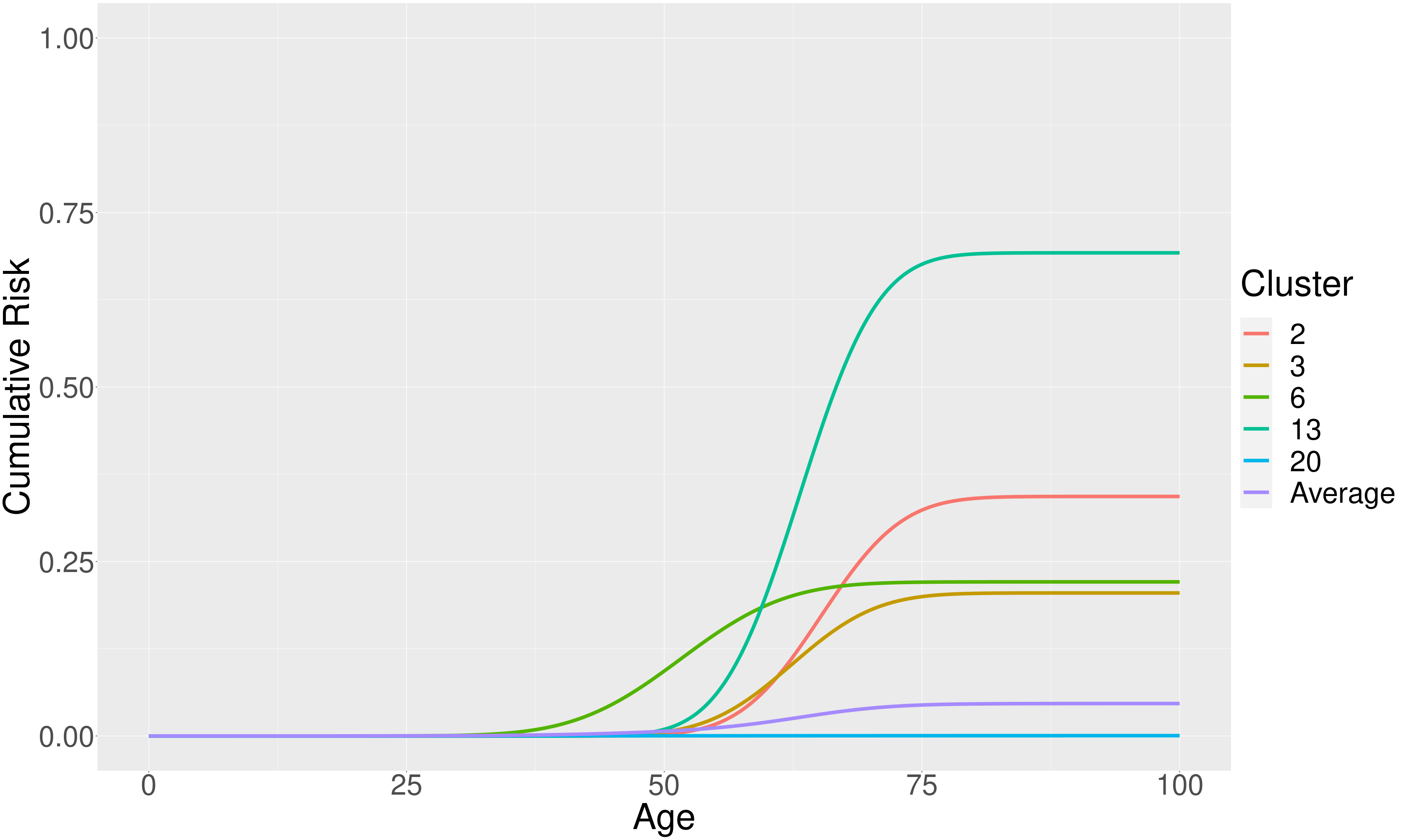}
\caption{The cumulative risk of \textsc{COPD} in  different clusters. The average cumulative risk is displayed in black.
}
\label{fig:cumulUKB}
\end{minipage}
\end{figure}

{\promote} was applied assuming a Gaussian distribution for $t\g d=1$ and we used the same prior as in Section \ref{sec:simex}. Additionally, for the purposes of censoring we treat two conditions as sex-specific, occurring in men only, and seven conditions as lifelong conditions, only occurring at birth. These conditions are considered fully observed in women and people over zero years of age respectively. A table of the disease coding, including which conditions are sex-specific and lifelong, can be found in Appendix \ref{discodes}. We fitted the model with between $K=20$ and $K=100$ clusters using $80\%$ of the data as training set and used the remaining $20\%$ of the data as test set to evaluate the model. We selected $K=50$ using an elbow plot shown in Appendix \ref{kselect}.

When applied to the test data we observe that individuals typically have $3-5$ most probable clusters as shown in Figure \ref{fig:boxUKB}. Despite this, we find that most individuals only have a moderately high probability of being assigned to one of these clusters, and only a very small number of individuals could be assigned to their most probable cluster with probability greater than $0.5$. This illustrates the uncertainty involved in longitudinal clustering of multimorbidity, particularly with unreliable and incomplete data, and highlights the need for probabilistic clustering which incorporates that uncertainty.

\paragraph{Cluster descriptions}  The clusters are well balanced, and with $50$ clusters we find that the smallest cluster from the test set had around 200 individuals for whom that cluster was the most probable, i.e. a little less than 1\% of the test data. The clusters on average contain $1.6$ conditions with high probability, i.e. $\bar{\pi}_{m,k}>0.5$. The mean posterior probability of a LTC being present in a cluster, i.e. $\bar{\pi}_{m,k}$ is shown graphically in Figure \ref{fig:bubbleUKB} in Appendix \ref{bubbleplot}. Typically, smaller clusters are associated with fewer high probability conditions. This is expected behaviour, since the presence of multiple LTCs provides more information to the model and hence individuals can be assigned to clusters with multiple high probability conditions with greater certainty.

We can identify several clusters containing high probability of only one condition, such as clusters $22$, $32$, and $38$ which contain only \textsc{hypertension} (onset: $47 \pm 12$ years), \textsc{osteoarthritis} (onset: $53 \pm 10$ years), and \textsc{allergies and chronic rhinitis} (onset: $34 \pm 12$ years) respectively. These clusters indicate LTCs which frequently are the only condition accrued by an individual by the end of follow up. We further identify clusters containing several conditions which are known to appear together often, such as clusters $7$, $25$, and $31$. In cluster 7 we see high probabilities of \textsc{Anxiety disorders} (onset: $60 \pm 11$ years) and \textsc{Depression}(onset: $57 \pm 12$ years); in cluster 25 we see high probabilities of \textsc{Hypertension} (onset: $44 \pm 8$ years) and \textsc{Type 2 Diabetes}(onset: $49 \pm 8$ years); and in cluster 31 we see high probabilities of \textsc{atrial fibrillation} (onset: $54 \pm 8$ years), \textsc{coronary heart disease} (onset: $52 \pm 8$ years), \textsc{heart valve disorders} (onset: $46 \pm 10$ years), and \textsc{hypertension} (onset: $48 \pm 9$ years).
\begin{figure*}[t!]
\centering
{\includegraphics[width=0.49\textwidth]{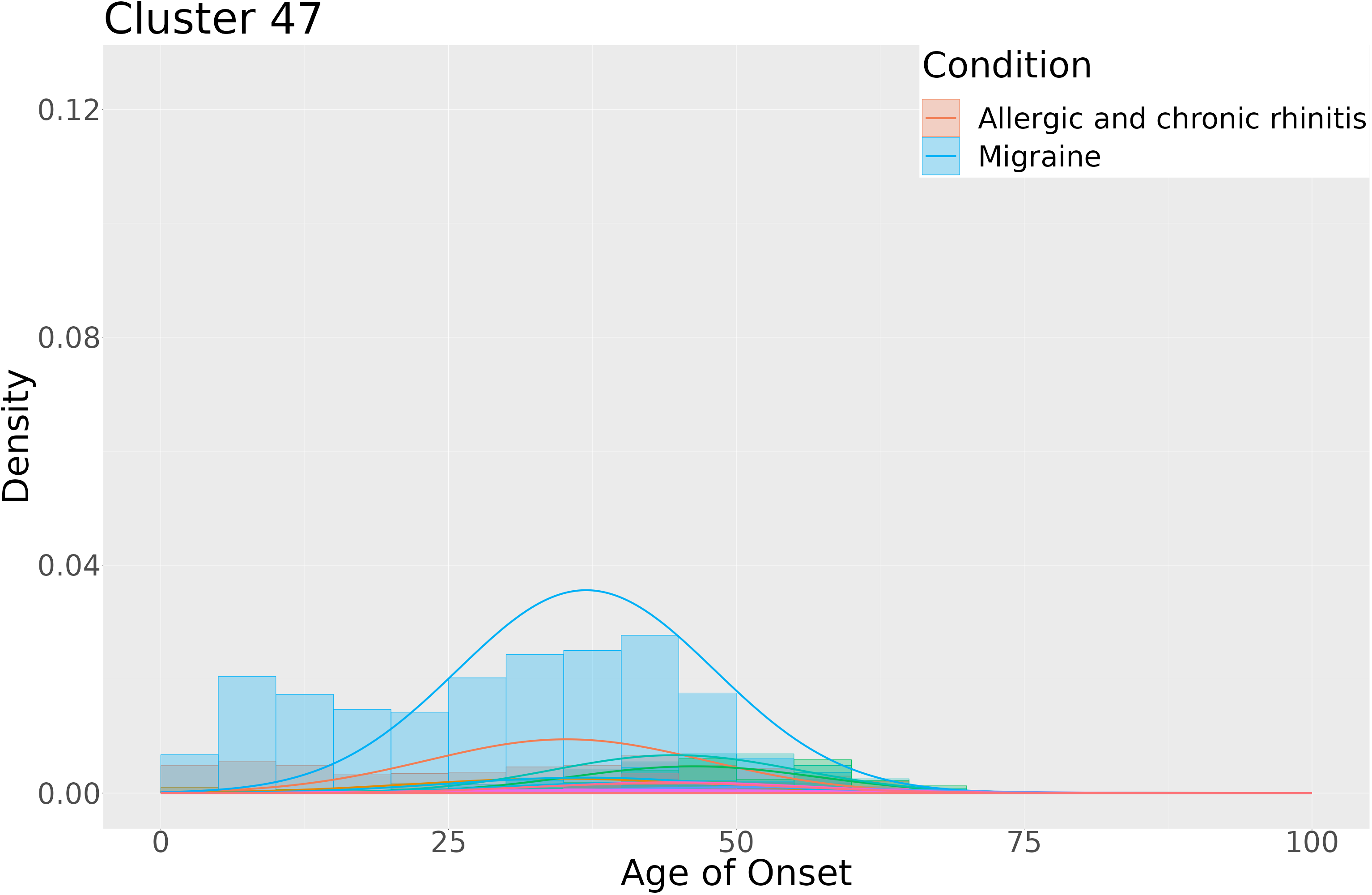}}
{\includegraphics[width=0.49\textwidth]{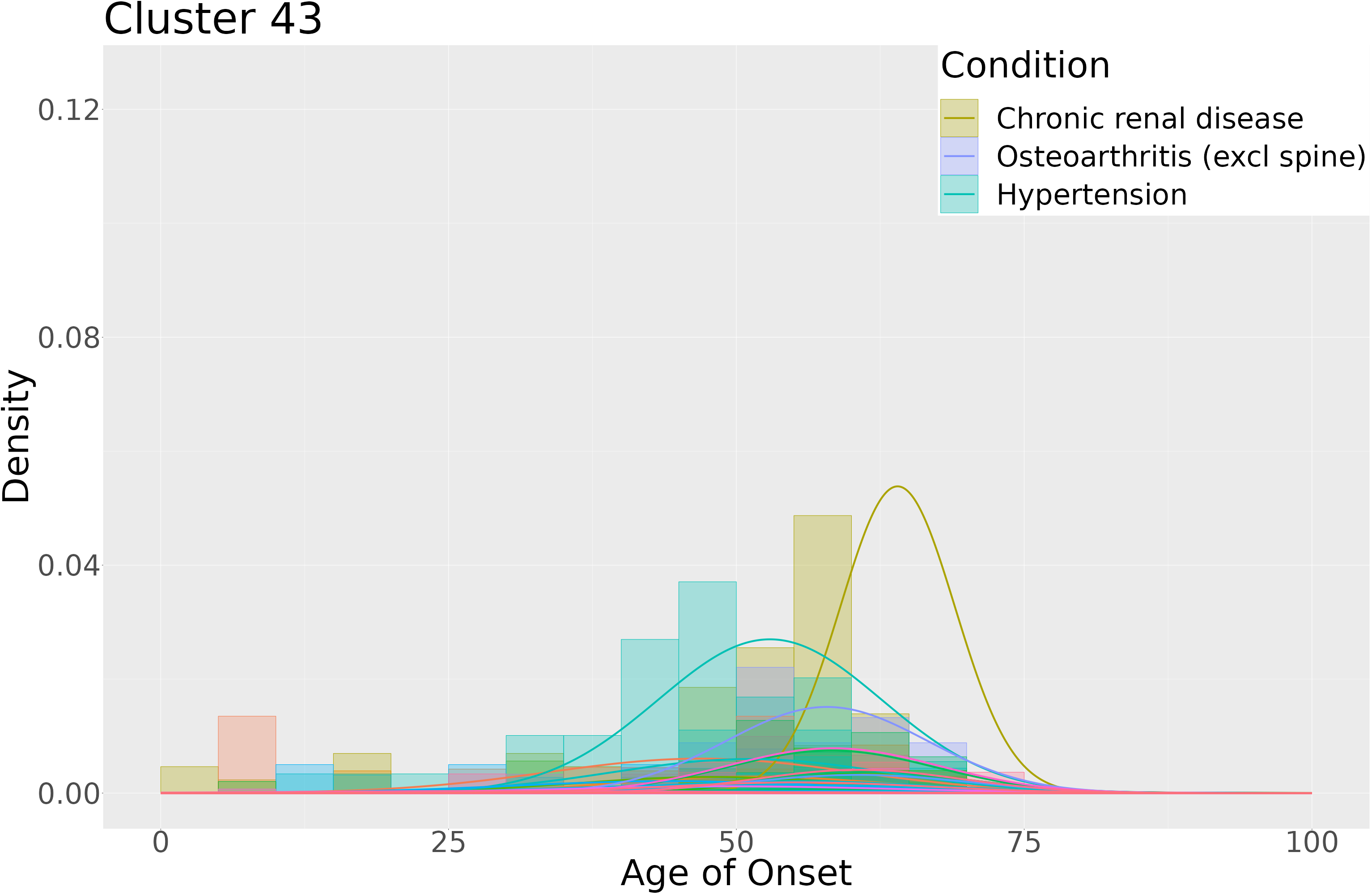}}
\caption{The histograms show empirical distributions of the observed diagnosis ages. The lines show the estimated distributions of diagnosis ages within the same cluster. All conditions are shown but the legend shows only conditions with total risk of 0.25 or higher for space. The full legend can be found in Appendix \ref{alltempprofs}.}
\label{fig:bardens}
\end{figure*}

We can identify when an individual who belongs to a cluster will be most vulnerable to a given condition. In Figure~\ref{fig:cumulUKB}, we show an example of the cumulative risk 
of \textsc{COPD} over time within several clusters and the average cumulative risk over population for comparison. We observe that different trajectories belonging to different clusters can have both different total lifetime risks (i.e. $\bar{\pi}_{m,k}$) and different periods in which the risk accumulates. For instance, we observe that Cluster $13$  accumulates risk from $50$ to $75$ years old. This risk is accumulated quickly, reaching a total lifetime risk of approximately $0.7$. Cluster $6$ on the other hand accumulates risk earlier, between the ages of $30$ and $63$ but accumulates a much lower total lifetime risk of approximately $0.2$. These clusters particularly contrast with Cluster $3$ which also accumulates a total lifetime risk of approximately $0.2$ but from $50$ to $75$ years of age.

\paragraph{Effect of censoring} We compare histograms of the empirical ages at diagnosis from the full test data with density plots of fitted ages at diagnosis for particular clusters in Figure~\ref{fig:bardens}. The full set of temporal risk profiles can be found in Appendix \ref{alltempprofs}. We observe in Figure~\ref{fig:bardens}(left) the impact of left censoring on the temporal risk profile. Most of the data used to construct the empirical temporal risk profile for \textsc{migraines} (onset: $37 \pm 10$ years) in cluster 47 are left censored. Hence the model constructs a temporal risk profile using the information that the condition is likely to occur before baseline, and shows large uncertainty about exactly when. In contrast Figure~\ref{fig:bardens}(right) shows an estimated temporal risk profile for \textsc{chronic renal disease} (onset: $65 \pm 5$ years) in cluster 43 where the model estimates heavy right-censoring. We observe that for this cluster the risk estimated by the model continues to remain high after the latest empirical onset data, and hence the model estimates that there is right-censoring present in this cluster.

\begin{figure*}[t!]
\centering
{\includegraphics[width=0.49\textwidth]{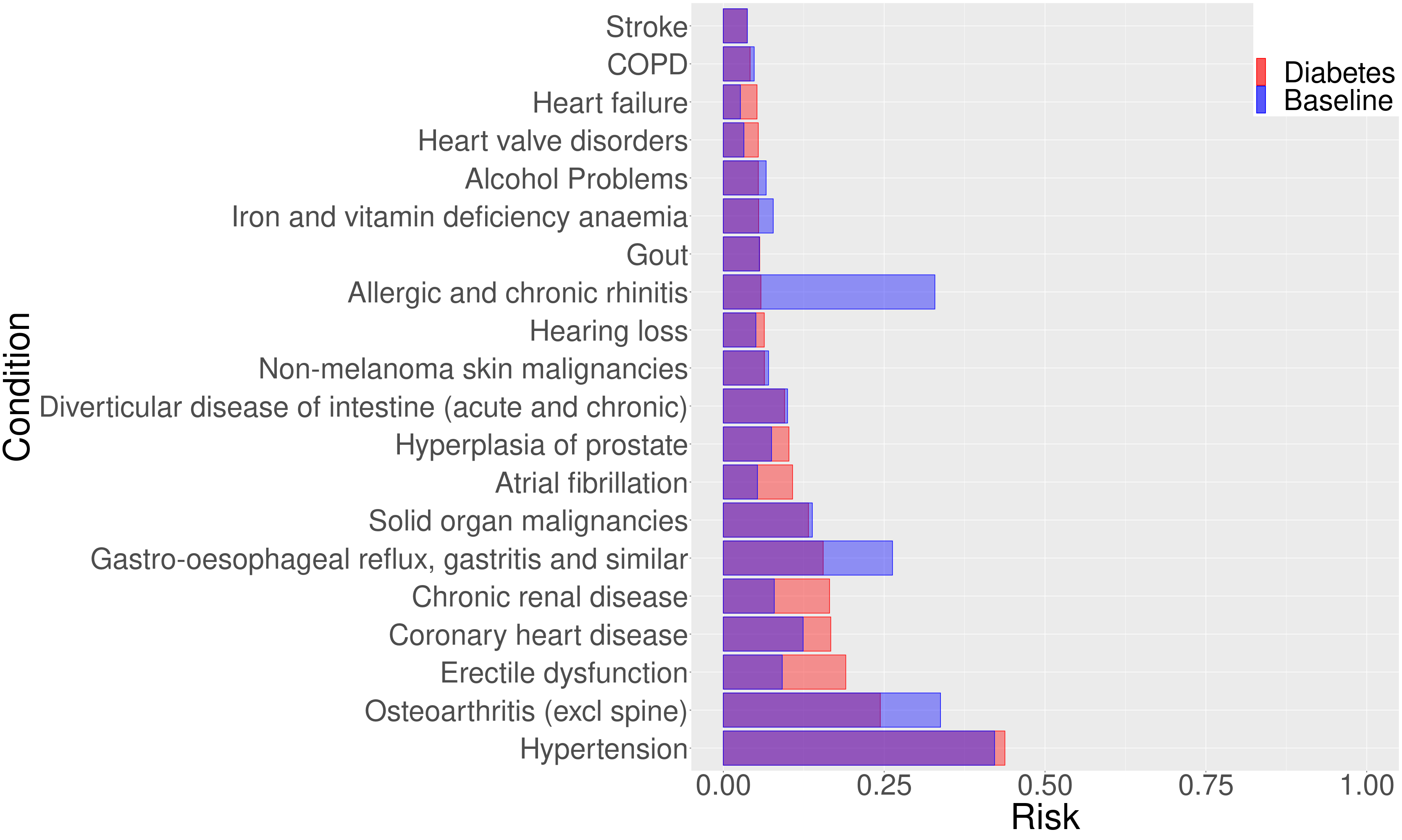}}
{\includegraphics[width=0.49\textwidth]{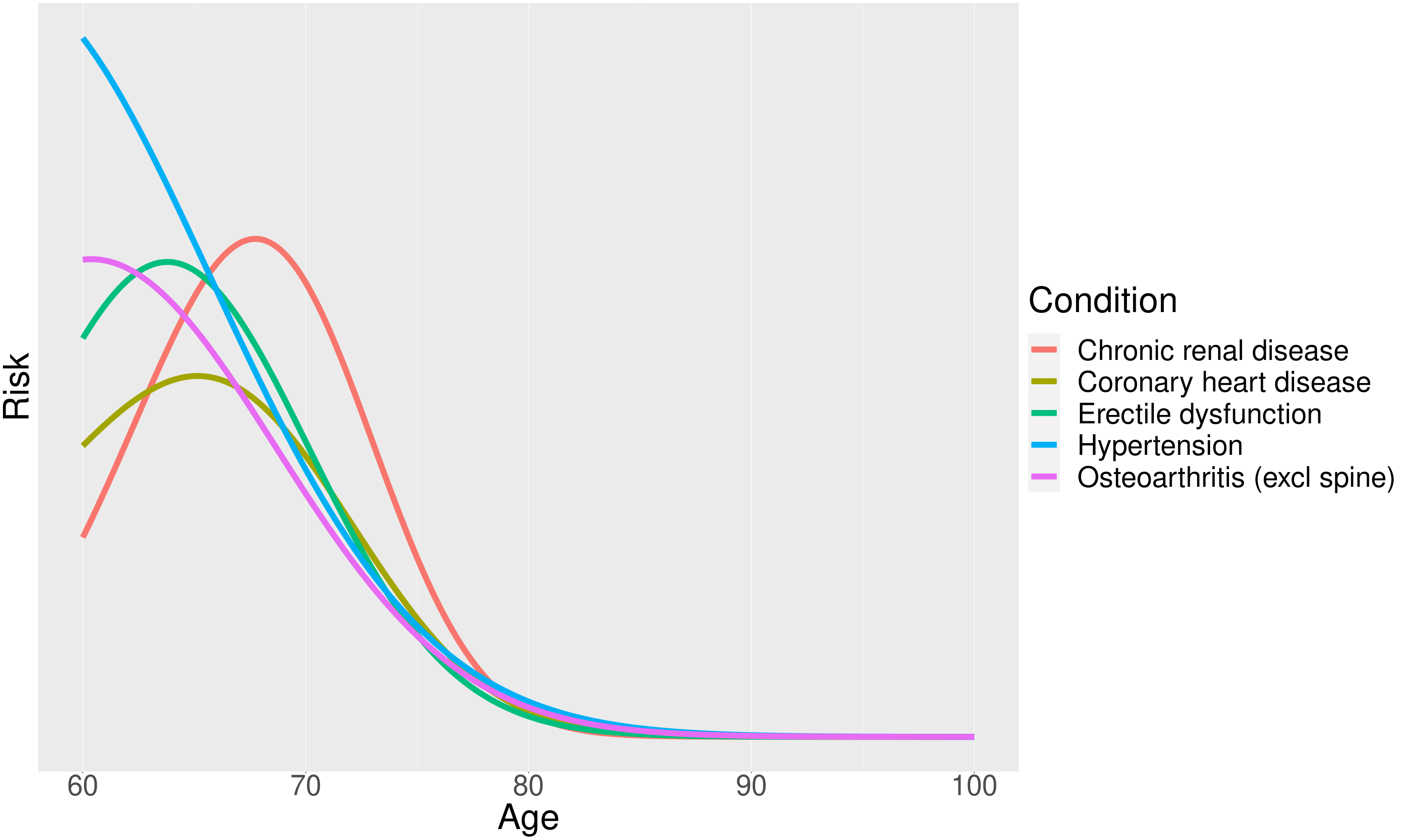}}
\caption{The risk profile for a hypothetical 60 year old man who was diagnosed with diabetes at 54 and no other conditions. The population presence is provided as a baseline. \textbf{(left)} Total rest of life risk of the twenty highest risk LTCs. \textbf{(right)} The risk over time of the five highest total risk conditions}
\label{fig:diabetesriskprofUKB}
\end{figure*}

\paragraph{Trajectory prediction} Given history of disease accrual for an individual we can forecast their future trajectory and build a risk profile which considers both rest of life total risk and also risks over different periods of time. Figure~\ref{fig:diabetesriskprofUKB}(left) presents the total risk of the highest risk conditions for a hypothetical $60$ year old man who was diagnosed with \textsc{type 2 diabetes} at $54$ years of age who has no other conditions. We observe that the highest risk conditions include both conditions which have high prevalence in population (show as baseline) such as \textsc{hypertension} and \textsc{osteoarthritis}; and also conditions which are known to co-occur with \textsc{type 2 diabetes} such as \textsc{chronic renal disease} \citep{gheith2016diabetic} and \textsc{coronary heart disease} \citep{petrie2018diabetes}.

The associated temporal risk for the hypothetical individual is  presented in Figure~\ref{fig:diabetesriskprofUKB}(right). We observe that some conditions such as \textsc{hypertension} accumulate risk immediately, indicating that the individual is currently at high risk. This is in contrast to some other conditions such as \textsc{chronic renal disease} which enter the high risk period later. This matches clinical understanding of these conditions, newly diagnosed diabetes leads to blood pressure checks and also \textsc{hypertension} shares many of the same risk factors as \textsc{type 2 diabetes} \citep{petrie2018diabetes}. In contrast, \textsc{chronic renal disease} is a complication of \textsc{type 2 diabetes} and, when it occurs, it is usually diagnosed $10-20$ years after a \textsc{type 2 diabetes} diagnosis \citep{gheith2016diabetic}. Moreover, we compared the estimated temporal risk profile using $50$ clusters with the profile using $100$ clusters and found similar results, indicating some robustness to the choice of $K$. The plots for this comparison are available in Appendix \ref{kselect}.

\paragraph{Predictive accuracy} Evaluated on the withheld UK Biobank test data, {\promote} demonstrates strong predictive performance. For each individual in the test data we use the data up until the last ten years to predict forward for those ten years. We then predict disease presence or absence and if exists then the respective age at diagnosis within those $10$ years and compare the predictions to the observed data. The disease presence predictions were adjusted to the ten year period and perform well, with an $0.86$ AUROC for the withheld disease predictions within those $10$ years indicating that positive and negative cases can be separated with high probability. Furthermore, taking the maximum a posteriori (MAP) predictions we observe $9.97$ MAE when compared with the observed age at diagnosis, indicating a $9.97$ year average difference between the predicted and observed diagnosis ages.

\section{Discussion}

We present a novel method, {\promote}, for longitudinal clustering of LTCs based upon both their presence and absence and their respective age at diagnosis. {\promote} can make both population and individual level inference, by finding clusters of trajectories and forecasting future individual trajectories as a predictive mixture of the underlying clusters. At the population level, the clusters can be used to inform genetic and socio-economic drivers while at the individual level, the expected future LTC trajectories can be used for clinical decision support (CDS).

A direct comparison of {\promote} and contemporary methods is difficult since they are applied on different datasets. However, 
in comparison with the ATM method \citep{jiang_age-dependent_2023} that also used UK Biobank data, we describe some differences. The ATM model was fitted on the trajectories of $282,957$ individuals compared to $159,145$ individuals used for {\promote}, and with different disease encoding of the conditions.
ATM focuses on the low-rank representation of EHRs through clustering. This contrasts with the approach of {\promote} which focuses on the integration of unreliable and incomplete EHR data for the clustering of individuals and the forecasting of future trajectories. Thus, we emphasize that the clusters are not directly comparable.

While the model introduces meaningful and useful inference we note that there are several limitations. The model makes several parametric choices when modelling the data. In particular, the parametric choice for the distribution of diagnosis ages can impact the model's ability to make useful forecasts at the individual level. However, our results allow any parametric choice within the exponential family, which affords the model considerable flexibility where appropriate care is taken. Moreover, our model attempts to mitigate the unreliability and incompleteness of the data by introducing left- and right-censoring but these are subject to parametric choices and may not fully remedy the problem. Within a cluster our model assumes conditional independence of LTC onset times, and while this greatly improves over a fully independent structure, our model does not capture the full dependence within the data. We also emphasize that the model does not attempt to infer causality.

Future work will attempt to address the model limitations and further improve inference. For example,  generalisation of the Dirichlet prior to a Dirichlet process prior will enable the model to automatically determine an appropriate number of clusters from the data \citep{ferguson1973bayesian}. Furthermore, allowing uncertainty arising from the age at diagnosis being different from actual disease onset that is not observed, will introduce considerable flexibility. Moreover, incorporating uncertainty about the time of death by potentially treating death as a condition where no other conditions can occur after may improve predictive performance. Other improvements to the method could include modelling the dependency between different LTC onset times within a cluster. Finally replicating {\promote} on additional data will add more value.

\section*{Acknowledgements}
This study is funded by the National Institute for Health and Care Research (NIHR) under its Artificial Intelligence for Multiple and Long-Term Conditions Programme (project references NIHR203982 and NIHR202639). The views expressed are those of the author and not necessarily those of the NIHR or the Department of Health and Social Care.

The study was conducted using the UK Biobank Resource under application number 57213. The authors would like to thank the UK Biobank participants and the UK Biobank staff for their contributions to this study.

Nick J. Reynolds' laboratory is funded in part by the NIHR Newcastle Biomedical Research Centre, the NIHR Newcastle HealthTech Research Centre in Diagnostic and Technology Evaluation and the NIHR Newcastle Patient Safety Research Collaboration.  Nick J. Reynolds is a NIHR Senior Investigator.

\printbibliography

\newpage
\appendix
\onecolumn

\newpage

\section{Table of Notation}\label{tableofnotation}

\begin{longtable}{|p{0.1\textwidth}|p{0.8\textwidth}|}
\hline
\textbf{Notation} & \textbf{Description}  \\ \hline
$\bfPhi$ & The data \\ \hline
$\bfd$ & A binary vector indicating presence/absence of the LTCs \\ \hline
$\bft$ & A vector of diagnosis ages of the LTCs \\ \hline
$\bfkappa$ & A vector indicating the degree of censoring affecting each condition \\ \hline
$N$ & The number of individuals \\ \hline
$M$ & The number of conditions \\ \hline
$K$ & The number of clusters \\ \hline
$\iota$ & An indicator variable which indicates whether an individual is alive or dead \\ \hline
$\rho$ & The baseline age \\ \hline
$\tau$ & The age at data extraction \\ \hline
$z$ & An indicator vector representing the cluster label\\ \hline
$\gamma$ & The model parameter for the multinomial distribution describing the clusters \\ \hline
$\theta$ & The hyperparameter for the dirichlet prior on $\gamma$ \\ \hline
$\pi$ & The model parameter for the Bernoulli distribution describing the presence/absence of a condition \\ \hline
$a$ and $b$ & The hyperparameters for the beta prior on $\pi$ \\ \hline
$\bfeta$ & The model parameters for the exponential family distribution describing the diagnosis ages \\ \hline
$\nu$ and $\bfchi$ & The hyperparameters for the conjugate prior on $\bfeta$ \\ \hline
$T$ & Sufficient statistics describing the diagnosis ages \\ \hline
$\theta^\ast$ & The variational parameters for the variational distribution describing $\gamma$ \\ \hline
$a^\ast$ and $b^\ast$ & The variational parameters for the variational distribution describing $\pi$ \\ \hline
$\nu^\ast$ and $\bfchi^\ast$ & The variational parameters for the variational distribution describing $\bfeta$ \\ \hline
$r^\ast$& The variational parameter for the variational distribution describing $z$ \\ \hline
$\bar{z}$, $\bar{n}$, $\bar{T}$, $\zeta$, $\lambda$, and $u$ & Intermediate quantities useful for simplifying the variational parameters\\ \hline
$\pi^\ast$& The variational parameter for the variational distribution describing $d$ \\ \hline
$\bfeta^\ast$& The variational parameters for the variational distribution describing $t$ \\ \hline
$\Phi^\prime$ & Data describing a new individual \\ \hline
$\bfd^\prime$ & A binary vector indicating presence/absence of the LTCs in a new individual\\ \hline
$\bft^\prime$ & A vector of diagnosis ages of the LTC for a new individual \\ \hline
$\bfkappa^\prime$ & A vector indicating the degree of censoring affecting each condition for a new individual \\ \hline
$\rho^\prime$ & The baseline age for a new individual \\ \hline
$\tau^\prime$ & The age at data extraction for a new individual\\ \hline
$\bar{\theta}$ & The parameters for the predictive posterior distribution describing the clusters \\ \hline
$\bar{\pi}$ & The parameter for the predictive posterior distribution describing the presence/absence of a condition \\ \hline
$\bar{\eta}$ & The parameters for the predictive posterior distribution describing the diagnosis ages of a condition \\ \hline
$\tilde{\pi}$ & The parameter for the predictive posterior distribution describing the presence/absence of a condition given the condition has not occurred by time $\tau^\prime$ \\ \hline
\end{longtable}

\newpage

\section{Disease Cluster Prevalence}\label{bubbleplot}
\begin{figure*}[!ht]
\begin{center}
\includegraphics[height=0.85\textheight]{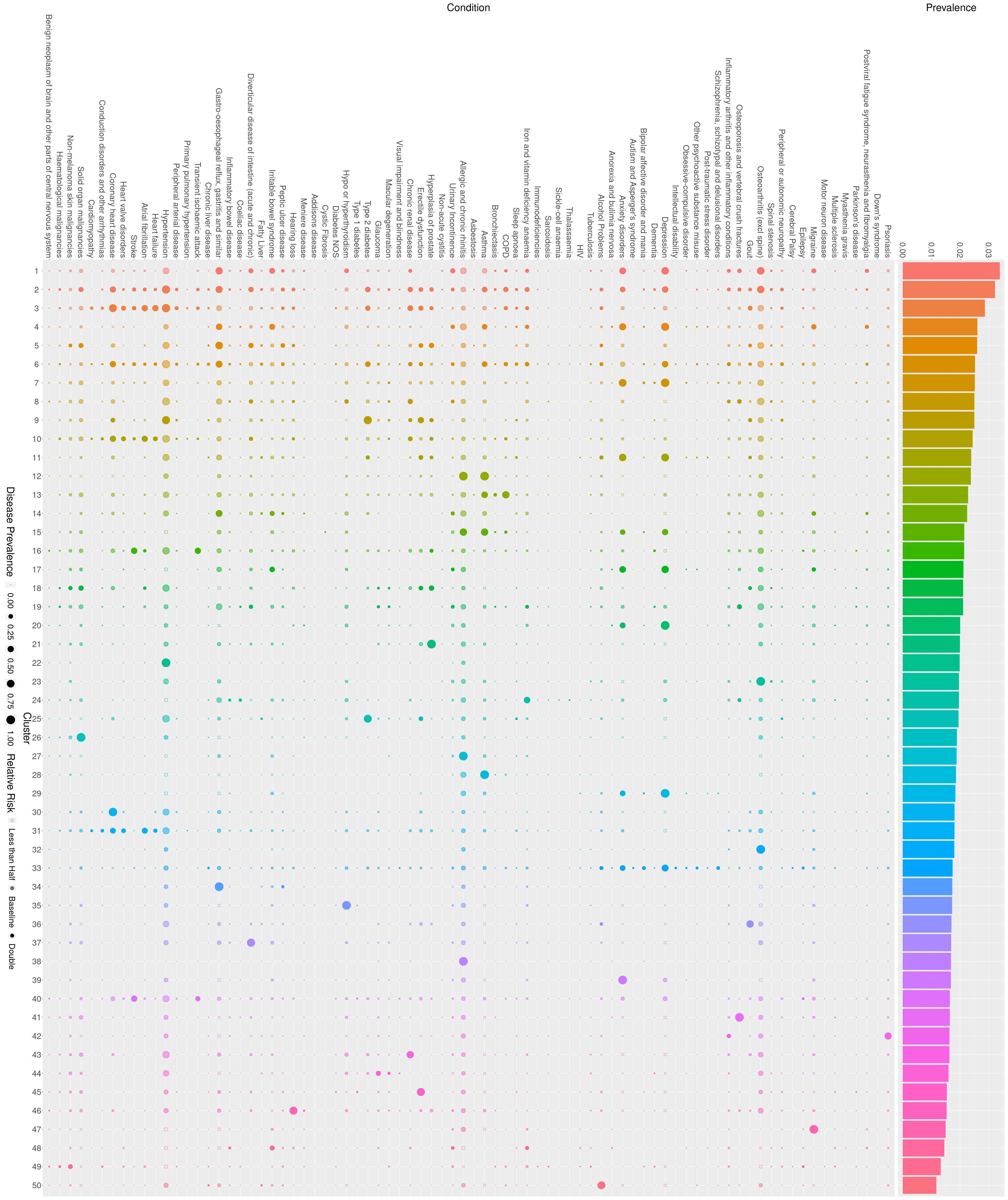}
\end{center}
\caption{We show posterior disease presence probabilities for each condition in each cluster. The log relative risk of each condition in a cluster relative to the same condition in the whole population is shown as the opacity of the bubbles. The clusters are ordered in descending order of prevalence in the test data, with the prevalence shown in the bar chart to the right of the bubble plot.}
\label{fig:bubbleUKB}
\end{figure*}

\newpage

\section{Derivation of Variational Bayes}\label{DeriveVB}

We define 
\begin{alignat}{1}
\bar{z}_{k} &= \sum_{n=1}^N \E_q\left(z_{k}^{(n)}\right)\nonumber\\
&= \sum_{n=1}^N \gamma_{k}^{(n)\ast}\\
\bar{n}_{m,k} &=\sum_{n=1}^{N} \E_q\left(d_{m}^{\left(n\right)}\right)\E_q\left(z_{k}^{\left(n\right)}\right)\nonumber\\
&= \sum_{n:\kappa_{m}^{(n)}={\obs}} d_{m}^{\left(n\right)}\E_q\left(z_{k}^{\left(n\right)}\right) + \sum_{n:\kappa_{m}^{(n)}={\part}} \E_q\left(z_{k}^{\left(n\right)}\right) + \sum_{n:\kappa_{m}^{(n)}={\unobs}} \E_q\left(d_{m}^{\left(n\right)}\right)\E_q\left(z_{k}^{\left(n\right)}\right)\nonumber\\
& = \sum_{n:\kappa_{m}^{(n)}={\obs}} d_{m}^{\left(n\right)}\bfgamma_{k}^{(n)\ast} + \sum_{n:\kappa_{m}^{(n)}={\part}} \bfgamma_{k}^{(n)\ast} + \sum_{n:\kappa_{m}^{(n)}={\unobs}} \pi_{m}^{(n)\ast}\bfgamma_{k}^{(n)\ast}\\
\bar{\bfT}_{m,k} &= \sum_{n : \kappa_{m}^{(n)} = {\obs}}\bfgamma_{k}^{(n)\ast}d_{m}^{(n)}\E\left(\suffstat_m^{(n)}\right)+\sum_{n : \kappa_{m}^{(n)} = {\part}}\bfgamma_{k}^{(n)\ast}\E\left(\suffstat_m^{(n)}\right)\nonumber\\
& \qquad\qquad + \sum_{n : \kappa_{m}^{(n)} = {\unobs}}\bfgamma_{k}^{(n)\ast}\pi_{m}^{(n)\ast}\E\left(\suffstat_m^{(n)} \g d_{m}^{(n)} = 1\right)
\end{alignat}

and note that for $m : \kappa_{m}^{(n)} = {\unobs}$,
\begin{alignat}{1}
\E_q\left(d_{m}^{\left(n\right)}\suffstat_{m}^{(n)}\right) & = \sum_{d=0}^{1}\int d_{m}^{(n)} \suffstat_m^{(n)}q\left(d_{m}^{(n)}, t_{m}^{(n)}\right) \d t \nonumber\\
& = \int \suffstat_m^{(n)}q\left(d_{m}^{(n)}=1, t_{m}^{(n)}\right) \d t + 0\nonumber\\
& = q\left(d_{m}^{(n)} = 1\right) \int \suffstat_m^{(n)}q\left(t_{m}^{(n)} \g d_{m}^{(n)} = 1\right) \d t\nonumber\\
& = \E_q\left(d_{m}^{\left(n\right)}\right)\E_q\left(\suffstat_{m}^{(n)}\mid d_{m}^{\left(n\right)}=1\right)\nonumber\\
& = \pi_{m}^{(n)\ast}\E_q\left(\suffstat_{m}^{(n)}\mid d_{m}^{\left(n\right)}=1\right).
\end{alignat}

We can write the full likelihood up to a constant of proportionality in a separable form:
\begin{alignat}{1}
p(\Phi,\bfZ, \bfG, \bfP, \bfE \g \bfK) = &\prod_{n=1}^N \left[ p(\bfz^{(n)} \g \bfgamma) \prod_{m:\kappa^{(n)}_m={\obs}} p(d^{(n)}_m,t^{(n)}_m \g \bfP, \bfE, \bfz^{(n)}) \right. \nonumber\\
& \qquad \prod_{m:\kappa^{(n)}_m={\part}} p(d^{(n)}_m,t^{(n)}_m \g \bfP, \bfE, \bfz^{(n)}) \nonumber\\
&\qquad \left. \prod_{m:\kappa^{(n)}_m={\unobs}} p(d^{(n)}_m,t^{(n)}_m \g \bfP, \bfE, \bfz^{(n)})  \right] \nonumber\\
&p(\bfgamma \g \bftheta) \prod_{k=1}^K\prod_{m=1}^M p(\bfpi_{m,k} \g \alpha_{m,k}, \beta_{m,k}) \prod_{k=1}^K\prod_{m=1}^M p(\bfeta_{m,k} \g \nu_{m,k}, \bfchi_{m,k})\nonumber\\
\propto & \prod_{k=1}^{K}\left[\gamma_{k}^{\theta_{k}-1}\prod_{n=1}^{N}\gamma_{k}^{z_{k}^{\left(n\right)}}\right]\times\nonumber \\
 & \text{ }\prod_{k=1}^{K}\left[\prod_{m=1}^{M}\left[\pi_{m,k}^{a_{m,k}-1}\left(1-\pi_{m,k}\right)^{b_{m,k}-1}\prod_{n:\kappa_{m}^{(n)}={\obs}}\pi_{m,k}^{d_{m}^{\left(n\right)}z_{k}^{\left(n\right)}}\left(1-\pi_{m,k}\right)^{\left(1-d_{m}^{\left(n\right)}\right)z_{k}^{\left(n\right)}}\right.\right.\nonumber \\
 & \text{ }\left.\left.\prod_{n:\kappa_{m}^{(n)}={\part}}\pi_{m,k}^{d_{m}^{\left(n\right)}z_{k}^{\left(n\right)}}\left(1-\pi_{m,k}\right)^{\left(1-d_{m}^{\left(n\right)}\right)z_{k}^{\left(n\right)}}\prod_{n:\kappa_{m}^{(n)}={\unobs}}\pi_{m,k}^{d_{m}^{\left(n\right)}z_{k}^{\left(n\right)}}\left(1-\pi_{m,k}\right)^{\left(1-d_{m}^{\left(n\right)}\right)z_{k}^{\left(n\right)}}\right]\right]\times\nonumber\\
 & \text{ }\prod_{k=1}^{K}\left[\prod_{m=1}^{M}\prod_{n:\kappa_{m}^{(n)}={\obs}}f_\expfam\left(t^{(n)}_{m}\g\bfeta_{m,k} \right)^{d_{m}^{\left(n\right)}z_{k}^{\left(n\right)}}\prod_{n:\kappa_{m}^{(n)}={\part}}f_{\expfamt_r}\left(t^{(n)}_{m}\g\bfeta_{m,k} \right)^{d_{m}^{\left(n\right)}z_{k}^{\left(n\right)}}\right.\nonumber\\
 & \text{ }\left.\prod_{n:\kappa_{m}^{(n)}={\unobs}}f_{\expfamt_l}\left(t^{(n)}_{m}\g\bfeta_{m,k} \right)^{d_{m}^{\left(n\right)}z_{k}^{\left(n\right)}} \right].\label{eq:postseparable}
\end{alignat}
We define the VB posterior as:
\begin{alignat}{1}
    q\left(\bfgamma, \bfpi, \bfeta, \bfZ, \Phi\right) & \propto q\left(\bfgamma\right)
 \left(\prod_{n=1}^{N}q\left(z^{(n)}\right)\right) \nonumber\\
    & \qquad \left(\prod_{m=1}^{M}\prod_{k=1}^{K}q\left(\pi_{m,k}\right)q\left(\bfeta_{m,k}\right)\right) \left(\prod_{n=1}^{N}\prod_{m : \kappa_{m}^{(n)} = {\part}}q\left(t_{m}^{(n)} \g d_{m}^{(n)} = 1\right)\right) \nonumber\\
    & \qquad \left(\prod_{n=1}^{N}\prod_{m : \kappa_{m}^{(n)} = {\unobs}}q\left(d_{m}^{(n)}, t_{m}^{(n)}\right)\right)\label{eq:postvariational}
\end{alignat}

\paragraph{\textcolor{red}{$\bfgamma$}} So following from \eqref{eq:postseparable} and \eqref{eq:postvariational} we have that:
\begin{alignat}{1}
    q\left(\bfgamma\right) &\propto \exp\left(\E_{-\gamma_{k}}\left(\log\left(\frac{\Gamma\left(\sum_{k=1}^{K}\theta_{k}\right)}{\prod_{k=1}^{K}\Gamma\left(\theta_{k}\right)}\prod_{k=1}^{K}\left[\gamma_{k}^{\theta_{k}-1}\prod_{n=1}^{N}\gamma_{k}^{z_{k}^{(n)}}\right]\right)\right)\right)\nonumber\\
    & \propto \frac{\Gamma\left(\sum_{k=1}^{K}\theta_{k}\right)}{\prod_{k=1}^{K}\Gamma\left(\theta_{k}\right)}\exp\left(\E_{-\gamma_{k}}\left(\log\left(\prod_{k=1}^{K}\left[\gamma_{k}^{\theta_{k}-1}\prod_{n=1}^{N}\gamma_{k}^{z_{k}^{(n)}}\right]\right)\right)\right) \nonumber\\
    & \propto \exp\left(\E_{-\gamma_{k}}\left(\sum_{k=1}^{K}\log\left(\gamma_{k}^{\theta_{k} + \sum_{n=1}^{N}z_{k}^{(n)} - 1}\right)\right)\right) \nonumber\\
    & \propto \prod_{k=1}^{K}\exp\left(\left(\theta_{k} + \sum_{n=1}^{N}\E_q\left(z_{k}^{(n)}\right) - 1\right)\log\left(\gamma_{k}\right)\right) \nonumber\\
    & \propto \prod_{k=1}^{K}\gamma_{k}^{\theta_{k} + \sum_{n=1}^{N}\bfgamma_{k}^{(n)\ast} - 1}
\end{alignat}
from which we can identify 
\begin{equation}
    \bfgamma \distas \Dirichlet\left(\theta^{*}\right)
\end{equation} 
where 

\begin{alignat}{1}
    \theta_{k}^{*} &= \theta_{k} + \bar{z}_k\\
    \bar{z}_k &= \sum_{n=1}^{N}\E_q\left(z_{k}^{(n)}\right)
\end{alignat} 

\paragraph{\textcolor{red}{$\pi_{m,k}$}}
We also have that:
\begin{alignat}{1}
    q\left(\pi_{m,k}\right) & \propto \exp\left(\E_{-\pi_{m,k}}\left(\log\left(\pi_{m,k}^{a_{m,k}-1}\left(1-\pi_{m,k}\right)^{b_{m,k}-1}\prod_{n : \kappa_{m}^{(n)} = {\obs}}\pi_{m,k}^{d_{m}^{\left(n\right)}z_{k}^{\left(n\right)}}\left(1-\pi_{m,k}\right)^{\left(1-d_{m}^{\left(n\right)}\right)z_{k}^{\left(n\right)}}\right.\right.\right. \nonumber\\
    & \qquad\qquad \left.\left.\left.\prod_{n : \kappa_{m}^{(n)} = {\part}}\pi_{m,k}^{d_{m}^{\left(n\right)}z_{k}^{\left(n\right)}}\left(1-\pi_{m,k}\right)^{\left(1-d_{m}^{\left(n\right)}\right)z_{k}^{\left(n\right)}}\prod_{n : \kappa_{m}^{(n)} = {\unobs}}\pi_{m,k}^{d_{m}^{\left(n\right)}z_{k}^{\left(n\right)}}\left(1-\pi_{m,k}\right)^{\left(1-d_{m}^{\left(n\right)}\right)z_{k}^{\left(n\right)}}\right)\right)\right)\nonumber\\
    & \propto \exp\Bigg(\left(a_{m,k} + \bar{n}_{m,k} - 1\right)\log\left(\pi_{m,k}\right) + \nonumber\\
    &\left.\qquad\qquad\left(b_{m,k} + \sum_{n=1}^{N} \E_{q}\left(z_{k}^{(n)}\right) - \bar{n}_{m,k} - 1\right)\log\left(1 - \pi_{m,k}\right)\right)\nonumber\\
    & \propto \pi_{m,k}^{a_{m,k}^{*}-1}\left(1-\pi_{m,k}\right)^{b_{m,k}^{*}-1}
\end{alignat}
for 
\begin{alignat}{1}
    a_{m,k}^{*} & =a_{m,k}+\bar{n}_{m,k};\\
b_{m,k}^{*} & =b_{m,k}+\bar{z}_k - \bar{n}_{m,k}
\end{alignat}
and we identify:
\begin{equation}
    \pi_{m,k} \distas \betarand\left(a_{m,k}^{*}, b_{m,k}^{*}\right)
\end{equation}

\paragraph{\textcolor{red}{$\bfeta_{m,k}$}}Next we have that

\begin{alignat}{1}
    q\left(\bfeta_{m,k}\right) & \propto \exp\left(\E_{-\bfeta_{m,k}}\left(\log\left(\prod_{n : \kappa_{m}^{(n)} = {\obs}}f_\expfam\left(t^{(n)}_{m}\g\bfeta_{m,k} \right)^{d_{m}^{\left(n\right)}z_{k}^{\left(n\right)}}\right.\right.\right. \nonumber\\
    & \left.\left.\left.\qquad \qquad\prod_{n : \kappa_{m}^{(n)} = {\part}}f_{\expfamt_r}\left(t^{(n)}_{m}\g\bfeta_{m,k} \right)^{d_{m}^{\left(n\right)}z_{k}^{\left(n\right)}}\prod_{n : \kappa_{m}^{(n)} = {\unobs}}f_{\expfamt_l}\left(t^{(n)}_{m}\g\bfeta_{m,k} \right)^{d_{m}^{\left(n\right)}z_{k}^{\left(n\right)}}\right)\right)\right) \nonumber\\
    & \propto \exp\Bigg(\E_{-\bfeta_{m,k}}\Bigg(\log\Bigg(g\left(\bfeta_{m,k}\right)^{\nu_{m,k}}\exp\left(\bfeta_{m,k}^\top\bfchi_{m,k}\right)\nonumber\\
    &\left.\left.\left.\qquad\qquad\prod_{n : \kappa_{m}^{(n)} = {\obs}}\left(g\left(\bfeta_{m,k}\right)\exp\left(\left(\bfeta_{m,k}\right)^\top \suffstat_m^{(n)}\right)\right)^{d_{m}^{(n)}z_{k}^{(n)}}\right.\right.\right.\nonumber\\
    & \qquad\qquad \prod_{n : \kappa_{m}^{(n)} = {\part}}\left(g\left(\bfeta_{m,k}\right)\exp\left(\left(\bfeta_{m,k}\right)^\top \suffstat_m^{(n)}\right)\right)^{d_{m}^{(n)}z_{k}^{(n)}}\nonumber\\
    & \qquad \qquad\left.\left.\left.\prod_{n : \kappa_{m}^{(n)} = {\unobs}}\left(g\left(\bfeta_{m,k}\right)\exp\left(\left(\bfeta_{m,k}\right)^\top \suffstat_m^{(n)}\right)\right)^{d_{m}^{(n)}z_{k}^{(n)}}\right)\right)\right)\nonumber\\
    & \propto \exp\left(\left(\nu_{m,k}+\bar{n}_{m,k}\right)\log\left(g\left(\bfeta_{m,k}\right)\right)+\bfeta_{m,k}^\top\left(\bfchi_{m,k}+\sum_{n : \kappa_{m}^{(n)} = {\obs}}\bfgamma_{k}^{(n)\ast}d_{m}^{(n)}\suffstat_m^{(n)}\right.\right.\nonumber\\
    &  \qquad\qquad \left.\left. +\sum_{n : \kappa_{m}^{(n)} = {\part}}\bfgamma_{k}^{(n)\ast}d_{m}^{(n)}\E\left(\suffstat_m^{(n)}\right) + \sum_{n : \kappa_{m}^{(n)} = {\obs}}\bfgamma_{k}^{(n)\ast}\E\left(d_{m}^{(n)}\suffstat_m^{(n)}\right)\right)\right) \nonumber\\
    & \propto g\left(\bfeta_{m,k}\right)^{\nu_{m,k}^{*}}\exp\left(\bfeta_{m,k}^\top\bfchi_{m,k}^{*}\right)
\end{alignat}
for
\begin{alignat}{1}
    \nu_{m,k}^{*} & =\nu_{m,k}+\bar{n}_{m,k};\\
\bfchi_{m,k}^{*} & =\bfchi_{m,k} + \bar{\bfT}_{m,k};
\end{alignat}
and we identify:
\begin{equation}
    \bfeta_{m,k} \distas \expfamc\left(\nu_{m,k}^{*},\bfchi_{m,k}^{*}\right)
\end{equation}

\paragraph{\textcolor{red}{$\bfz^{(n)}$}}Next we have that:
\begin{alignat}{1}
q\left(\bfz^{(n)}\right) & \propto \exp\left(\E_q\left(\log\left(\prod_{k=1}^{K}\left[\gamma_{k}\prod_{m:\kappa_{m}^{(n)}={\obs}}^{M}\left[\left(1-\pi_{m,k}\right)^{\left(1-d_{m}^{\left(n\right)}\right)}\left(\pi_{m,k}\prod_{n=1}^{N}f_{\expfam}\left(t^{(n)}_{m}\g\bfeta_{m,k} \right)\right)^{d_{m}^{\left(n\right)}}\right]\right.\right.\right.\right. \nonumber\\
& \qquad\qquad\qquad\qquad\qquad\qquad\prod_{m:\kappa_{m}^{(n)}={\part}}^{M}\left[\left(1-\pi_{m,k}\right)^{\left(1-d_{m}^{\left(n\right)}\right)}\left(\pi_{m,k}\prod_{n=1}^{N}f_{\expfamt_r}\left(t^{(n)}_{m}\g\bfeta_{m,k} \right)\right)^{d_{m}^{\left(n\right)}}\right]\nonumber\\
& \qquad\qquad\qquad\qquad\qquad\quad\text{ }\left.\left.\left.\left.\prod_{m:\kappa_{m}^{(n)}={\unobs}}^{M}\left[\left(1-\pi_{m,k}\right)^{\left(1-d_{m}^{\left(n\right)}\right)}\left(\pi_{m,k}\prod_{n=1}^{N}f_{\expfamt_l}\left(t^{(n)}_{m}\g\bfeta_{m,k} \right)\right)^{d_{m}^{\left(n\right)}}\right]\right]^{z_{k}^{\left(n\right)}}\right)\right)\right) \nonumber\\
& \propto \exp\left(\sum_{k=1}^{K} z_{k}^{(n)}\left(\E\left(\log\gamma_{k}\right) + \sum_{m : \kappa_{m}^{(n)} = {\obs}}d_{m}^{\left(n\right)}\E\left(\log\pi_{m,k}\right) + \sum_{m : \kappa_{m}^{(n)} = {\part}}d_{m}^{\left(n\right)}\E\left(\log\pi_{m,k}\right) \right.\right.\nonumber \\
& \qquad + \sum_{m : \kappa_{m}^{(n)} = {\unobs}}\E\left(d_{m}^{\left(n\right)}\right)\E\left(\log\pi_{m,k}\right) - \sum_{m : \kappa_{m}^{(n)} = {\obs}}d_{m}^{\left(n\right)}\left(\E\left(\log\left(g\left(\bfeta_{m,k}\right)\right)\right)+ \E\left(\bfeta_{m,k}\right)^\top \suffstat\right)\nonumber\\
& \qquad  - \sum_{m : \kappa_{m}^{(n)} = {\part}}d_{m}^{\left(n\right)}\left(\E\left(\log\left(g\left(\bfeta_{m,k}\right)\right)\right) + \E_{r}\left(\log\left(h\left(t_{m}^{(n)}\right)\right)\right) + \E\left(\bfeta_{m,k}\right)^\top \E_{r}\left(\suffstat\right)\right)\nonumber\\
& \qquad - \sum_{m : \kappa_{m}^{(n)} = {\unobs}}\E\left(d_{m}^{\left(n\right)}\right)\left(\E\left(\log\left(g\left(\bfeta_{m,k}\right)\right)\right) + \E_{l}\left(\log\left(h\left(t_{m}^{(n)}\right)\right)\right) + \E\left(\bfeta_{m,k}\right)^\top \E_{l}\left(\suffstat\right)\right)\nonumber \\
& \qquad + \sum_{m : \kappa_{m}^{(n)} = {\obs}}\left(1 - d_{m}^{\left(n\right)}\right)\E\left(\log\left(1 - \pi_{m,k}\right)\right) + \sum_{m : \kappa_{m}^{(n)} = {\part}}\left(1 - d_{m}^{\left(n\right)}\right)\E\left(\log\left(1 - \pi_{m,k}\right)\right)\nonumber\\
& \qquad \left.\left. + \sum_{m : \kappa_{m}^{(n)} = {\unobs}}\E\left(1 - d_{m}^{\left(n\right)}\right)\E\left(\log\left(1 - \pi_{m,k}\right)\right)\right)\right)\nonumber\\
& \propto \prod_{k=1}^{K} \exp\left(c_{k}\right)^{z_{k}^{(n)}}
\end{alignat}
where 
\begin{alignat}{1}
c_{k} & = \E\left(\log\gamma_{k}\right) + \sum_{m : \kappa_{m}^{(n)} = {\obs}}d_{m}^{\left(n\right)}\E\left(\log\pi_{m,k}\right) + \sum_{m : \kappa_{m}^{(n)} = {\part}}d_{m}^{\left(n\right)}\E\left(\log\pi_{m,k}\right) \nonumber \\
& \qquad + \sum_{m : \kappa_{m}^{(n)} = {\unobs}}\E\left(d_{m}^{\left(n\right)}\right)\E\left(\log\pi_{m,k}\right) - \sum_{m : \kappa_{m}^{(n)} = {\obs}}d_{m}^{\left(n\right)}\left(\E\left(\log\left(g\left(\bfeta_{m,k}\right)\right)\right)+ \E\left(\bfeta_{m,k}\right)^\top \suffstat\right)\nonumber\\
& \qquad  - \sum_{m : \kappa_{m}^{(n)} = {\part}}d_{m}^{\left(n\right)}\left(\E\left(\log\left(g\left(\bfeta_{m,k}\right)\right)\right) + \E_{r}\left(\log\left(h\left(t_{m}^{(n)}\right)\right)\right) + \E\left(\bfeta_{m,k}\right)^\top \E_{r}\left(\suffstat\right)\right)\nonumber\\
& \qquad - \sum_{m : \kappa_{m}^{(n)} = {\unobs}}\E\left(d_{m}^{\left(n\right)}\right)\left(\E\left(\log\left(g\left(\bfeta_{m,k}\right)\right)\right) + \E_{l}\left(\log\left(h\left(t_{m}^{(n)}\right)\right)\right) + \E\left(\bfeta_{m,k}\right)^\top \E_{l}\left(\suffstat\right)\right)\nonumber \\
& \qquad + \sum_{m : \kappa_{m}^{(n)} = {\obs}}\left(1 - d_{m}^{\left(n\right)}\right)\E\left(\log\left(1 - \pi_{m,k}\right)\right) + \sum_{m : \kappa_{m}^{(n)} = {\part}}\left(1 - d_{m}^{\left(n\right)}\right)\E\left(\log\left(1 - \pi_{m,k}\right)\right)\nonumber\\
& \qquad + \sum_{m : \kappa_{m}^{(n)} = {\unobs}}\E\left(1 - d_{m}^{\left(n\right)}\right)\E\left(\log\left(1 - \pi_{m,k}\right)\right)\nonumber\\
&= \zeta_{k} + \sum_{m : \kappa_{m}^{(n)} = {\obs}}  d_{m}^{(n)}\left(\lambda_{m,k}^{(n)} + u_{\obs, m,k}^{(n)}\right) + \sum_{m : \kappa_{m}^{(n)} = {\part}} \left(\lambda_{m,k}^{(n)} + u_{\part,m,k}^{(n)}\right)\nonumber\\
& \qquad\qquad + \sum_{m : \kappa_{m}^{(n)} = {\unobs}} \pi_{m}^{(n)\ast}\left(\lambda_{m,k}^{(n)} + u_{\unobs,m,k}^{(n)}\right)
\end{alignat}
where
\begin{alignat}{1}
\zeta_k &= \E\left(\log\gamma_k\right) + \E\left(\log\left(1 - \pi_{m,k}\right)\right)\nonumber\\
&= \psi\left(\theta_{k}^{\ast}\right) - \psi\left(\sum_{j=1}^{K} \theta_{j}^{\ast}\right) + \sum_{m=1}^{M} \psi\left(b_{m,k}^{\ast}\right) -  \psi\left(a_{m,k}^{\ast} + b_{m,k}^{\ast}\right)\\
\lambda_{m,k} &= \E\left(\log\pi_{m,k}\right) - \E\left(\log\left(1 - \pi_{m,k}\right)\right)\nonumber\\
&= \psi\left(a_{m,k}^{\ast}\right) - \psi\left(b_{m,k}^{\ast}\right)\\
u_{\obs, m,k}^{(n)} &= \E\left(\log\left(g\left(\bfeta_{m,k}\right)\right)\right)+ \E\left(\bfeta_{m,k}\right)^\top \suffstat\\
u_{\part,m,k}^{(n)}&= \E\left(\log\left(g\left(\bfeta_{m,k}\right)\right)\right) + \E_{r}\left(\log\left(h\left(t_{m}^{(n)}\right)\right)\right) + \E\left(\bfeta_{m,k}\right)^\top \E_{r}\left(\suffstat\right)\\
u_{\unobs,m,k}^{(n)} &= \E\left(\log\left(g\left(\bfeta_{m,k}\right)\right)\right) + \E_{l}\left(\log\left(h\left(t_{m}^{(n)}\right)\right)\right) + \E\left(\bfeta_{m,k}\right)^\top \E_{l}\left(\suffstat\right)
\end{alignat}
with $\psi$ denoting the digamma function, and we identify 
\begin{equation}
z_{k}^{\left(n\right)} \distas \multinomial\left(1, \bfgamma_{k}^{(n)\ast}\right)
\end{equation}
where
\begin{equation}
\bfgamma_{k}^{(n)\ast} = \frac{c_{k}}{\sum_{k=1}^{K}c_{k}}
\end{equation}

\paragraph{Left Censored}
\paragraph{\textcolor{red}{$t^{(n)}_m$}}

For the left censored data we further have 
\begin{alignat}{1}
q\left(t_{m}^{(n)}\mid d_{m}^{(n)}=1\right) & \propto \exp\left(\E_{-t_{m}^{(n)}}\left(\log\left(\prod_{k=1}^{K}\left[\left(f_{\expfamt_r}\left(t^{(n)}_{m}\g\bfeta_{m,k} \right)\right)^{d_{m}^{\left(n\right)}}\right]^{z_{k}^{\left(n\right)}}\right)\right)\right)\nonumber\\
& \propto \exp\left(\log\left(\bfone\left(t_{m}^{(n)} < \rho^{(n)}\right)\right)+\sum_{k=1}^{K}\E_q\left(z_{k}^{(n)}\right)\log\left(h\left(t^{(n)}_{m}\right)\right)\right.\nonumber\\
&\left. \qquad\qquad+ \left(\sum_{k=1}^{K}\E_q\left(z_{k}^{(n)}\right)\E\left(\bfeta_{m,k}\right)\right)^\top \suffstat_m^{(n)}\right)\nonumber\\
& \propto \bfone\left(t_{m}^{(n)} < \rho^{(n)}\right) h\left(t^{(n)}_{m}\right)^{\sum_{k=1}^{K}\bfgamma_{k}^{(n)\ast}}\exp\left(\left(\sum_{k=1}^{K}\bfgamma_{k}^{(n)\ast}\E\left(\bfeta_{m,k}\right)\right)^\top\suffstat_m^{(n)}\right)\nonumber\\
& \propto h\left(t^{(n)}_{m}\right)\exp\left(\left(\bfeta_{m}^{(n)*}\right)^\top\suffstat_m^{(n)}\right)\bfone\left(t_{m}^{(n)} < \rho^{(n)}\right)
\end{alignat}
where
\begin{equation}
    \bfeta_{m}^{(n)*} = \sum_{k=1}^{K}\bfgamma_{k}^{(n)\ast}\E\left(\bfeta_{m,k}\right)
\end{equation}
and we can identify the distribution as 
\begin{equation}
t_{m}^{(n)}\mid d_{m}^{(n)}=1 \distas \expfamt_{r}\left(\bfeta_{m}^{(n)*}\right).
\end{equation}

\paragraph{Right censored}

For the right censored data we instead have

\paragraph{\textcolor{red}{$d_{m}^{(n)}$}}

\begin{alignat}{1}
q\left(d_{m}^{(n)} \right) & \propto \exp\left(\E_{-d_{m}^{(n)}}\left(\log\left(\prod_{k=1}^{K}\left[\left(1-\pi_{m,k}\right)^{1-d_{m}^{(n)}} \left(\pi_{m,k}f_{\expfamt_l}\left(t^{(n)}_{m}\g\bfeta_{m,k} \right)\right)^{d_{m}^{\left(n\right)}}\right]^{z_{k}^{\left(n\right)}}\right)\right)\right)\nonumber\\
& \propto \exp\left(\E_{-d_{m}^{(n)}}\left(\sum_{k=1}^{K}d_{m}^{\left(n\right)}z_{k}^{\left(n\right)}\log\left(\pi_{m,k}\right) - \sum_{k=1}^{K}d_{m}^{\left(n\right)}z_{k}^{\left(n\right)}\log\left(1 - \pi_{m,k}\right) \right. \right.\nonumber\\
& \qquad\qquad\qquad\qquad\quad \left. \left. + \sum_{k=1}^{K}d_{m}^{\left(n\right)}z_{k}^{\left(n\right)}\log\left(f_{\expfamt_l}\left(t^{(n)}_{m}\g\bfeta_{m,k} \right)\right)\right)\right) \nonumber\\
& \propto \exp\left(d_{m}^{(n)}\left(\sum_{k=1}^{K}\bfgamma_{k}^{(n)\ast}\E\left(\log\left(\pi_{m,k}\right)\right) - \sum_{k=1}^{K}\bfgamma_{k}^{(n)\ast}\E\left(\log\left(1 - \pi_{m,k}\right)\right)\right.\right.\nonumber\\
& \qquad\qquad\qquad\qquad\quad+\sum_{k=1}^{K}\bfgamma_{k}^{(n)\ast}\E\left(\log\left(g\left(\bfeta_{m,k}\right)\right) + \log\left(h\left(t_{m}^{(n)}\right)\right)\right)\nonumber\\
& \left.\left. \qquad\qquad\qquad\qquad\quad  + \left(\sum_{k=1}^{K}\bfgamma_{k}^{(n)\ast}\E\left(\bfeta_{m,k}\right)\right)^\top \E\left(\suffstat_m^{(n)}\right) \right)\right)\nonumber\\
& \propto \exp\left(d_{m}^{(n)}\left(p_{m}^{\left(n\right)*} + \left(\bfeta_{m}^{(n)*}\right)^\top\E\left(\suffstat_m^{(n)}\right)\right)\right) \nonumber\\
& \propto \exp\left(c_{m}\right)^{d_{m}^{(n)}}
\end{alignat}
where 
\begin{alignat}{1}
p_{m}^{\left(n\right)*} & = \sum_{k=1}^{K}\bfgamma_{k}^{(n)\ast}\E\left(\log\left(\pi_{m,k}\right)\right) - \sum_{k=1}^{K}\bfgamma_{k}^{(n)\ast}\E\left(\log\left(1 - \pi_{m,k}\right)\right) \nonumber\\
& \qquad +\sum_{k=1}^{K}\bfgamma_{k}^{(n)\ast}\E\left(\log\left(g\left(\bfeta_{m,k}\right)\right) + \log\left(h\left(t_{m}^{(n)}\right)\right)\right)\\
\bfeta_{m}^{(n)*} & = \sum_{k=1}^{K}\bfgamma_{k}^{(n)\ast}\E\left(\bfeta_{m,k}\right)\\
& c_{m} = p_{m}^{\left(n\right)*} + \left(\bfeta_{m}^{(n)*}\right)^\top\E\left(\suffstat_m^{(n)}\right)
\end{alignat}
and we can identify as 
\begin{equation}
    d_{m}^{(n)} \distas \bernoulli\left(\pi_{m}^{(n)\ast}\right)
\end{equation}
where 
\begin{equation}
    \pi_{m}^{(n)\ast} = \frac{c_{m}}{1 + c_{m}}
\end{equation}

\paragraph{\textcolor{red}{$t_{m}^{(n)}$}}

furthermore we have 
\begin{alignat}{1}
q\left(t_{m}^{(n)}\g d_{m}^{(n)}=1\right) & \propto \exp\left(\E_{-t_{m}^{(n)}}\left(\log\left(\prod_{k=1}^{K}\left[\left(f_{\expfamt_l}\left(t^{(n)}_{m}\g\bfeta_{m,k} \right)\right)^{d_{m}^{\left(n\right)}}\right]^{z_{k}^{\left(n\right)}}\right)\right)\right)\nonumber\\
& \propto \exp\left(\log\left(\bfone\left(t_{m}^{(n)} > \tau^{(n)}\right)\right) + \sum_{k=1}^{K}\E_q\left(z_{k}^{(n)}\right)\log\left(h\left(t^{(n)}_{m}\right)\right)\right.\nonumber \\
& \left.\qquad\qquad + \left(\sum_{k=1}^{K}\E_q\left(z_{k}^{(n)}\right)\E\left(\bfeta_{m,k}\right)\right)^\top \suffstat_m^{(n)}\right)\nonumber\\
& \propto \bfone\left(t_{m}^{(n)} > \tau^{(n)}\right) h\left(t^{(n)}_{m}\right)^{\sum_{k=1}^{K}\bfgamma_{k}^{(n)\ast}}\exp\left(\left(\sum_{k=1}^{K}\bfgamma_{k}^{(n)\ast}\E\left(\bfeta_{m,k}\right)\right)^\top\suffstat_m^{(n)}\right)\nonumber\\
& \propto h\left(t^{(n)}_{m}\right)\exp\left(\left(\bfeta_{m}^{(n)*}\right)^\top\suffstat_m^{(n)}\right)\bfone\left(t_{m}^{(n)} > \tau^{(n)}\right)
\end{alignat}
where
\begin{equation}
    \bfeta_{m}^{(n)*} = \sum_{k=1}^{K}\bfgamma_{k}^{(n)\ast}\E\left(\bfeta_{m,k}\right)
\end{equation}
and we can identify the distribution as 
\begin{equation}
t_{m}^{(n)}\mid d_{m}^{(n)}=1 \distas \expfamt_{l}\left(\bfeta_{m}^{(n)*}\right).
\end{equation}

\newpage
\section{Predictive distribution}\label{DerivePred}
We can write the predictive posterior for a new individual $\Delta' = \{\alive,0,0,\varnothing\}$  who is age $0$ and has no observations as:
\begin{alignat}{1}
    p\left(z^{\prime}, \primebfd, \primebft \g \bfPhi\right) & = \int p\left( z^{\prime}, \primebfd, \primebft \g \bfgamma, \bfpi, \bfeta \right) p\left(\bfgamma, \bfpi, \bfeta \g \bfPhi\right) \d \bfgamma \d \bfpi \d \bfeta\nonumber\\
    & = \int p\left( z^{\prime} \g \bfgamma\right) p\left(\bfgamma \g \bfPhi \right) \d \bfgamma \int p\left( \primebfd \g  \bfpi, z^{\prime} \right) p\left( \bfpi \g \bfPhi \right) \d \bfpi \int p\left( \primebft \g \bfeta, z^{\prime}, \primebfd \right) p\left(\bfeta \g \bfPhi \right) \d \bfeta\nonumber\\
    & =  p\left(z^{\prime} \g \bfPhi \right)  p\left(\primebfd \g \bfPhi, z^{\prime} \right)  p\left( \primebft \g \bfPhi, z^{\prime}, \primebfd \right)
\end{alignat}
Now, we can derive the probability that the individual belongs to each cluster:
\begin{alignat}{1}
    p\left(z^{\prime}_{i} = 1 \g \bfPhi \right) & = \int p\left( z^{\prime}_{i} = 1 \g \bfgamma\right) p\left(\bfgamma \g \bfPhi \right) \d \bfgamma\nonumber\\
    & \propto \int \prod_{k=1}^{K}\left( \gamma_{k}^{z^{\prime}_k}\right) \frac{\Gamma\left(\sum_{k=1}^{K}\theta_{k}^{\ast}\right)}{\prod_{k=1}^{K}\Gamma\left(\theta_{k}^{\ast}\right)}\prod_{k=1}^{K}\left( \gamma_{k}^{\theta_{k} - 1}\right) \d \bfgamma\nonumber\\
    & \propto \frac{\Gamma\left(\sum_{k=1}^{K}\theta_{k}^{\ast}\right)}{\prod_{k=1}^{K}\Gamma\left(\theta_{k}^{\ast}\right)} \int \prod_{k=1}^{K}\left( \gamma_{k}^{\bfone\left(k=i\right) + \theta_{k} - 1}\right) \d \bfgamma\nonumber\\
    & \propto \frac{\Gamma\left(\sum_{k=1}^{K}\theta_{k}^{\ast}\right)}{\prod_{k=1}^{K}\Gamma\left(\theta_{k}^{\ast}\right)} \frac{\prod_{k=1}^{K}\Gamma\left(\bfone\left(k=i\right) + \theta_{k}^{\ast}\right)}{\Gamma\left(1 + \sum_{k=1}^{K}\theta_{k}^{\ast}\right)}\nonumber\\
    & \propto \frac{\theta_{i}^{\ast}}{\sum_{k=1}^{K}\theta_{k}^{\ast}}
\end{alignat}
where we use $\Gamma\left(n+1\right) = n\Gamma\left(n\right)$ to simplify the second last line and we can identify the distribution as
\begin{equation}
    z^{\prime} \distas \multinomial\left(1, \bar{\bftheta}\right)
\end{equation}
where
\begin{equation}
    \bar{\theta}_i= \frac{\theta_{i}^{\ast}}{\sum_{k=1}^{K}\theta_{k}^{\ast}}
\end{equation}
Next we have that the probability that a condition is present in the individual's trajectory given that they belong to a particular cluster is:
\begin{alignat}{1}
    p\left(d^{\prime}_{m} = 1 \g \bfPhi, z^{\prime}_{k}=1\right) & = \int p\left( d^{\prime}_{m} \g  \pi_{m,k}, z^{\prime}_{k}=1 \right) p\left( \pi_{m,k} \g \bfPhi \right) \d \pi_{m,k} \nonumber\\
    & = \int \pi_{m,k} \pi_{m,k}^{a_{m,k}^{\ast} - 1} \left(1 - \pi_{m,k}\right)^{b_{m,k}^{\ast} - 1} \d \pi_{m,k}\nonumber\\
    & = \frac{a_{m,k}^{\ast}}{a_{m,k}^{\ast} + b_{m,k}^{\ast}}
\end{alignat}
from which we identify 
\begin{equation}
    d^{\prime}_{m} \g z^{\prime}_{k} = 1 \distas \bernoulli\left(\bar{\pi}_{m,k}\right)
\end{equation}
where
\begin{equation}
    \bar{\pi}_{m,k}=\frac{a_{m,k}^{\ast}}{a_{m,k}^{\ast} + b_{m,k}^{\ast}}
\end{equation}
Finally we have the predictive  posterior density function for the age at diagnosis of a condition given that the condition is present and the individual belongs to a particular cluster:
\begin{alignat}{1}
    f_{\mathcal{P}}(t_{m}^{\prime} \g \nu_{m,k}^{*},\bfchi_{m,k}^{*}) &= p\left( t^{\prime}_{m} \g \bfPhi, z^{\prime}_{k}=1, d^{\prime}_{m}=1 \right)\nonumber\\ & = \int p\left( t^{\prime}_{m} \g \bfeta_{m,k}, z^{\prime}_{k}=1, d^{\prime}_{m} = 1 \right) p\left(\bfeta_{m,k} \g \bfPhi \right) \d \bfeta_{m,k}\nonumber\\
    & = \int g\left(\bfeta_{m,k}\right)  h\left(t^{\prime}_{m}\right)\exp\left(\bfeta_{m,k}^\top \suffstat\left(t_{m}^{\prime}\right)\right) g\left(\bfeta_{m,k}\right)^{\nu_{m,k}^{\ast}} f\left(\nu_{m,k}^{\ast},\bfchi_{m,k}^{\ast}\right)\exp\left(\bfeta_{m,k}^\top\bfchi_{m,k}^{\ast}\right) \d \bfeta_{m,k}\nonumber \\
    & = h\left(t^{\prime}_{m}\right) f\left(\nu_{m,k}^{\ast},\bfchi_{m,k}^{\ast}\right) \int g\left(\bfeta_{m,k}\right)^{\nu_{m,k}^{\ast}+1} \exp\left(\bfeta_{m,k}^\top\left(\suffstat\left(t_{m}^{\prime}\right) + \bfchi_{m,k}^{\ast}\right)\right) \d \bfeta_{m,k}\nonumber\\
    & = \frac{h\left(t^{\prime}_{m}\right) f\left(\nu_{m,k}^{\ast},\bfchi_{m,k}^{\ast}\right)}{f\left(\nu_{m,k}^{\ast} + 1,\bfchi_{m,k}^{\ast} + \suffstat\left(t_{m}^{\prime}\right)\right)}
\end{alignat}

From this probability density function we also find the cumulative and complimentary cumulative predictive density functions for $t_{m}^{\prime}$ given disease presence $d_{m}^{\prime} = 1$ and a cluster $z_{k}^{\prime} = 1.$

\begin{alignat}{1}
F_{\mathcal{P}}(\rho^{\prime} \g \nu_{m,k}^{*},\bfchi_{m,k}^{*}) & = \int_{0}^{\rho^{\prime}} f_{\mathcal{P}}(t_{m}^{\prime} \g \nu_{m,k}^{*},\bfchi_{m,k}^{*}) \d t_{m}^{\prime}\\
S_{\mathcal{P}}(\tau^{\prime} \g \nu_{m,k}^{*},\bfchi_{m,k}^{*}) & = 1 - F_{\mathcal{P}}(\tau^{\prime} \g \nu_{m,k}^{*},\bfchi_{m,k}^{*})
\end{alignat}
Now we consider the more typical case of a new individual 
$\Delta' = \{\alive,\rho^\prime,\tau^\prime,\Phi'\}$ who is age $\tau^{\prime}$ and who can also be represented as  
\begin{equation}
\Phi_{\obstar}^{\prime} = \left(\bfd_{\obstar}^{\prime},\bft_{\obstar}^{\prime},\bfkappa_{\obstar}^{\prime}\right) =  \left(d^{\prime}_{m},t^{\prime}_{m},\kappa^{\prime}_{m}\right)_{m:\kappa_{m}^{\prime}={\obstar}}
\end{equation}
where $\obstar = \obs, \part$ or $\unobs$.
For this new individual we want to derive the predictive posterior over the missing quantities $\bfd_{\unobs}^{\prime}$ and $\bft_{\unobs}^{\prime}$ given the available data:
\begin{equation}
p\left(\bfd_{\unobs}^{\prime},\bft_{\unobs}^{\prime} \g \Phi_{\obs}^{\prime}, \bfd_{\part}^{\prime},\bfkappa_{\part}^{\prime}, \bfkappa_{\unobs}^{\prime}, \bfPhi,\rho^\prime,\tau^\prime\right)
=\sum_{k=1}^Kp(\bfd_{\unobs}^{\prime},\bft_{\unobs}^{\prime} \g z_k^{\prime},\bfkappa_{\unobs}^{\prime}, \bfPhi,\tau^\prime)
p(z_k^{\prime}\g\Phi^\prime_{\obs},  \bfd_{\part}^{\prime},\bfkappa_{\part}^{\prime},\bfkappa_{\unobs}^{\prime},\bfPhi,\rho^\prime,\tau^\prime)
\end{equation}

First we have the probability that the individual belongs to a particular cluster:

\begin{alignat}{1}
    \tilde{\phi}_{k} & = p(z_k\g\Phi^\prime_{\obs},  \bfd_{\part}^{\prime},\bfkappa_{\part}^{\prime},\bfkappa_{\unobs}^{\prime},\bfPhi,\rho^\prime,\tau^\prime)\nonumber \\
    & \propto p\left(z_k^{\prime}\g \bfPhi\right)p\left(\bfPhi_{\obs}^{\prime}\g z_k^{\prime}, \bfPhi\right)p\left(\bfd_{\part}^{\prime}, \bfkappa_{\part}^{\prime} \g z_k^{\prime}, \bfPhi, \rho^{\prime}\right)p\left(\bfkappa_{\unobs}^{\prime}\g z_k^{\prime}, \bfPhi, \tau^{\prime}\right)\nonumber\\
    & \propto \bar{\theta}_k \prod_{m:\kappa^\prime=\obs} \bar{\pi}_{m,k} f_{\mathcal{P}}(t^\prime_m)\prod_{m:\kappa^\prime=\part} \bar{\pi}_{m,k}F_{\mathcal{P}}\left(\rho^\prime\right)\prod_{m:\kappa^\prime=\unobs} \left(1-\bar{\pi}_{m,k}F_{\mathcal{P}}\left(\tau^\prime\right)\right)
\end{alignat}

which is the second term in the desired predictive posterior. Next we have the probability that a disease is present in the trajectory given that it has not been observed thus far:

\begin{alignat}{1}
\tilde{\pi}_{m,k} & = p\left(d^{\prime}_{m}=1 \g z^{\prime}_{k}=1, \kappa_{m}^{\prime} = {\unobs},\bfPhi,\tau^\prime\right)\nonumber\\
& = \frac{p\left(\kappa_{m}^{\prime} = {\unobs}, d^{\prime}_{m}=1 \g \bfPhi, z^{\prime}_{k}=1,\tau^\prime\right)}{p\left( \kappa_{m}^{\prime} = {\unobs} \g \bfPhi, z^{\prime}_{k}=1,\tau^\prime\right)}\nonumber\\
& \propto p\left(d^{\prime}_{m}=1 \g \bfPhi, z^{\prime}_{k}=1\right)p\left( \kappa_{m}^{\prime} = {\unobs} \g \bfPhi, z^{\prime}_{k}=1, d^{\prime}_{m}=1,\tau^\prime\right)\nonumber\\
& \propto \bar{\pi}_{m,k}S_{\mathcal{P}}(\tau^{\prime} \g \nu_{m,k}^{*},\bfchi_{m,k}^{*})
\end{alignat}

Now we can derive the first term in the predictive posterior:

\begin{align}{}
&p\left(\bfd_{\unobs}^{\prime},\bft_{\unobs}^{\prime} \g z_k^{\prime},\bfkappa_{\unobs}^{\prime}, \bfPhi,\tau^\prime\right) \nonumber\\
& = \prod_{m:\kappa_{m}^{\prime} = \unobs} \left(p\left(t_{m}^{\prime} \g z_k^{\prime},\kappa_{m}^{\prime}, \bfPhi,\tau^\prime,d_{m}^{\prime}=1\right) p\left(d_{m}^{\prime}=1 \g z_k^{\prime},\kappa_{m}^{\prime}, \bfPhi,\tau^\prime\right)\right)^{d_{m}^{\prime}} \left( p\left(d_{m}^{\prime}=0 \g z_k^{\prime},\kappa_{m}^{\prime}, \bfPhi,\tau^\prime\right)\right)^{1 - d_{m}^{\prime}}\nonumber\\
& = \prod_{m:\kappa_{m}^{\prime} = \unobs} \left(f_{\mathcal{P}}(t_{m}^{\prime} \g t_{m}^{\prime} > \tau^{\prime}) \tilde{\pi}_{m,k}\right)^{d_{m}^{\prime}}\left(1 - \tilde{\pi}_{m,k}\right)^{1 - d_{m}^{\prime}}
\end{align}

Finally we can construct the predictive posterior density function:

\begin{alignat}{1}
p\left(\bfd_{\unobs}^{\prime},\bft_{\unobs}^{\prime} \g \Phi_{\obs}^{\prime}, \bfd_{\part}^{\prime},\bfkappa_{\part}^{\prime}, \bfkappa_{\unobs}^{\prime}, \bfPhi,\rho^\prime,\tau^\prime\right)
& =\sum_{k=1}^Kp(\bfd_{\unobs}^{\prime},\bft_{\unobs}^{\prime} \g z_k^{\prime},\bfkappa_{\unobs}^{\prime}, \bfPhi,\tau^\prime)
p(z_k^{\prime}\g\Phi^\prime_{\obs},  \bfd_{\part}^{\prime},\bfkappa_{\part}^{\prime},\bfkappa_{\unobs}^{\prime},\bfPhi,\rho^\prime,\tau^\prime)\nonumber\\
& \propto \sum_{k=1}^{K} \phi_{k} \prod_{m:\kappa_{m}^{\prime} = \unobs} \left(f_{\mathcal{P}}(t_{m}^{\prime} \g t_{m}^{\prime} > \tau^{\prime}) \tilde{\pi}_{m,k}\right)^{d_{m}^{\prime}}\left(1 - \tilde{\pi}_{m,k}\right)^{1 - d_{m}^{\prime}}
\end{alignat}

\newpage
\section{Gaussian Variational Bayes Updating Equations}\label{gaussianVBder}

For the Gaussian case with mean $\mu_{m,k}$ and variance $\sigma_{m,k}^2$we have natural parameters $\eta_{m,k,1} = \frac{\mu_{m,k}}{\sigma_{m,k}^2}$ and $\eta_{m,k,2} = -\frac{1}{2\sigma_{m,k}^2}$ and sufficient statistics $T_{m,1} = t_{m}$ and $T_{m,2} = t_{m}^2.$ We also have that the exponential form of the Gaussian probability density function can be written as:
\begin{alignat}{1}
    f_\expfam(t_m \g \bfeta_{m,k}) =   g\left(\bfeta_{m,k}\right)  h\left(t_m\right)\exp\left(\bfeta_{m,k}^\top \suffstat_m\right)
\end{alignat}
where
\begin{alignat}{1}
    g\left(\bfeta_{m,k}\right) & = \exp\left(-\frac{\eta_{m,k,1}^2}{4\eta_{m,k,2}} - \frac{1}{2}\log\left(-2\eta_{m,k,2}\right)\right)\nonumber\\
    & = \exp\left(\frac{\mu_{m,k}^2}{2\sigma_{m,k}^2} - \frac{1}{2}\log\left(\frac{1}{\sigma_{m,k}^2}\right)\right)\\
    h\left(t_m\right) & = \frac{1}{\sqrt{2\pi}}\\
    \bfeta_{m,k}^\top \suffstat_m & = \eta_{m,k,1}T_{m,1}^{(n)} + \eta_{m,k,2}T_{m,2}^{(n)}\nonumber\\
    & = \frac{\mu_{m,k}^2t_{m}^{(n)}}{\sigma_{m,k}^2} - \frac{\left(t_{m}^{(n)}\right)^2}{2\sigma_{m,k}^2}
\end{alignat}

We further have that the Normal-Inverse gamma conjugate posterior on $\mu_{m,k}$ and $\sigma_{m,k}^2$ has hyperparameters $u_{m,k}^\ast, v_{m,k}^\ast, \alpha_{m,k}^\ast,$ and $\beta_{m,k}^\ast.$ The normal-inverse gamma distribution also belongs to the exponential family with of natural statistics $\log\left(\frac{1}{\sigma_{m,k}^2}\right), \frac{1}{\sigma_{m,k}^2}, \frac{\mu_{m,k}}{\sigma_{m,k}^2}, \frac{\mu_{m,k}^2}{\sigma_{m,k}^2}.$ The expectations of these natural statistics can be found using the moment generating function \citep{wasserman2004parametric} and are as follows:

\begin{alignat}{1}
    \E\left(\log\left(\frac{1}{\sigma_{m,k}^2}\right)\right) & = \psi\left(\alpha_{m,k}^\ast\right) - \log\left(\beta_{m,k}^\ast\right)\\
    \E\left(\frac{1}{\sigma_{m,k}^2}\right) & = \frac{\alpha_{m,k}^\ast}{\beta_{m,k}^\ast}\\
    \E\left(\frac{\mu_{m,k}}{\sigma_{m,k}^2}\right) & = \frac{u_{m,k}^\ast\alpha_{m,k}^\ast}{\beta_{m,k}^\ast}\\
    \E\left(\frac{\mu_{m,k}^2}{\sigma_{m,k}^2}\right) & = \frac{\left(u_{m,k}^\ast\right)^2\alpha_{m,k}^\ast}{v_{m,k}^\ast\beta_{m,k}^\ast}\\
\end{alignat}

We have the following expectations:

\begin{alignat}{1}
    \E\left(\log\left(g\left(\bfeta_{m,k}\right)\right)\right) & = \E\left(\frac{\mu_{m,k}^2}{2\sigma_{m,k}^2} + \frac{1}{2}\log\left(\sigma_{m,k}^2\right)\right)\nonumber\\
    & = \frac{1}{2}\E\left(\frac{\mu_{m,k}^2}{\sigma_{m,k}^2}\right) - \frac{1}{2}\E\left(\log\left(\frac{1}{\sigma_{m,k}^2}\right)\right)\nonumber\\
    & = \frac{u_{m,k}^2\alpha_{m,k}}{2v_{m,k}\beta_{m,k}} - \frac{1}{2}\left(\psi\left(\alpha_{m,k}\right) - \log\left(\beta_{m,k}\right)\right)\\
    \E\left(\eta_{m,k,1}\right) & = \frac{u_{m,k}^\ast\alpha_{m,k}^\ast}{\beta_{m,k}^\ast}\\
    \E\left(\eta_{m,k,2}\right) & = -\frac{\alpha_{m,k}^\ast}{2\beta_{m,k}^\ast}\\
\end{alignat}

We write $f_Z, F_Z$ and $S_Z$ for the probability density, cumulative density and survival functions of the standard normal distribution. We  transform the Gaussian parameters to the mean $\mu_{m}^\ast = \frac{\eta_{m,1}^\ast}{2\eta_{m,2}^\ast}$ and standard deviation $\sigma_{m}^\ast = \sqrt{-\frac{1}{2\eta_{m,2}^\ast}}$. We  transform the truncation thresholds such that $\Dot{\tau}_{m} = \frac{\tau - \mu_{m}^\ast}{\sigma_{m}^\ast}$ and $\Dot{\rho}_{m} = \frac{\rho - \mu_{m}^\ast}{\sigma_{m}^\ast}$. Then we can write the truncated expectations as perturbations of the untruncated moments \citep{johnson1995continuous}:

\begin{alignat}{1}
    \E_l\left(T_{m,1}\right) = \E_l\left(t_{m}\right) & = \mu_{m}^\ast + \sigma_{m}^\ast\frac{f_Z\left(\Dot{\tau}_{m}\right)}{S_Z\left(\Dot{\tau}_{m}\right)}\\
    \E_r\left(T_{m,1}\right) = \E_l\left(t_{m}\right) & = \mu_{m}^\ast - \sigma_{m}^\ast\frac{f_Z\left(\Dot{\rho}_{m}\right)}{F_Z\left(\Dot{\rho}_{m}\right)}\\
    \E_l\left(T_{m,2}\right) = \E_l\left(t_{m}^2\right) & = \left(\sigma_{m}^\ast\right)^2\left(1 + \frac{\Dot{\tau}_{m}f_Z\left(\Dot{\tau}_{m}\right)}{S_Z\left(\Dot{\tau}_{m}\right)} - \left(\frac{f_Z\left(\Dot{\tau}_{m}\right)}{S_Z\left(\Dot{\tau}_{m}\right)}\right)^2\right) + \left(\mu_{m}^\ast + \sigma_{m}^\ast\frac{f_Z\left(\Dot{\tau}_{m}\right)}{S_Z\left(\Dot{\tau}_{m}\right)}\right)^2\\
    \E_r\left(T_{m,2}\right) = \E_r\left(t_{m}^2\right) & = \left(\sigma_{m}^\ast\right)^2\left(1 - \frac{\Dot{\rho}_{m}f_Z\left(\Dot{\rho}_{m}\right)}{F_Z\left(\Dot{\rho}_{m}\right)} - \left(\frac{f_Z\left(\Dot{\rho}_{m}\right)}{F_Z\left(\Dot{\rho}_{m}\right)}\right)^2\right) + \left(\mu_{m}^\ast - \sigma_{m}^\ast\frac{f_Z\left(\Dot{\rho}_{m}\right)}{F_Z\left(\Dot{\rho}_{m}\right)}\right)^2
\end{alignat}

Finally, since $h\left(t\right)$ is a constant, we simply have: 

\begin{equation}
    \E_l\left(\log\left(h\left(t_m\right)\right)\right) = \E_r\left(\log\left(h\left(t_m\right)\right)\right) = \log\left(\frac{1}{\sqrt{2\pi}}\right)
\end{equation}

\newpage
\section{Full Variational Bayes Algorithm}\label{cenalg}
\begin{algorithm*}[htb!]
    \caption{Full Variational Bayes Updates}\label{Algorithm:VBcen}
    \algorithmicinput Prior Hyperparameters $\left(\bftheta,a_{m,k},b_{m,k},\text{ and }\nu_{m,k},\bfchi_{m,k}\right)$, Data $\mathcal{D}$, Initialisation Values $\left(\gamma_{k}^{(n)*}, \pi_{m}^{(n)*}, \eta_{m}^{(n)*}\right)$, Number of Clusters $K$, Stopping Condition $\epsilon$, Initial Difference $\delta$
    \begin{algorithmic}[1]
    \State $\left(\theta_{k},a_{m,k},b_{m,k},\nu_{m,k}, \bfchi_{m,k}\right) \leftarrow \left(\theta_{k}^{(0)},a_{m,k}^{(0)},b_{m,k}^{(0)},\nu_{m,k}, \bfchi_{m,k}^{(0)}\right)$
    \State $i \leftarrow 1$
    \While{$\delta > \epsilon$}
    \State $\bar{\bfz}\leftarrow \sum_{n=1}^N \bfgamma^{(n)\ast}$
    \State $\bar{n}_{m,k} \leftarrow \sum_{n:\kappa_{m}^{(n)}={\obs}, {\part}} d_{m}^{\left(n\right)}\bfgamma_{k}^{(n)\ast} + \sum_{n:\kappa_{m}^{(n)}={\unobs}} \pi_{m}^{(n)\ast}\bfgamma_{k}^{(n)\ast}$
    \State $\bar{T}_{m,k} \leftarrow \sum_{n:\kappa_{m}^{(n)}={\obs}} d_{m}^{\left(n\right)}\bfgamma_{k}^{(n)\ast}\suffstat_m^{(n)} + \sum_{n:\kappa_{m}^{(n)}={\part}} \bfgamma_{k}^{(n)\ast}\E_{r}\left(\suffstat_m^{(n)}\right) + \sum_{n:\kappa_{m}^{(n)}={\unobs}} \pi_{m}^{(n)\ast}\bfgamma_{k}^{(n)\ast}\E_{l}\left(\suffstat_m^{(n)}\right)$
    \State $\bftheta^{(i)} \leftarrow\bftheta + \bar{\bfz}$
    \State $a_{m,k}^{(i)} \leftarrow a_{m,k}+\bar{n}_{m,k}$
    \State $b_{m,k}^{(i)} \leftarrow  b_{m,k}+\bar{z}_{k}-\bar{n}_{m,k}$
    \State $\nu_{m,k}^{(i)}\leftarrow\nu_{m,k}+\bar{n}_{m,k}$
    \State $\bfchi_{m,k}^{(i)}\leftarrow\bfchi_{m,k}+\bar{T}_{m,k}$
    \State $\zeta_{k} \leftarrow \psi\left(\theta_{k}^{(i)}\right) - \psi\left(\sum_{j=1}^{K} \theta_{j}^{(i)}\right) + \sum_{m=1}^{M} \psi\left(b_{m,k}^{(i)}\right) -  \psi\left(a_{m,k}^{(i)} + b_{m,k}^{(i)}\right)$
    \State $\lambda_{m,k} \leftarrow \psi\left(a_{m,k}^{(i)}\right) - \psi\left(b_{m,k}^{(i)}\right)$
    \State $u_{\obs,m,k}^{(n)} \leftarrow \E\left(\log\left(g\left(\bfeta_{m,k}\right)\right)\right) + \E\left(\bfeta_{m,k}\right)^\top \suffstat_m^{(n)}$
    \State $u_{\part,m,k}^{(n)} \leftarrow \E\left(\log\left(g\left(\bfeta_{m,k}\right)\right)\right) + \E_{r}\left(\log\left(h\left(t_{m}^{(n)}\right)\right)\right)$
    \State $u_{\unobs,m,k}^{(n)} \leftarrow \E\left(\log\left(g\left(\bfeta_{m,k}\right)\right)\right) + \E_{l}\left(\log\left(h\left(t_{m}^{(n)}\right)\right)\right) + \E\left(\bfeta_{m,k}\right)^\top \E_{l}\left(\suffstat_m^{(n)}\right)$
    \State $\begin{aligned}[t]
            \gamma_{k}^{(n)*} & \leftarrow  \exp\left(\zeta_{k} + \sum_{m : \kappa_{m}^{(n)} = {\obs}}  d_{m}^{(n)}\left(\lambda_{m,k}^{(n)} + u_{\obs, m,k}^{(n)}\right) + \sum_{m : \kappa_{m}^{(n)} = {\part}} \left(\lambda_{m,k}^{(n)} + u_{\part,m,k}^{(n)}\right)\right.\\
            & \qquad\qquad\left. + \sum_{m : \kappa_{m}^{(n)} = {\unobs}} \pi_{m}^{(n)\ast}\left(\lambda_{m,k}^{(n)} + u_{\unobs,m,k}^{(n)}\right)\right)
        \end{aligned}$
        \State $\gamma_{k}^{(n)*} \leftarrow \frac{\gamma_{k}^{(n)*}}{\sum_{k=1}^{K}\gamma_{k}^{(n)*}}$
        \For{$m : \kappa_{m}^{(n)} = {\unobs}$}
            \State $\begin{aligned}[t]
                \pi_{m}^{(n)*} &\leftarrow \frac{\exp\left(\sum_{k=1}^{K}\bfgamma_{k}^{(n)\ast}\left(\lambda_{m,k} + u_{\unobs,m,k}^{(n)}\right)\right)}{1 + \exp\left(\sum_{k=1}^{K}\bfgamma_{k}^{(n)\ast}\left(\lambda_{m,k} + u_{\unobs,m,k}^{(n)}\right)\right)}
            \end{aligned}$
            \State $\eta_{m}^{(n)*} =\sum_{k=1}^{K}\bfgamma_{k}^{(n)\ast}\E\left(\bfeta_{m,k}\right)$
        \EndFor
        \For{$m : \kappa_{m}^{(n)} = {\part}$}
            \State $\eta_{m}^{(n)*} = \sum_{k=1}^{K}\bfgamma_{k}^{(n)\ast}\E\left(\bfeta_{m,k}\right)$
        \EndFor
        \State $\delta \leftarrow \left\|\left(\theta_{k}^{(i)},a_{m,k}^{(i)},b_{m,k}^{(i)},\nu_{m,k}^{(i)},\bfchi_{m,k}^{(i)}\right) - \left(\theta_{k}^{(i-1)},a_{m,k}^{(i-1)},b_{m,k}^{(i-1)},\nu_{m,k}^{(i-1)},\bfchi_{m,k}^{(i-1)}\right)\right\|$
        \algstore{VB}
    \end{algorithmic}
\end{algorithm*}

\begin{algorithm*}[!t]
    \begin{algorithmic}
        \algrestore{VB}
        \State $i \leftarrow i+1$
    \EndWhile
    \State $\left(\theta_{k}^{*},a_{m,k}^{*},b_{m,k}^{*},\nu_{m,k}^{*},\bfchi_{m,k}^{*}\right) \leftarrow \left(\theta_{k}^{(i)},a_{m,k}^{(i)},b_{m,k}^{(i)},\nu_{m,k}^{(i)},\bfchi_{m,k}^{(i)}\right)$
    \end{algorithmic}
    \algorithmicoutput $\left(\theta_{k}^{*},a_{m,k}^{*},b_{m,k}^{*},\nu_{m,k}^{*},\bfchi_{m,k}^{*}\right)$
\end{algorithm*}

\begin{figure*}
    \centering
    \vspace*{4.25in}
\end{figure*}

\clearpage

\section{All simulated risk profiles}\label{simrisks}

\qquad\newline
\begin{figure}[H]
\centering
\subfigure[][Cluster 1]{\includegraphics[height=0.15\textheight]{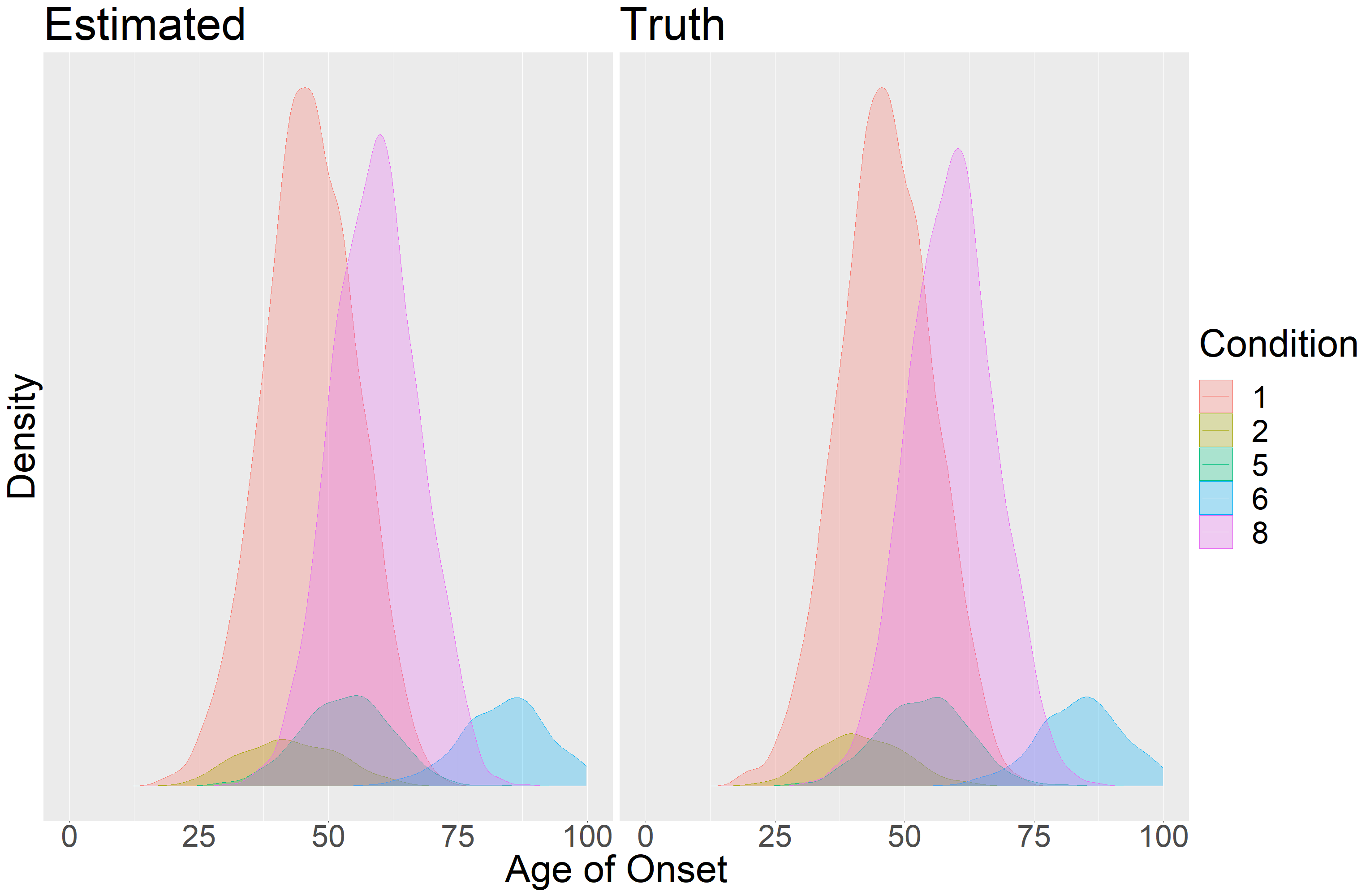}}
\subfigure[][Cluster 2]{\includegraphics[height=0.15\textheight]{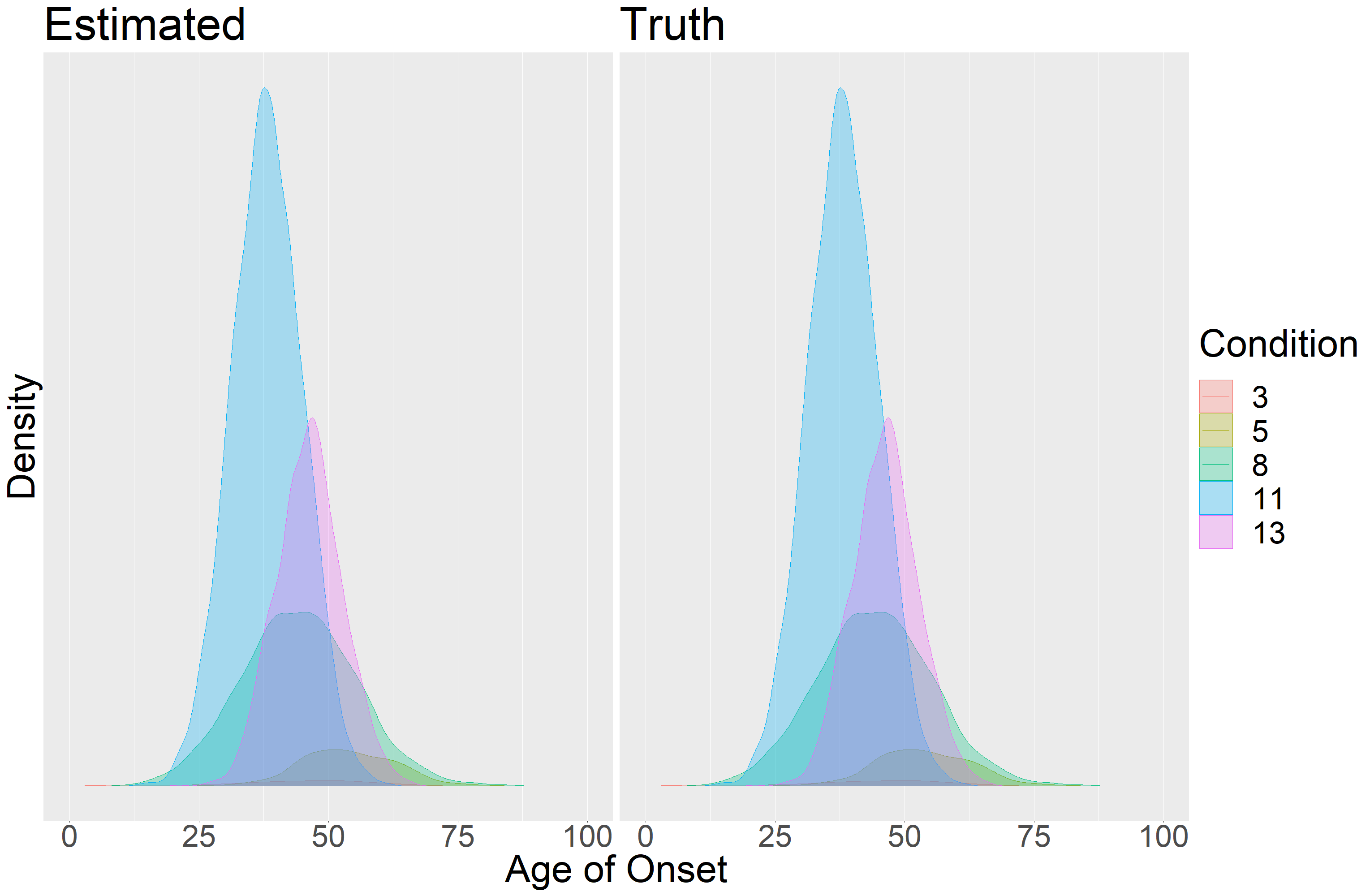}}
\subfigure[][Cluster 3]{\includegraphics[height=0.15\textheight]{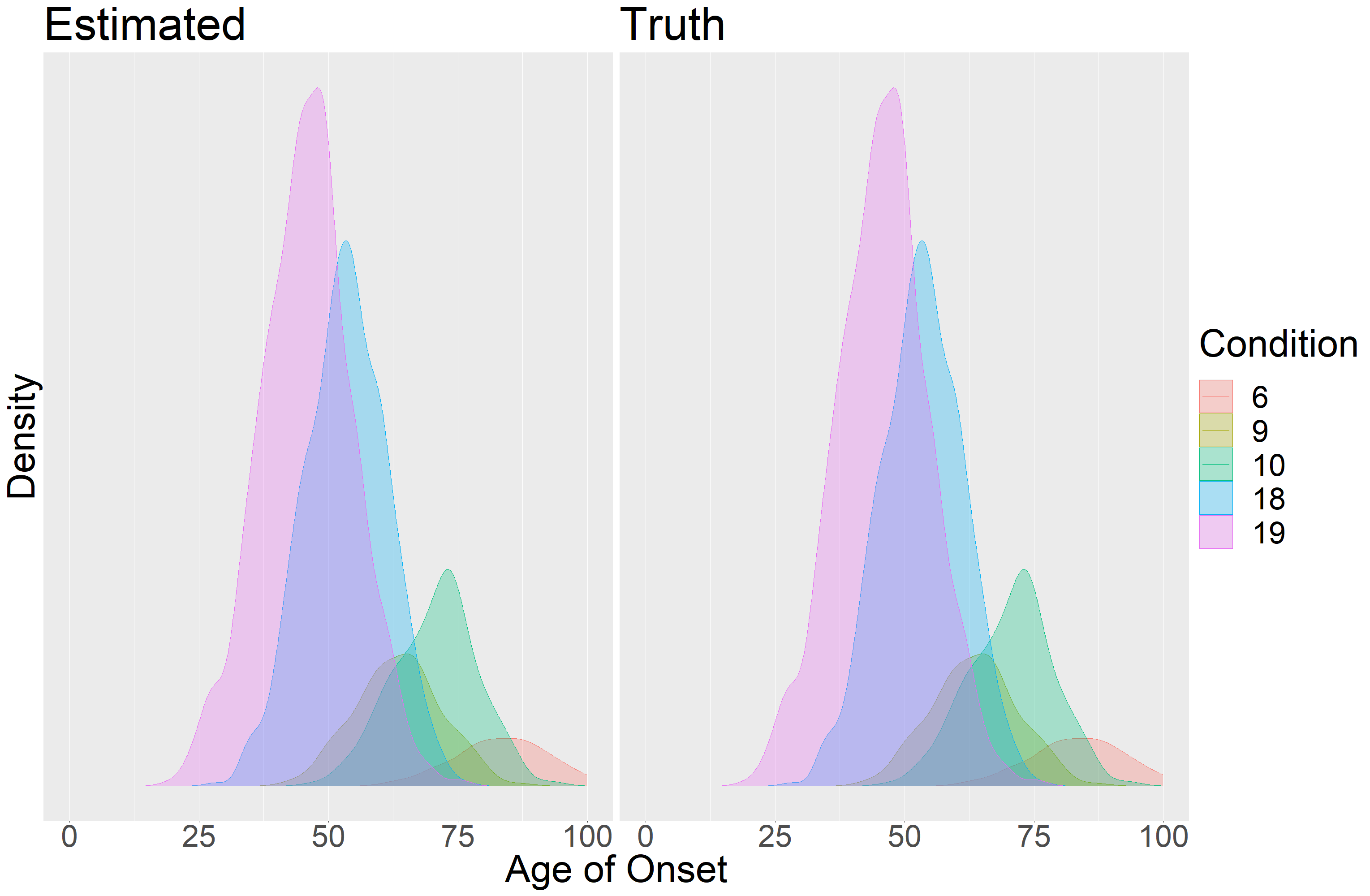}}
\subfigure[][Cluster 4]{\includegraphics[height=0.15\textheight]{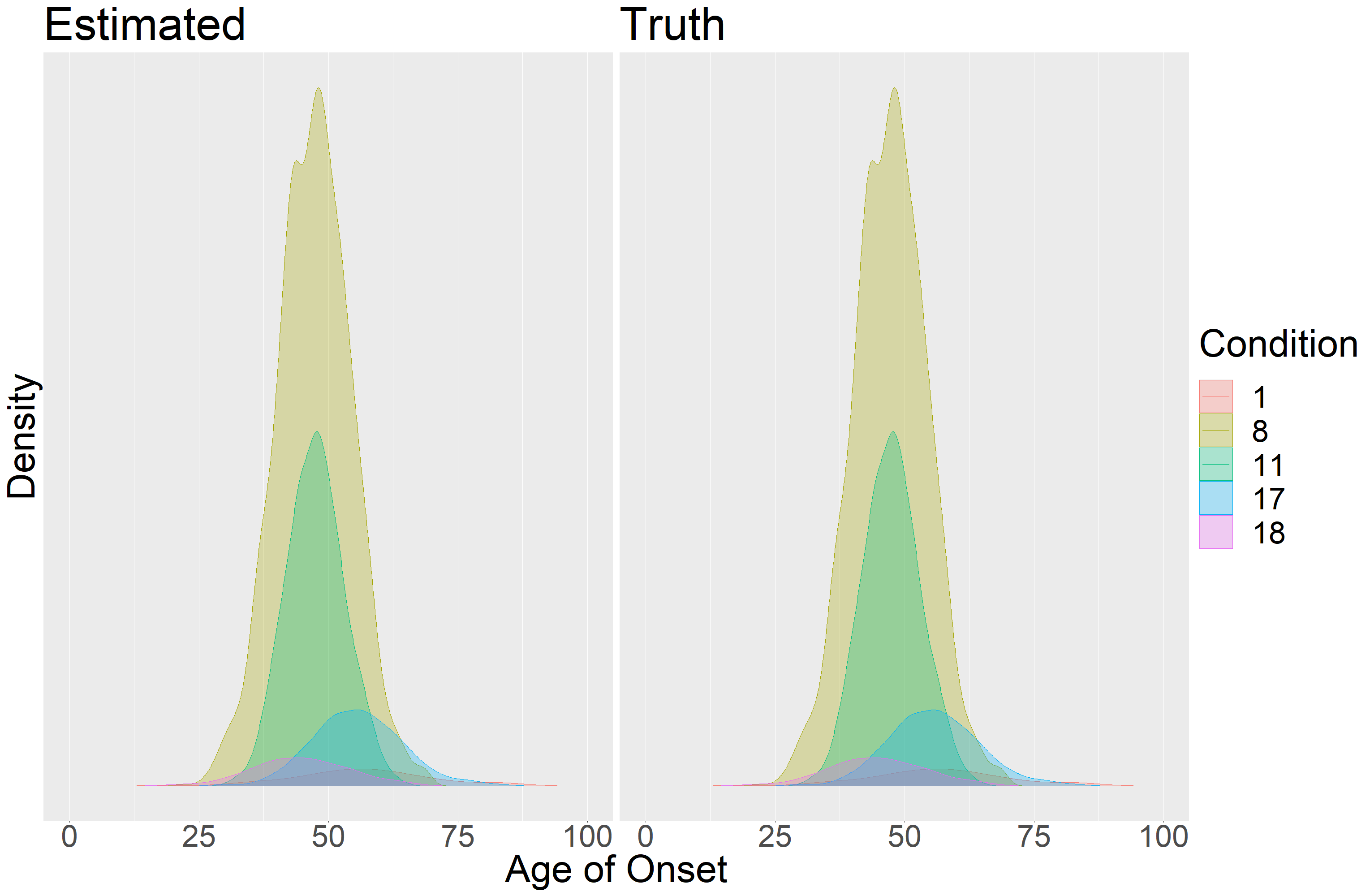}}
\subfigure[][Cluster 5]{\includegraphics[height=0.15\textheight]{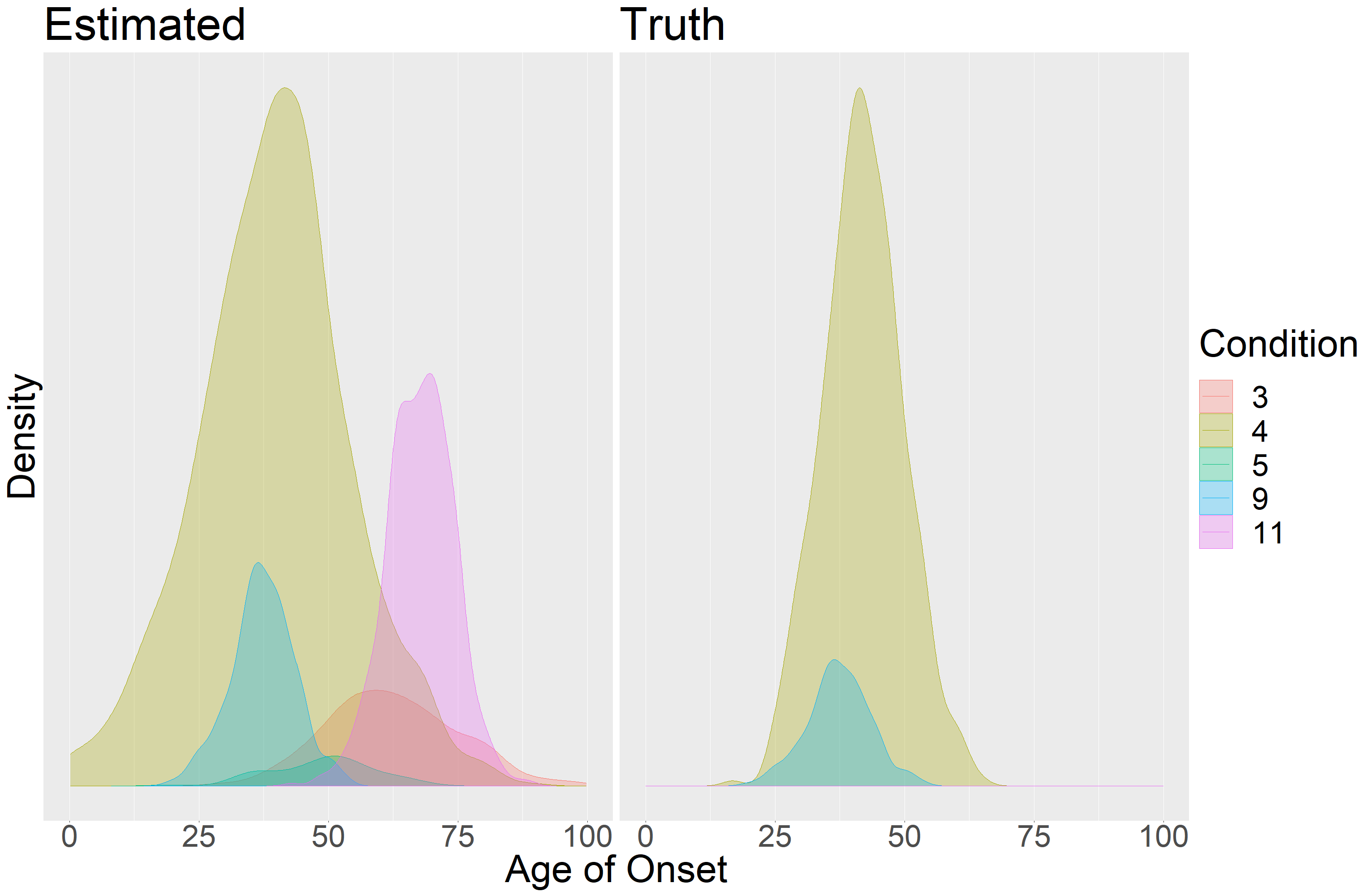}}
\subfigure[][Cluster 6]{\includegraphics[height=0.15\textheight]{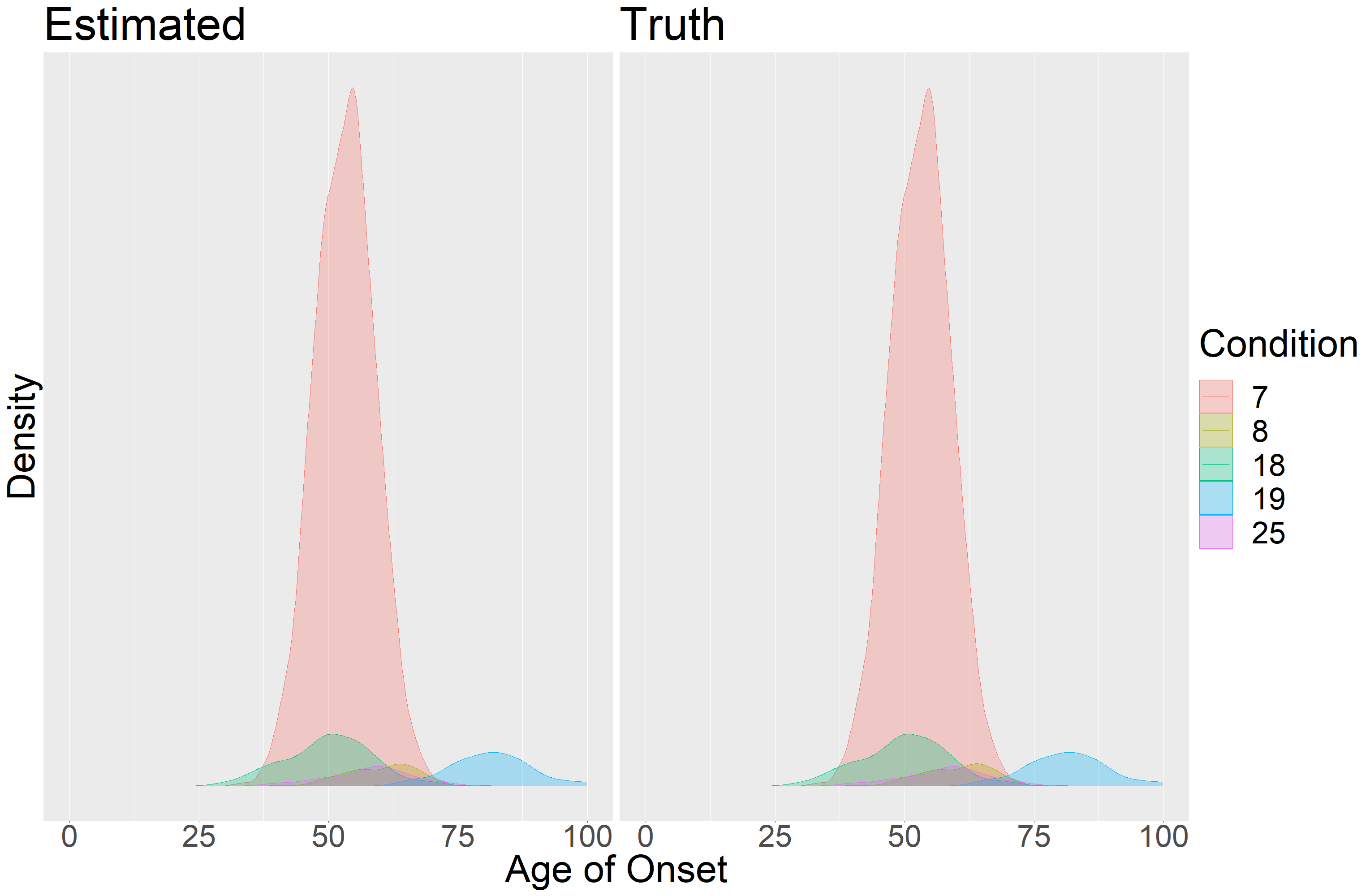}}
\subfigure[][Cluster 7]{\includegraphics[height=0.15\textheight]{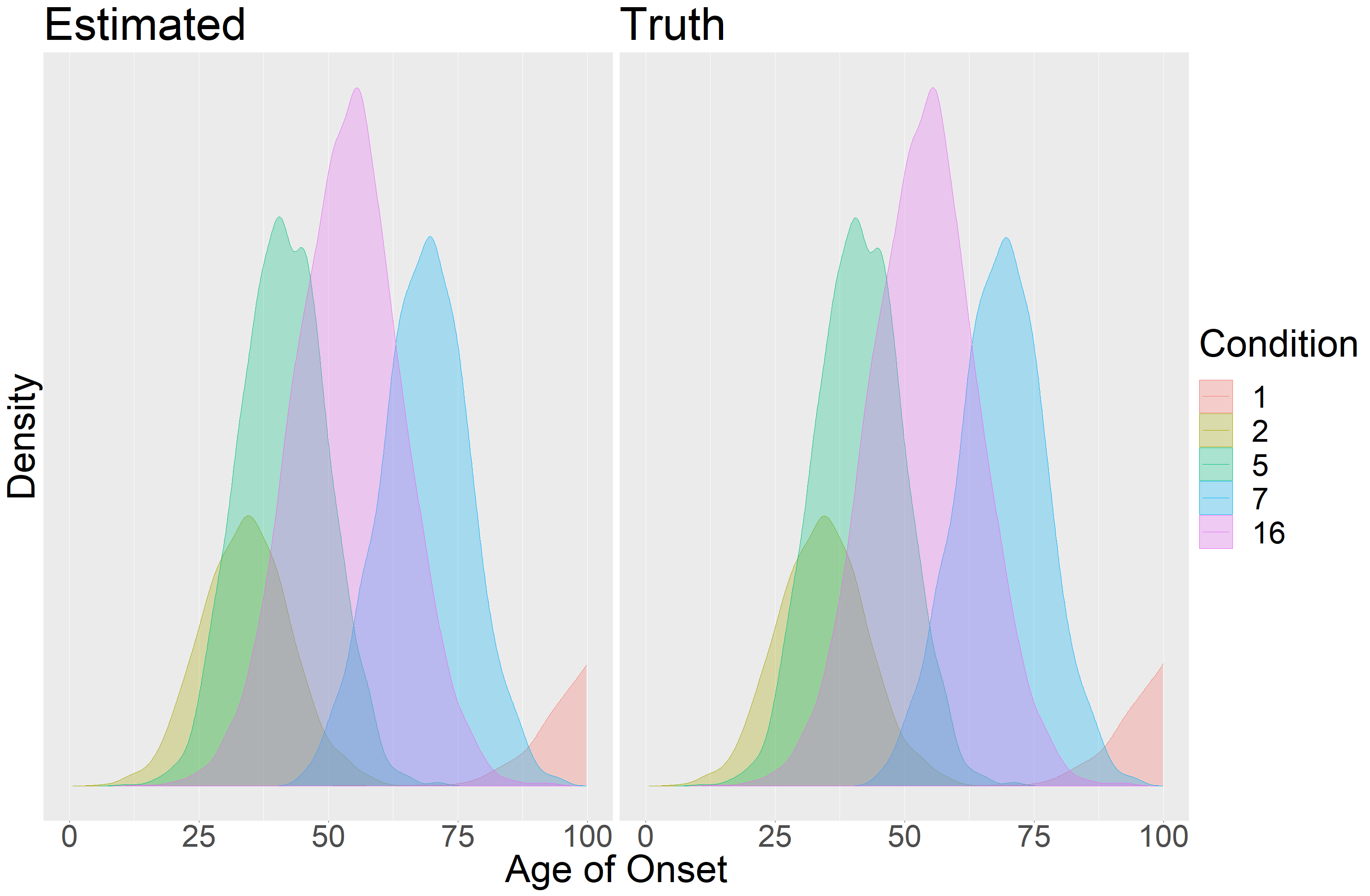}}
\subfigure[][Cluster 8]{\includegraphics[height=0.15\textheight]{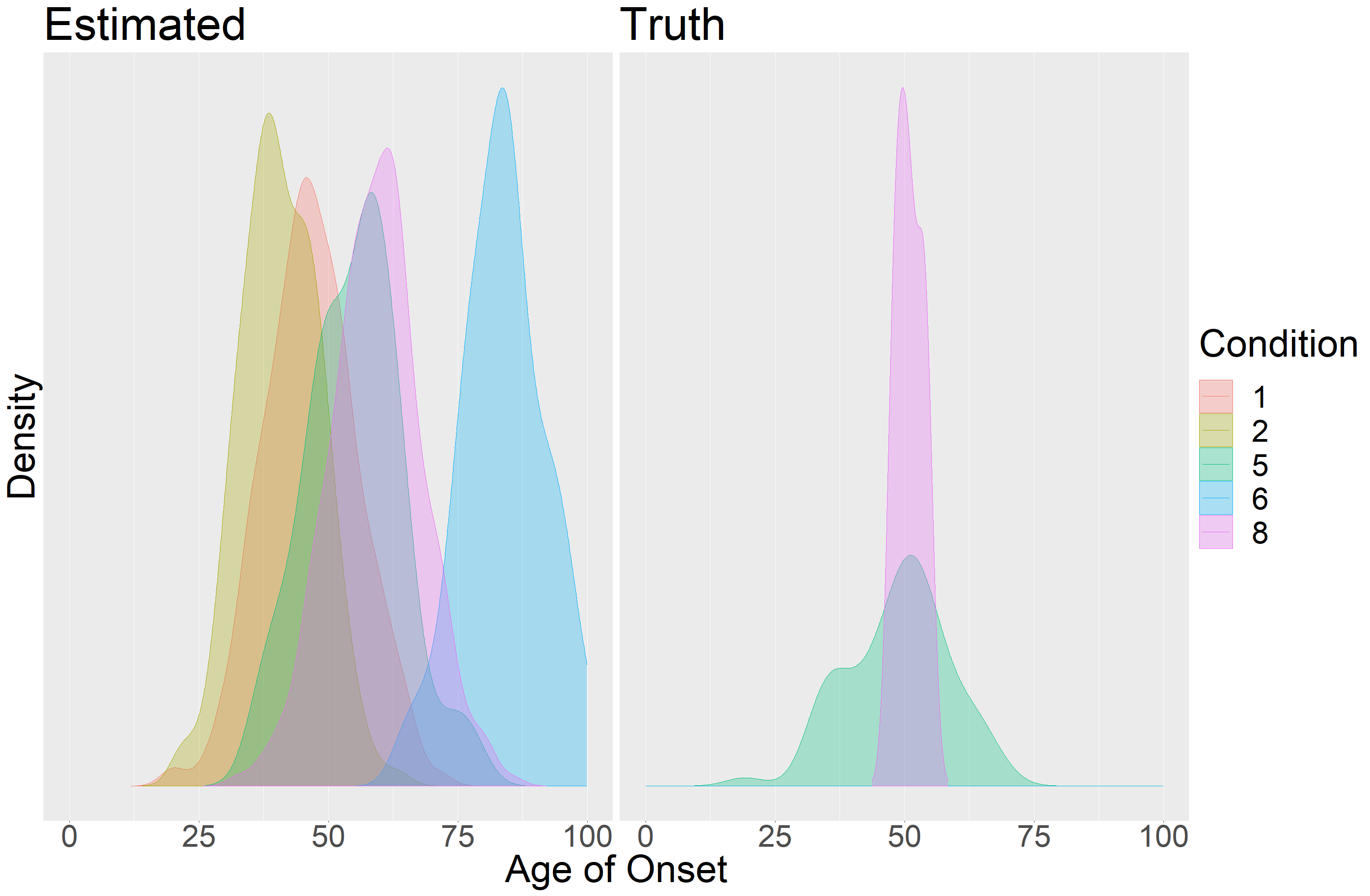}}
\subfigure[][Cluster 9]{\includegraphics[height=0.15\textheight]{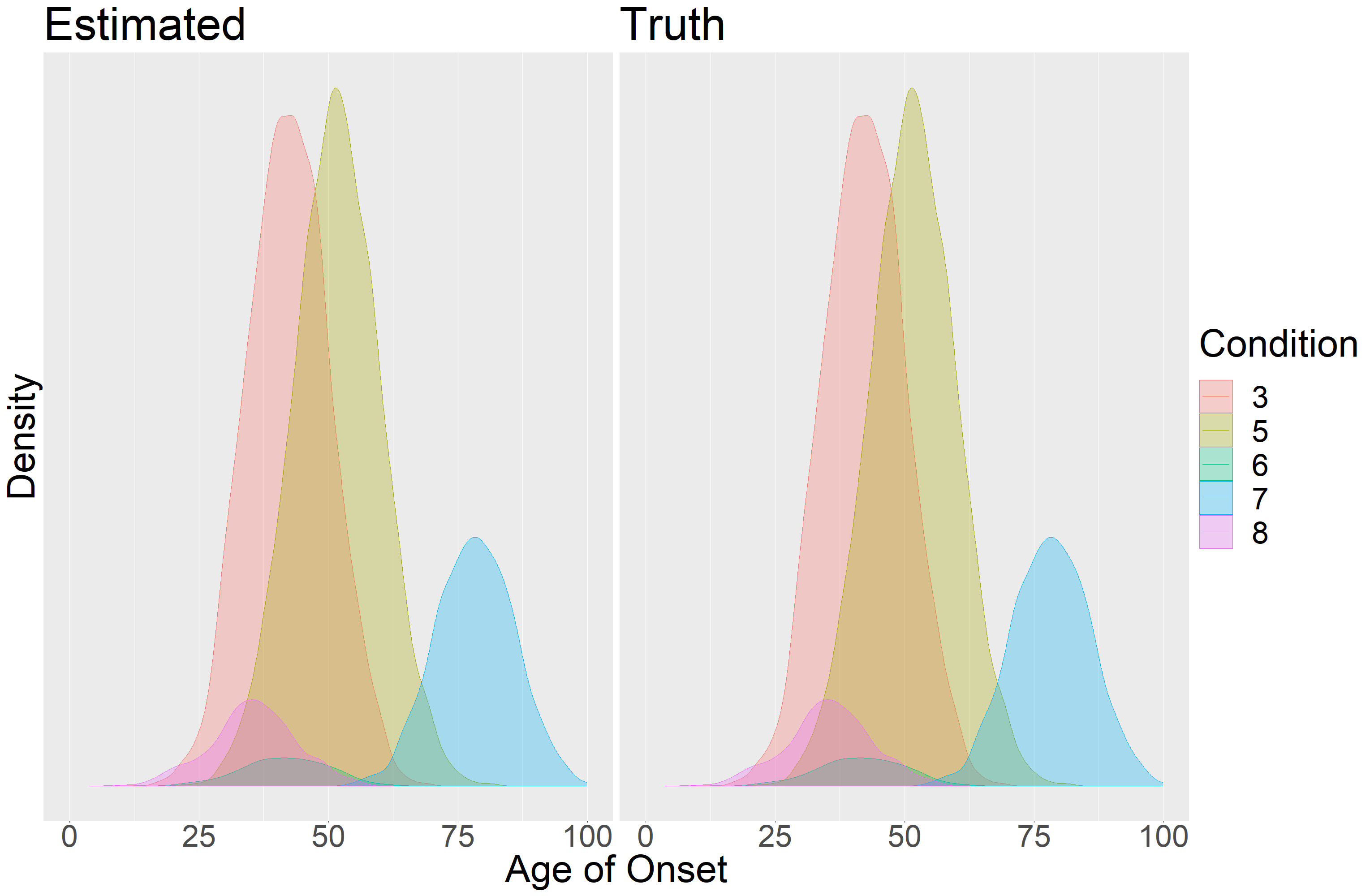}}
\subfigure[][Cluster 10]{\includegraphics[height=0.15\textheight]{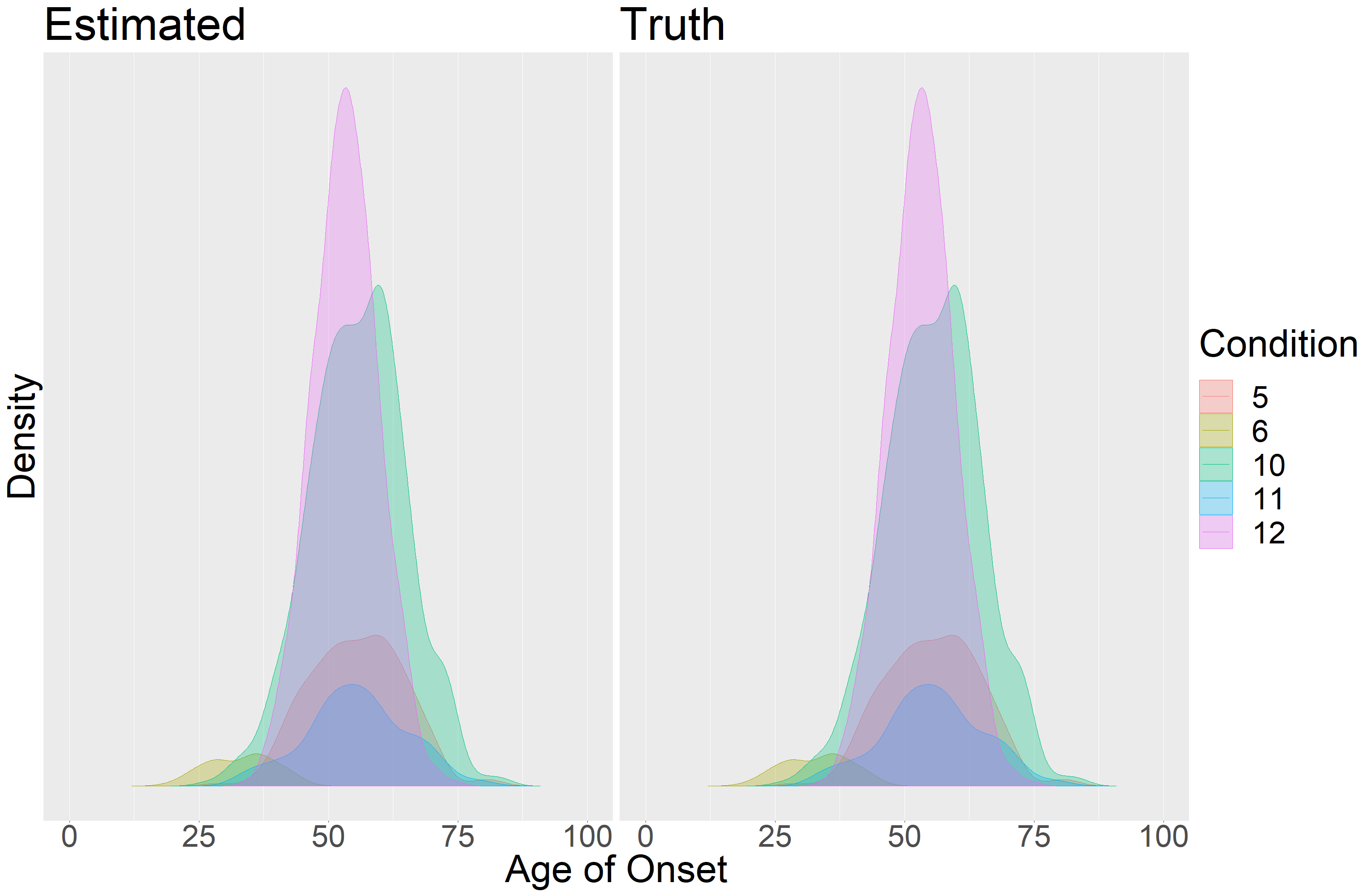}}
\end{figure}

\newpage

\section{Disease Coding}\label{discodes}
\begin{longtable}{|p{0.09\textwidth}|p{0.27\textwidth}|p{0.35\textwidth}|p{0.08\textwidth}|p{0.08\textwidth}|}
\hline
\textbf{Category} & \textbf{Body system}                      & \textbf{Condition} & \textbf{Sex} \textbf{Specific} & \textbf{Lifelong}                                                  \\ \hline
Physical          & Benign   Neoplasm/CIN                     & Benign neoplasm of   brain and other parts of central nervous system & No & No \\ \hline
Physical          & Cancers                                   & Haematological   malignancies    & No & No                                    \\ \hline
Physical          & Cancers                                   & Non-melanoma skin   malignancies   & No & No                                  \\ \hline
Physical          & Cancers                                   & Solid organ   malignancies     & No & No                                      \\ \hline
Physical          & Diseases of the   Circulatory System      & Cardiomyopathy        & No & No                                               \\ \hline
Physical          & Diseases of the   Circulatory System      & Conduction   disorders and other arrhythmias              & No & No           \\ \hline
Physical          & Diseases of the   Circulatory System      & Coronary heart   disease                                    & No & No         \\ \hline
Physical          & Diseases of the   Circulatory System      & Heart valve   disorders                                      & No & No        \\ \hline
Physical          & Diseases of the   Circulatory System      & Atrial fibrillation                                           & No & No       \\ \hline
Physical          & Diseases of the   Circulatory System      & Heart failure                                                   & No & No     \\ \hline
Physical          & Diseases of the   Circulatory System      & Hypertension                                                   & No & No      \\ \hline
Physical          & Diseases of the   Circulatory System      & Peripheral arterial   disease                                  & No & No      \\ \hline
Physical          & Diseases of the   Circulatory System      & Primary pulmonary   hypertension                              & No & No       \\ \hline
Physical          & Diseases of the   Circulatory System      & Stroke                                                         & No & No      \\ \hline
Physical          & Diseases of the   Circulatory System      & Transient ischaemic   attack                                    & No & No     \\ \hline
Physical          & Diseases of the   Digestive System        & Chronic liver   disease                                       & No & No       \\ \hline
Physical          & Diseases of the   Digestive System        & Gastro-oesophageal   reflux, gastritis and similar        & No & No           \\ \hline
Physical          & Diseases of the   Digestive System        & Inflammatory bowel   disease                                & No & No         \\ \hline
Physical          & Diseases of the   Digestive System        & Coeliac disease                                              & No & No        \\ \hline
Physical          & Diseases of the   Digestive System        & Diverticular   disease of intestine (acute and chronic)       & No & No       \\ \hline
Physical          & Diseases of the   Digestive System        & Fatty Liver                                                 & No & No         \\ \hline
Physical          & Diseases of the   Digestive System        & Irritable bowel   syndrome                                  & No & No         \\ \hline
Physical          & Diseases of the   Digestive System        & Peptic ulcer   disease                                     & No & No          \\ \hline
Physical          & Diseases of the   Ear                     & Hearing loss                                                 & No & No        \\ \hline
Physical          & Diseases of the   Ear                     & Meniere disease                                            & No & No          \\ \hline
Physical          & Diseases of the   Endocrine System        & Addisons disease                                         & No & No            \\ \hline
Physical          & Diseases of the   Endocrine System        & Cystic Fibrosis                                          & No & Yes           \\ \hline
Physical          & Diseases of the   Endocrine System        & Diabetes NOS                                               & No & No          \\ \hline
Physical          & Diseases of the   Endocrine System        & Hypo or   hyperthyroidism                                  & No & No          \\ \hline
Physical          & Diseases of the   Endocrine System        & Type 1 diabetes                                            & No & No          \\ \hline
Physical          & Diseases of the   Endocrine System        & Type 2 diabetes                                            & No & No          \\ \hline
Physical          & Diseases of the   Eye                     & Glaucoma                                                   & No & No          \\ \hline
Physical          & Diseases of the   Eye                     & Macular   degeneration                                   & No & No            \\ \hline
Physical          & Diseases of the   Eye                     & Visual impairment   and blindness                        & No & No            \\ \hline
Physical          & Diseases of the   Genitourinary system    & Chronic renal   disease                        & No & No                      \\ \hline
Physical          & Diseases of the   Genitourinary system    & Erectile   dysfunction                            & Male & No                   \\ \hline
Physical          & Diseases of the   Genitourinary system    & Hyperplasia of   prostate                                           & Male & No \\ \hline
Physical          & Diseases of the   Genitourinary system    & Non-acute cystitis                         & No & No                          \\ \hline
Physical          & Diseases of the   Genitourinary system    & Urinary   Incontinence                          & No & No                     \\ \hline
Physical          & Diseases of the   Respiratory System      & Allergic and   chronic rhinitis                            & No & No          \\ \hline
Physical          & Diseases of the   Respiratory System      & Asbestosis                                                & No & No           \\ \hline
Physical          & Diseases of the   Respiratory System      & Asthma                                                    & No & No           \\ \hline
Physical          & Diseases of the   Respiratory System      & Bronchiectasis                                            & No & No           \\ \hline
Physical          & Diseases of the   Respiratory System      & COPD                                                      & No & No           \\ \hline
Physical          & Diseases of the   Respiratory System      & Sleep apnoea                                              & No & No           \\ \hline
Physical          & Haematological/Immunological   conditions & Iron and vitamin   deficiency anaemia                     & No & No           \\ \hline
Physical          & Haematological/Immunological   conditions & Immunodeficiencies                                        & No & No           \\ \hline
Physical          & Haematological/Immunological   conditions & Sarcoidosis                                               & No & No           \\ \hline
Physical          & Haematological/Immunological   conditions & Sickle-cell anaemia                                      & No & Yes            \\ \hline
Physical          & Haematological/Immunological   conditions & Thalassaemia                                             & No & Yes            \\ \hline
Physical          & Infectious   Diseases                     & HIV                                                       & No & No           \\ \hline
Physical          & Infectious   Diseases                     & Tuberculosis                                             & No & No            \\ \hline
Mental            & Mental Health   Disorders                 & Alcohol Problems                                          & No & No           \\ \hline
Mental            & Mental Health   Disorders                 & Anorexia and   bulimia nervosa                            & No & No           \\ \hline
Mental            & Mental Health   Disorders                 & Anxiety disorders                                         & No & No           \\ \hline
Mental            & Mental Health   Disorders                 & Autism and   Asperger's syndrome                          & No & Yes           \\ \hline
Mental            & Mental Health   Disorders                 & Bipolar affective   disorder and mania                & No & No               \\ \hline
Mental            & Mental Health   Disorders                 & Dementia                                                  & No & No           \\ \hline
Mental            & Mental Health   Disorders                 & Depression                                               & No & No            \\ \hline
Mental            & Mental Health   Disorders                 & Intellectual   disability                                & No & Yes            \\ \hline
Mental            & Mental Health   Disorders                 & Obsessive-compulsive   disorder                          & No & No            \\ \hline
Mental            & Mental Health   Disorders                 & Other psychoactive   substance misuse                   & No & No             \\ \hline
Mental            & Mental Health   Disorders                 & Post-traumatic   stress disorder                         & No & No            \\ \hline
Mental            & Mental Health   Disorders                 & Schizophrenia,   schizotypal and delusional disorders   & No & No             \\ \hline
Physical          & Musculoskeletal   conditions              & Inflammatory   arthritis and other inflammatory conditions   & No & No        \\ \hline
Physical          & Musculoskeletal   conditions              & Osteoporosis and   vertebral crush fractures            & No & No             \\ \hline
Physical          & Musculoskeletal   conditions              & Gout                                                      & No & No           \\ \hline
Physical          & Musculoskeletal   conditions              & Osteoarthritis   (excl spine)                              & No & No          \\ \hline
Physical          & Musculoskeletal   conditions              & Spinal stenosis                                            & No & No          \\ \hline
Physical          & Neurological   conditions                 & Peripheral or   autonomic neuropathy                       & No & No          \\ \hline
Physical          & Neurological   conditions                 & Cerebral Palsy                                            & No & Yes           \\ \hline
Physical          & Neurological   conditions                 & Epilepsy                                                  & No & No           \\ \hline
Physical          & Neurological   conditions                 & Migraine                                                  & No & No           \\ \hline
Physical          & Neurological   conditions                 & Motor neuron   disease                                     & No & No          \\ \hline
Physical          & Neurological   conditions                 & Multiple sclerosis                                       & No & No            \\ \hline
Physical          & Neurological   conditions                 & Myasthenia gravis                                        & No & No            \\ \hline
Physical          & Neurological   conditions                 & Parkinson's disease                                       & No & No           \\ \hline
Physical          & Neurological   conditions                 & Postviral fatigue   syndrome, neurasthenia and fibromyalgia     & No & No     \\ \hline
Physical          & Perinatal   conditions                    & Down's syndrome                                             & No & Yes         \\ \hline
Physical          & Skin   conditions                         & Psoriasis                                                & No & No            \\ \hline
\end{longtable}

\newpage
\section{Selecting the number of clusters}\label{kselect}

\begin{figure}[ht]
\centering
\includegraphics[width=0.75\textwidth]
{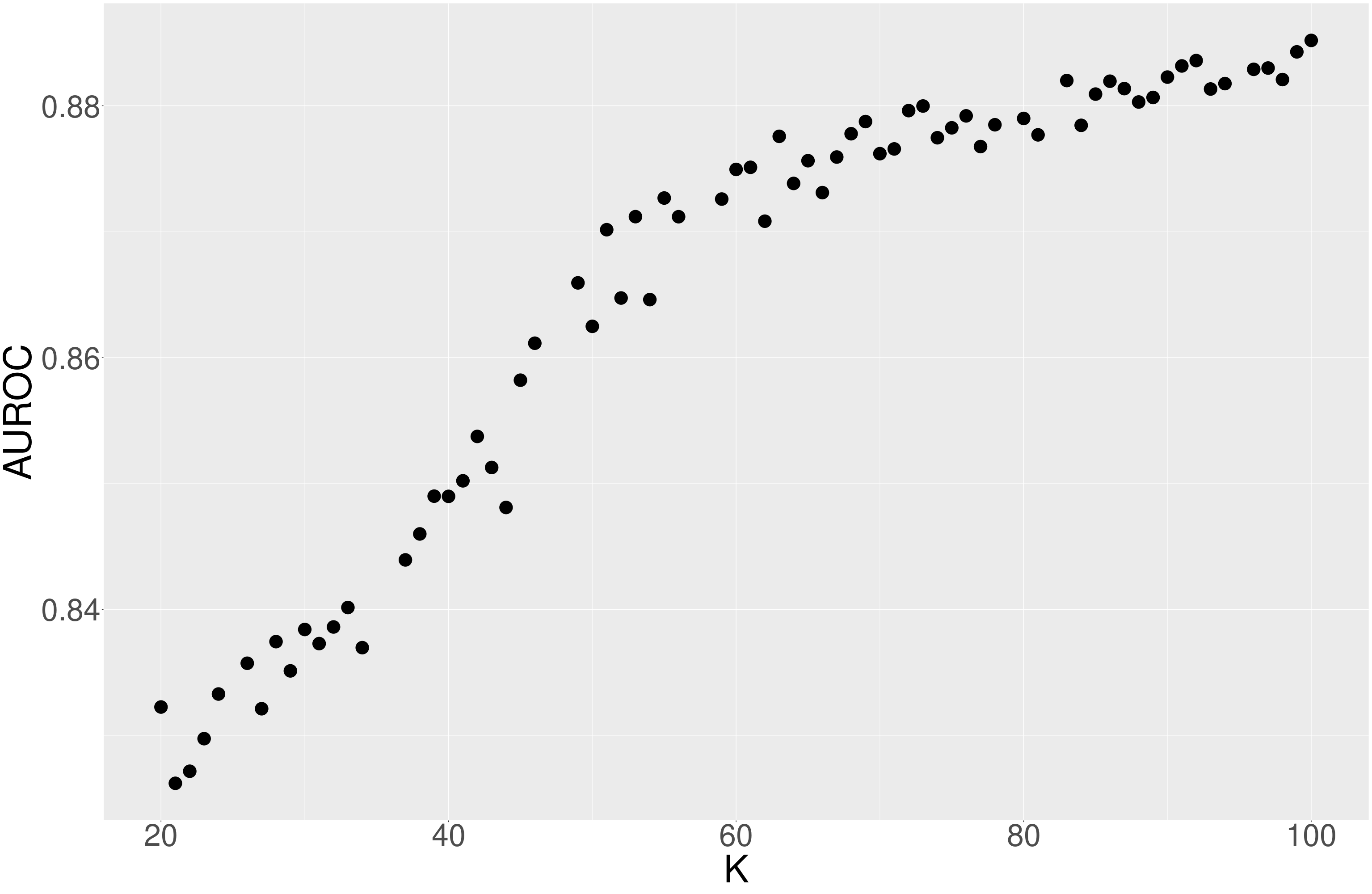}
\caption{We show AUROC for the disease presence predictions for different choices of the number of clusters, $K$.}
\label{fig:elbow}
\end{figure}

Figure~\ref{fig:elbow} shows the AUROC for the $10$ years of withheld disease presence predictions using different choices of $K$ between $20$ and $100$. We observe an elbow at approximately $K=50$ and therefore use $50$ clusters in our analysis. Figure~\ref{fig:K50100} below show hypothetical risk profiles for the same individual using $K=50$ and $K=100$. We note that although there are some differences, the profiles are largely the same and indicate robustness to the choice of $K$.

\begin{figure}[ht]
\centering
\subfigure[][$K=50$]{\includegraphics[width=0.4\textwidth]{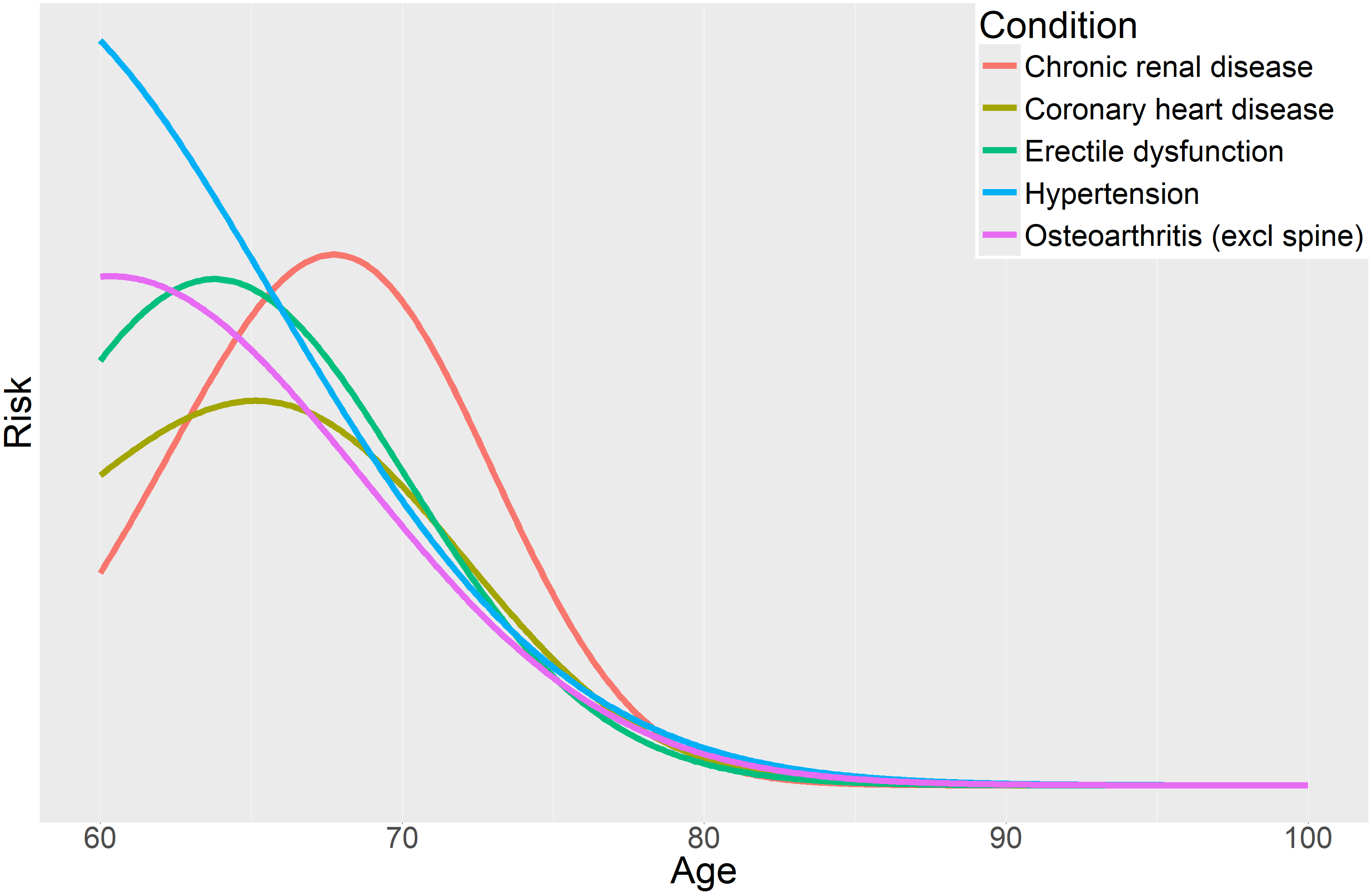}}
\subfigure[][$K=100$]{\includegraphics[width=0.4\textwidth]{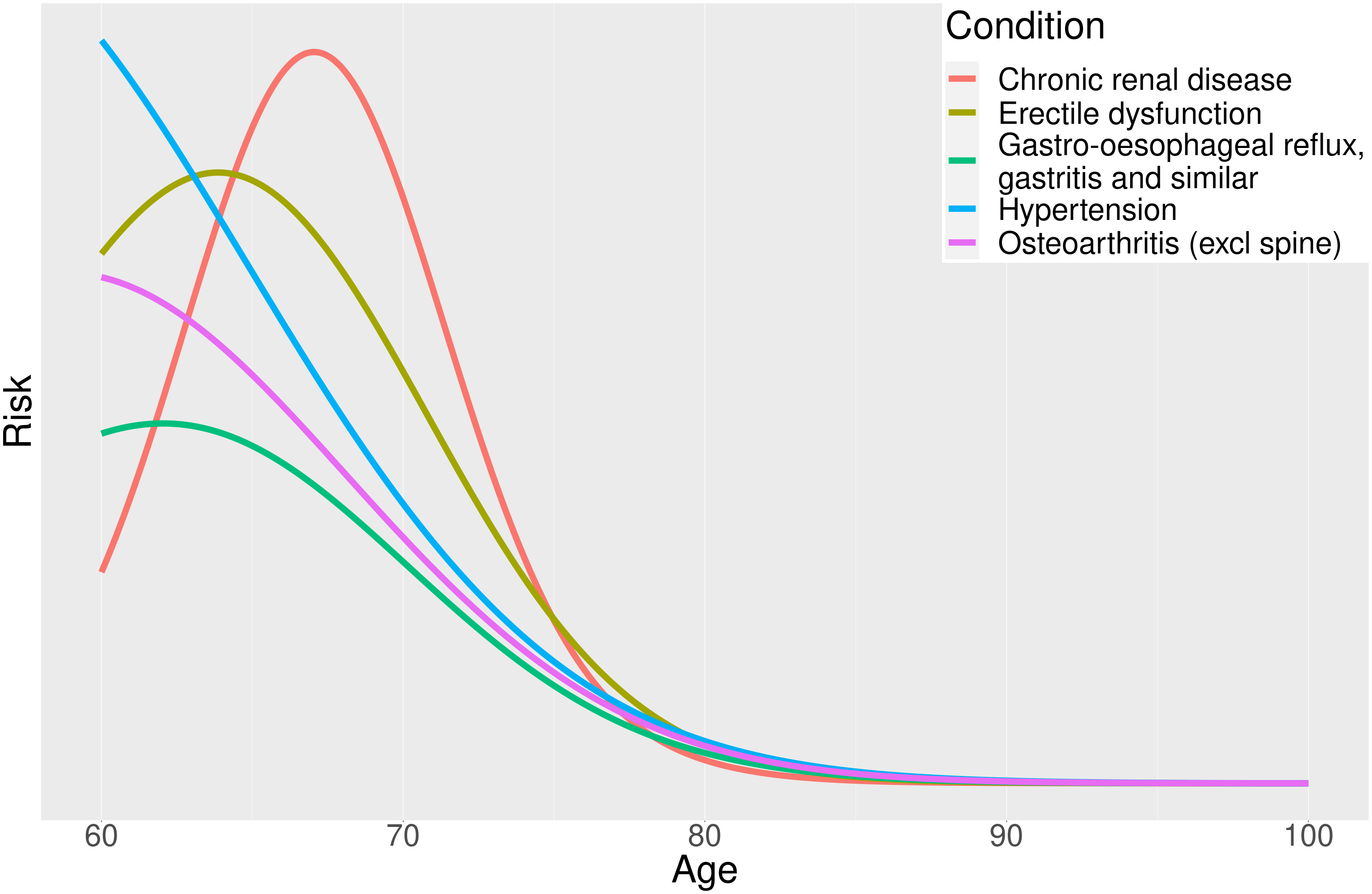}}
\caption{Temporal risk profiles for a $60$ year old man who was diagnosed with \textsc{type 2 diabetes} aged 54. Shown are the predicted risk profiles for the model using $K=50$ and $K=100$. Whilst there are some differences, the profiles are largely similar.}
\label{fig:K50100}
\end{figure}

\clearpage
\section{UK Biobank All Temporal Profiles}\label{alltempprofs}

We present all of the temporal profiles here, with the empirical ages at diagnosis shown as histograms with the fitted ages at diagnosis overlaid.

\begin{figure}[ht]
\centering
\includegraphics[width=\textwidth]{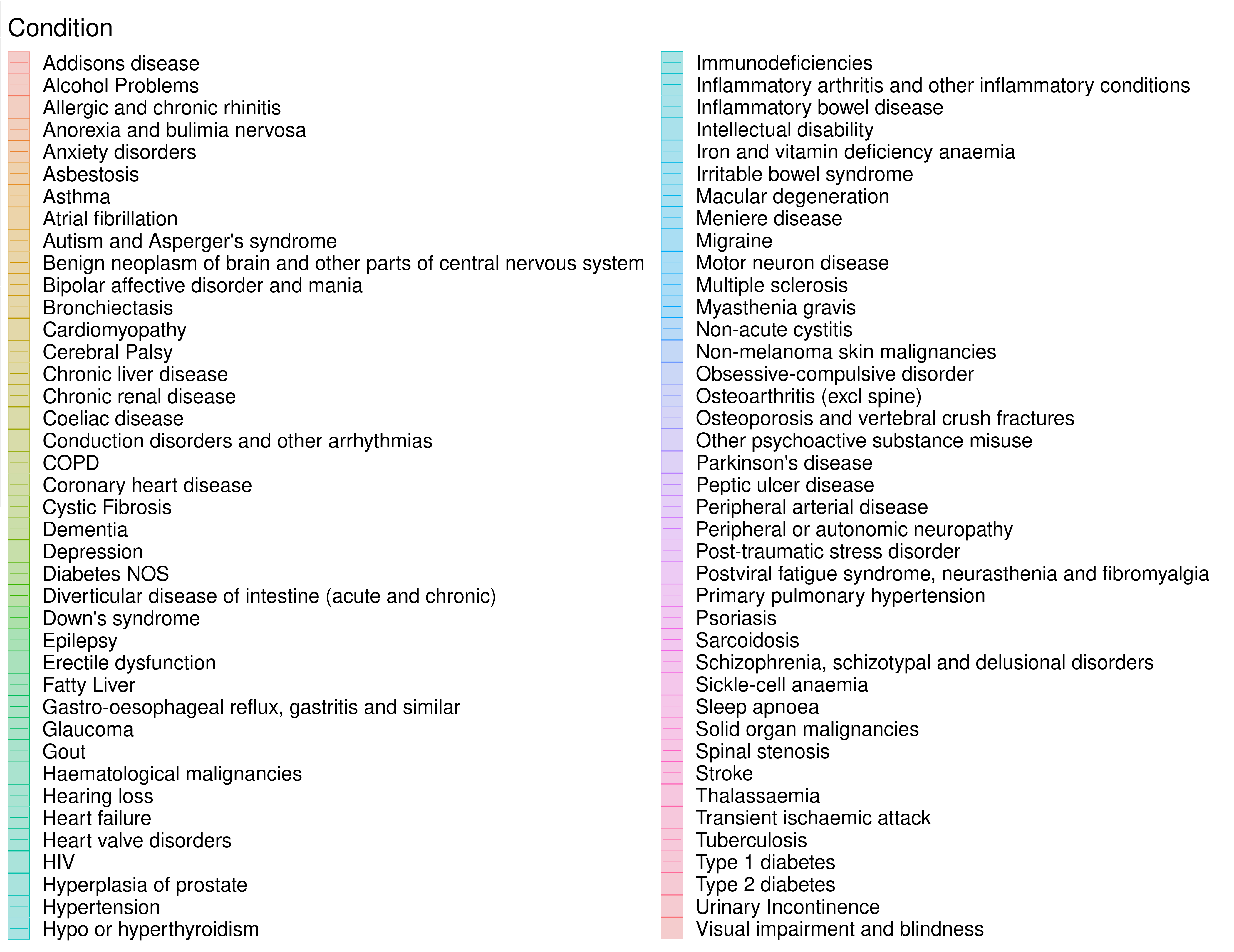}
\end{figure}

\iftrue

\begin{figure}[ht]
\centering
\includegraphics[width=0.45\textwidth]{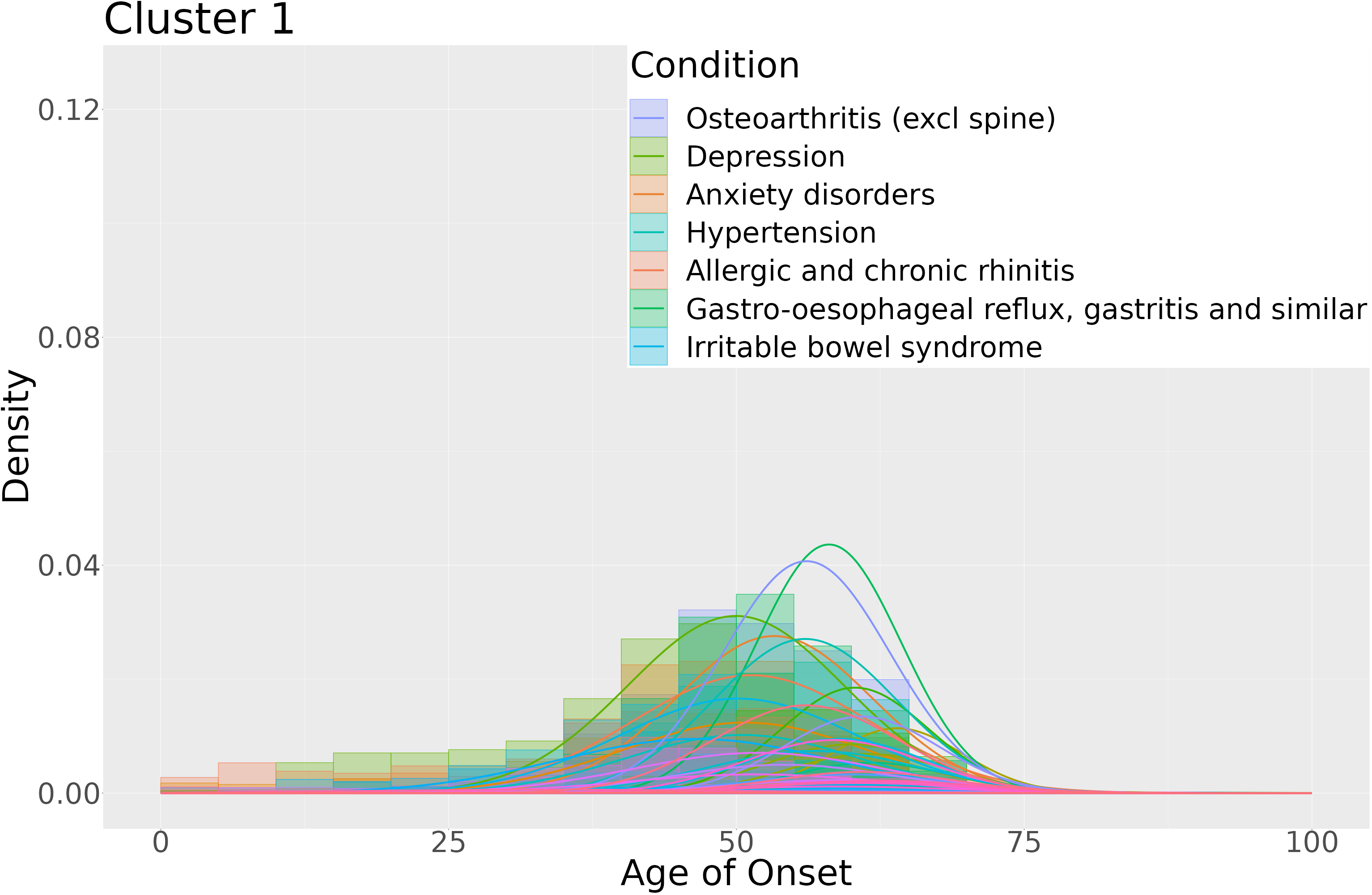}\qquad
\subfigure{\includegraphics[width=0.45\textwidth]{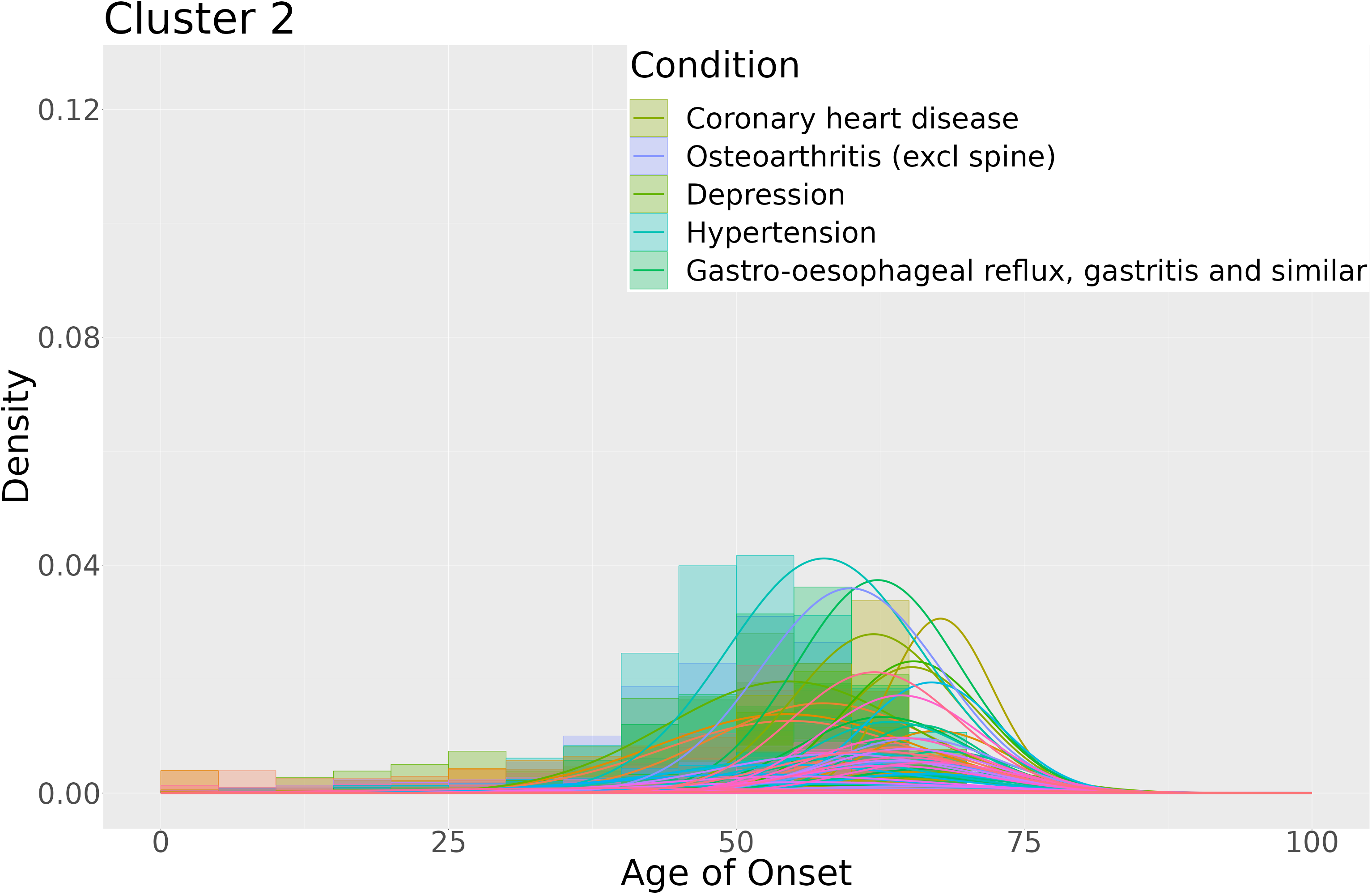}}
\end{figure}

\begin{figure}[ht]
\centering
\includegraphics[width=0.45\textwidth]{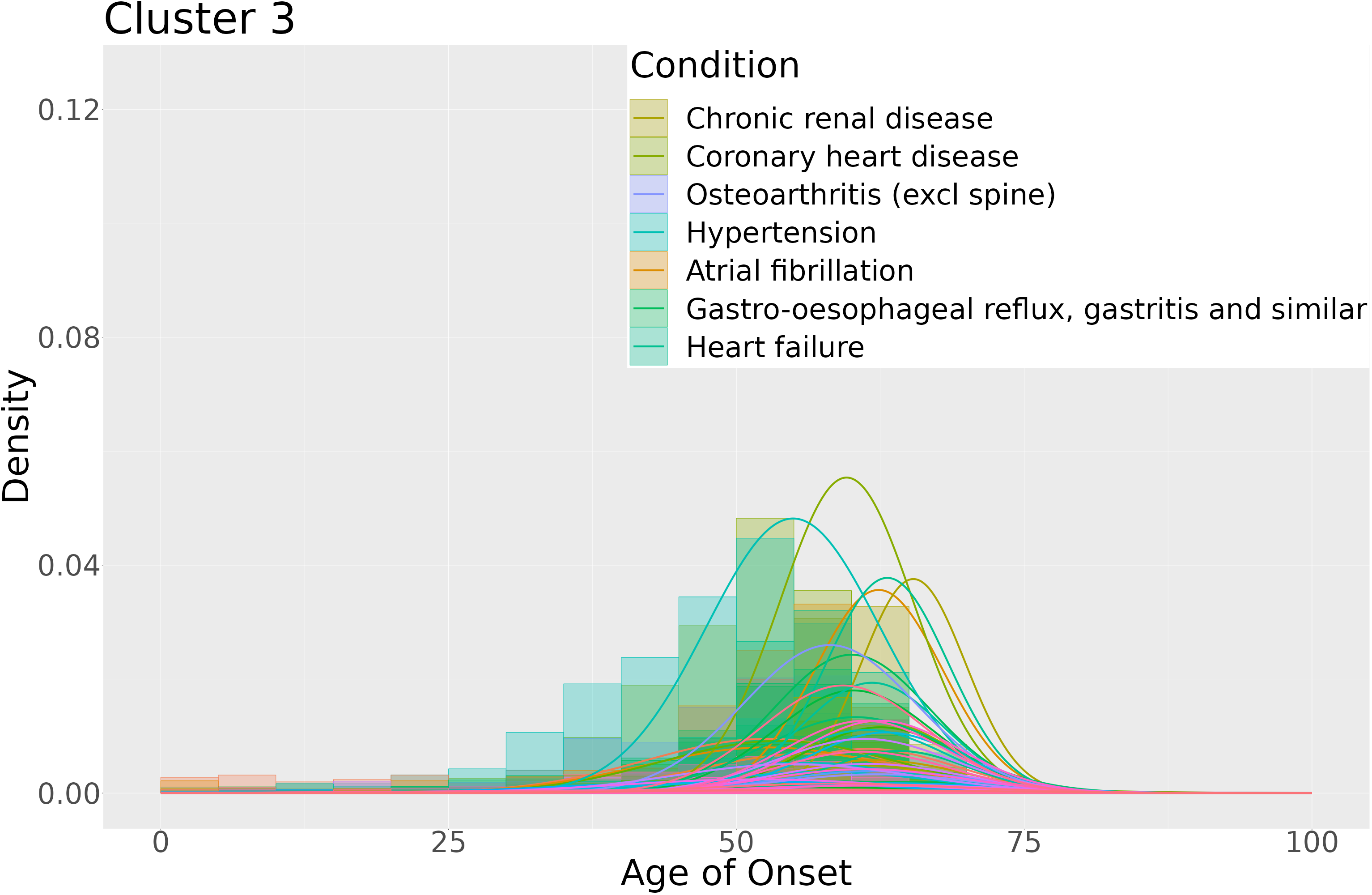}\qquad
\subfigure{\includegraphics[width=0.45\textwidth]{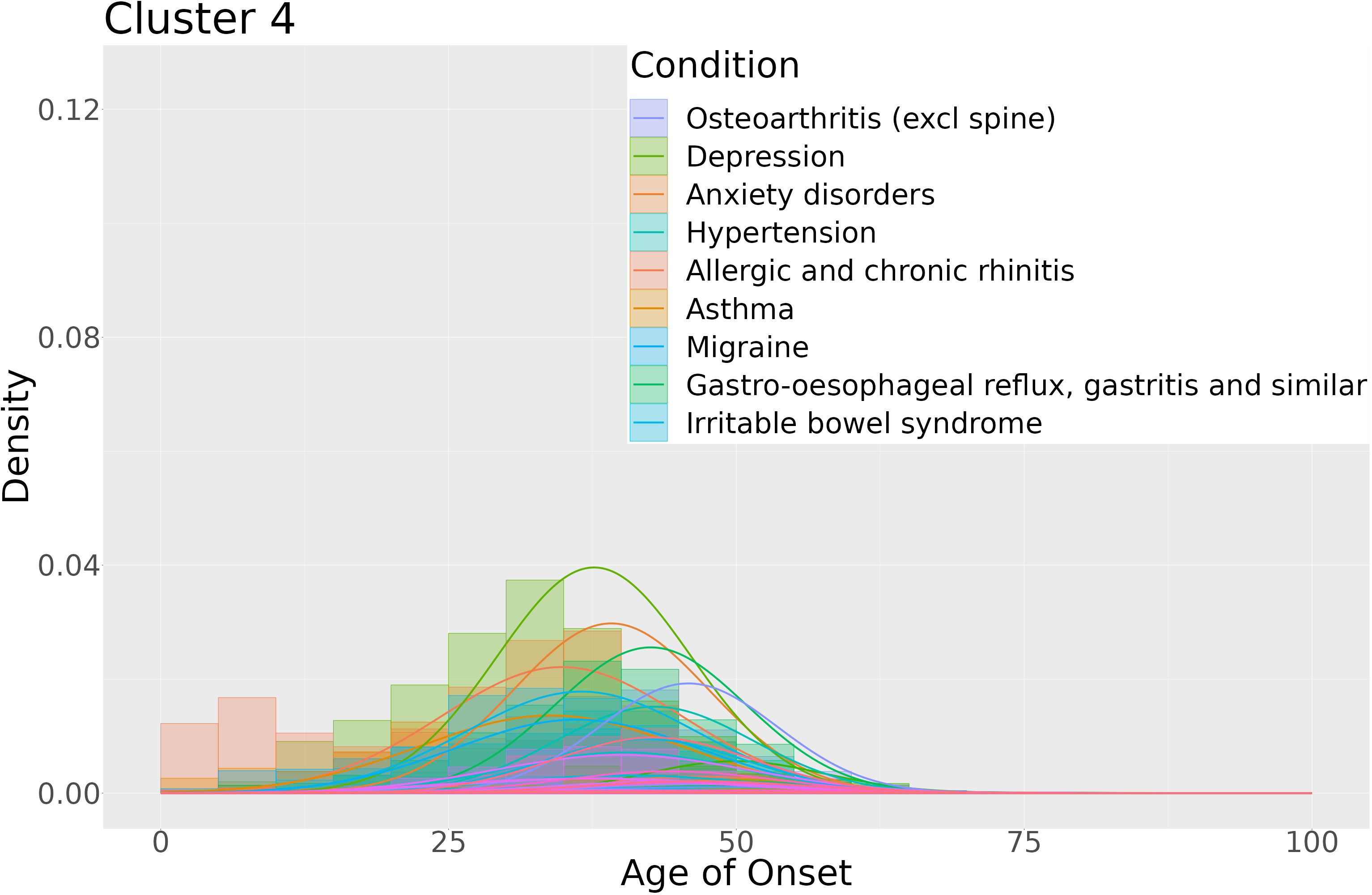}}
\end{figure}

\begin{figure}[ht]
\centering
\includegraphics[width=0.45\textwidth]{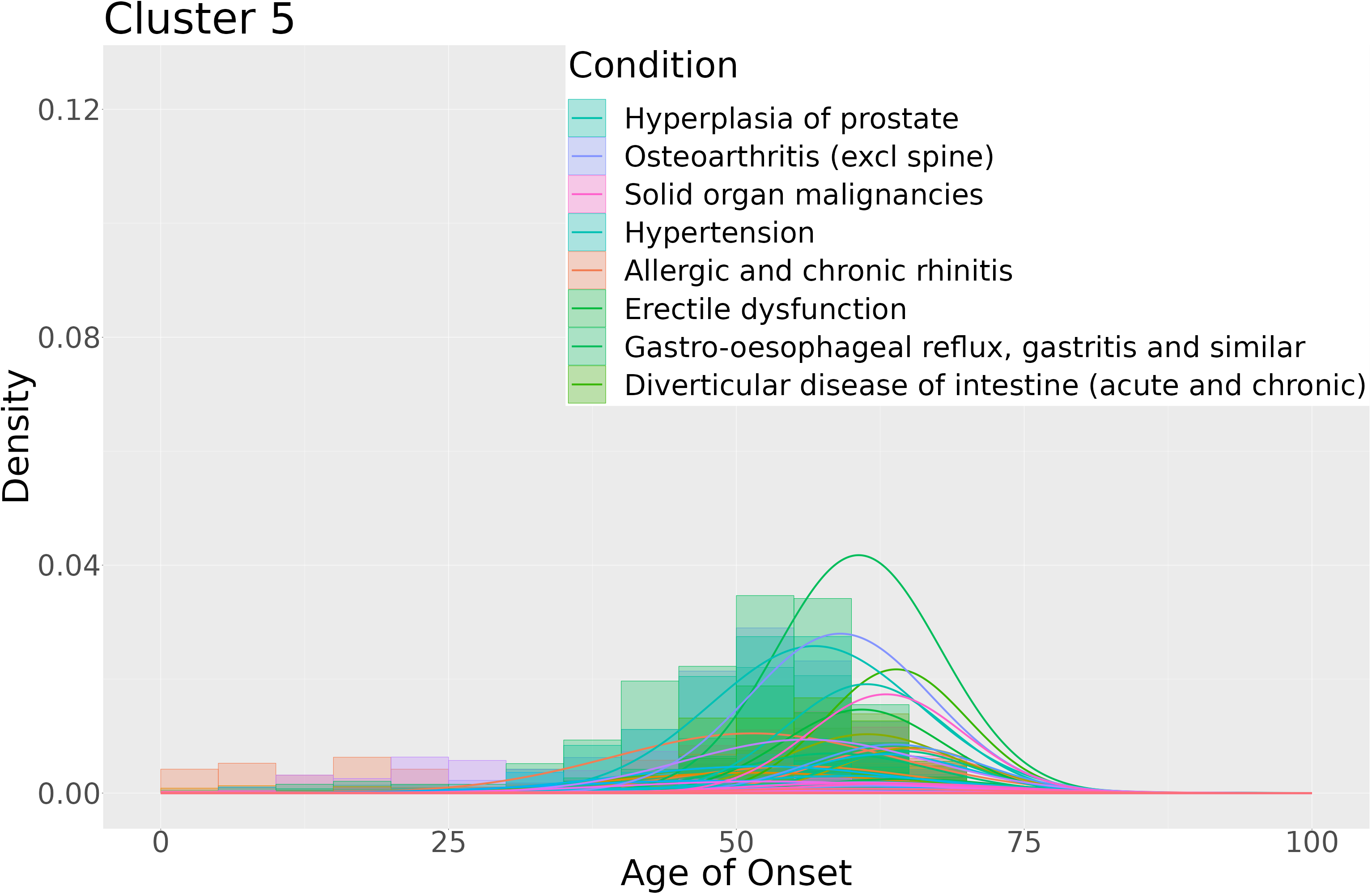}\qquad
\subfigure{\includegraphics[width=0.45\textwidth]{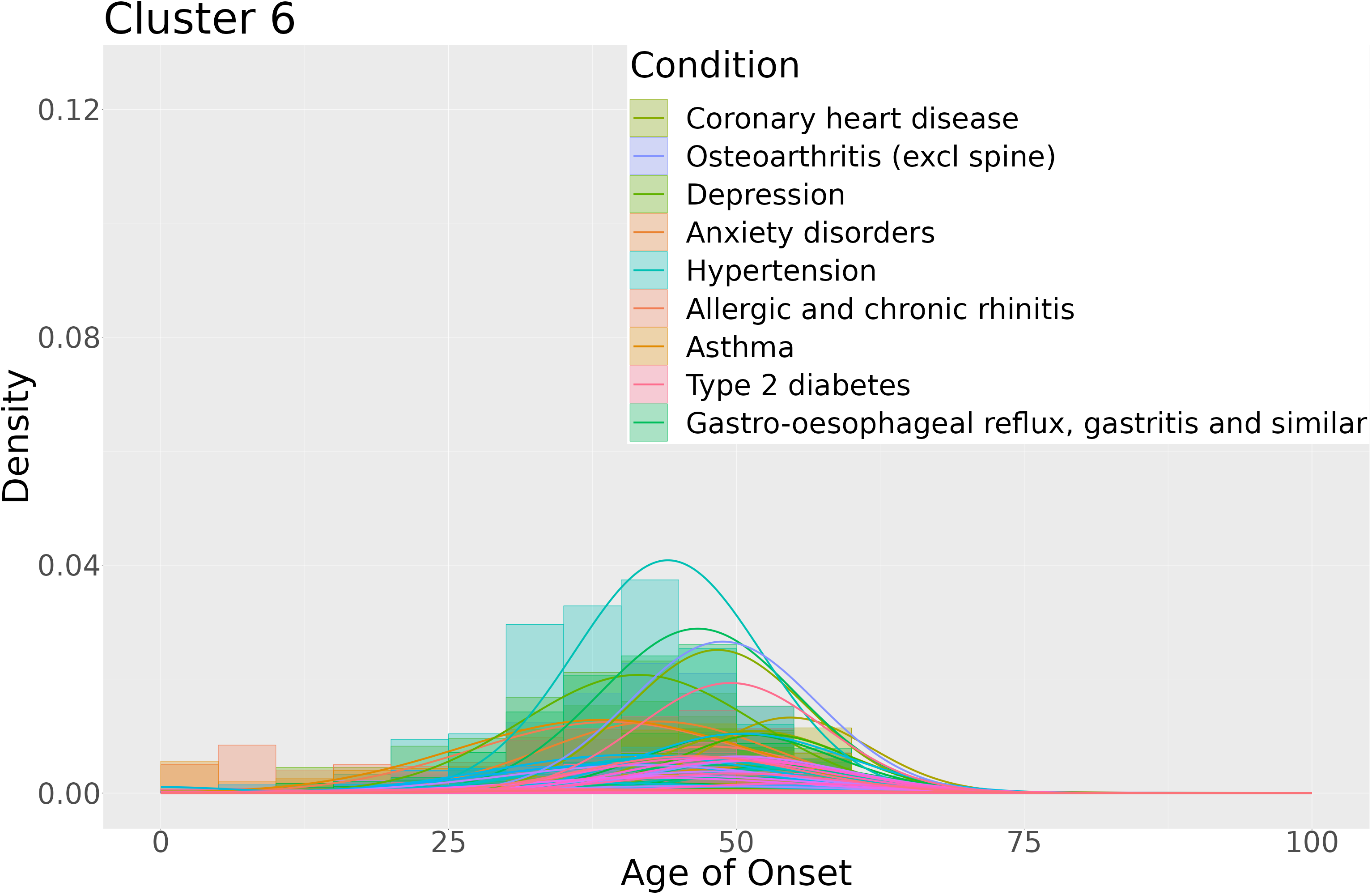}}
\end{figure}

\begin{figure}[ht]
\centering
\includegraphics[width=0.45\textwidth]{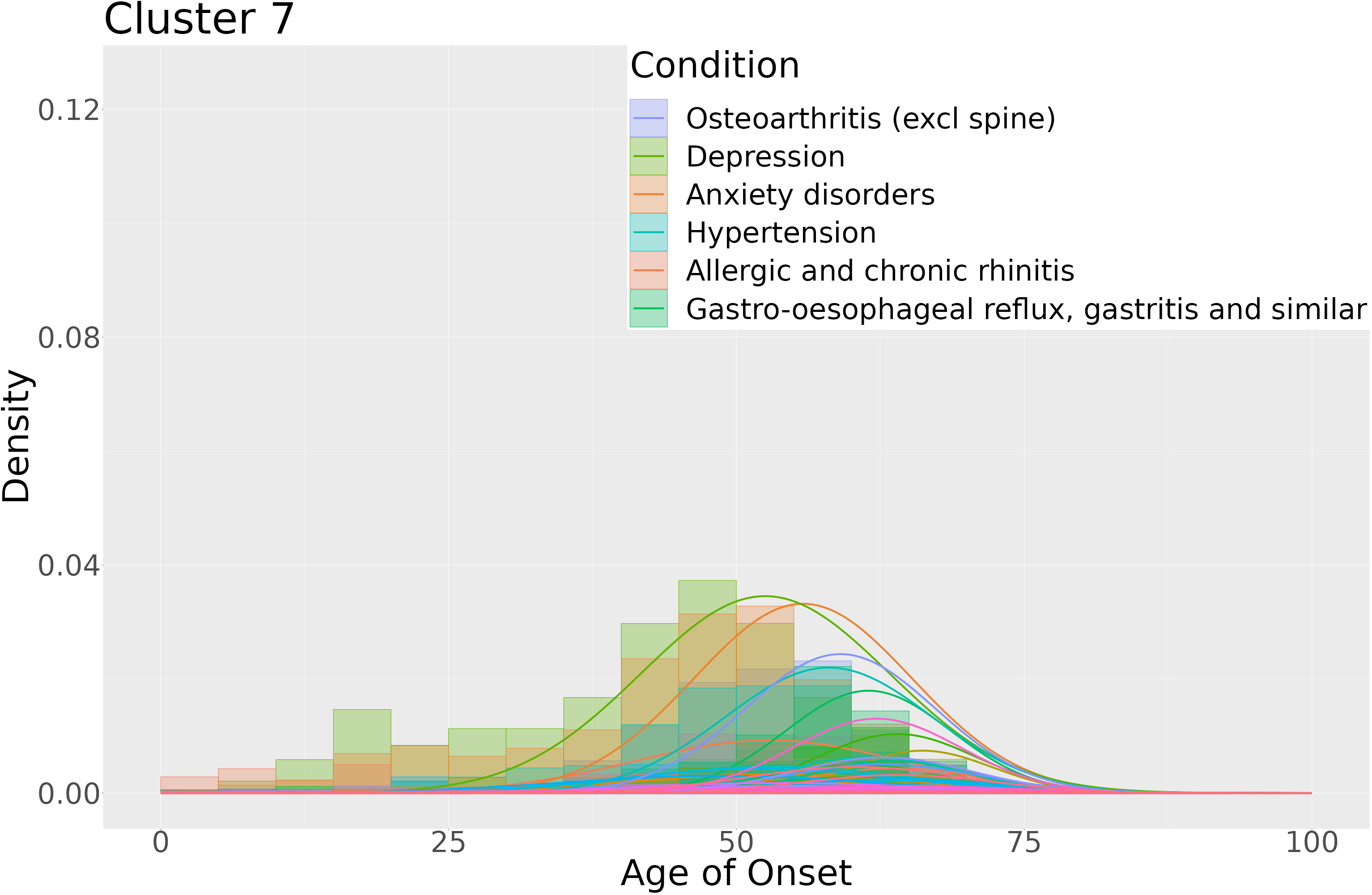}\qquad
\subfigure{\includegraphics[width=0.45\textwidth]{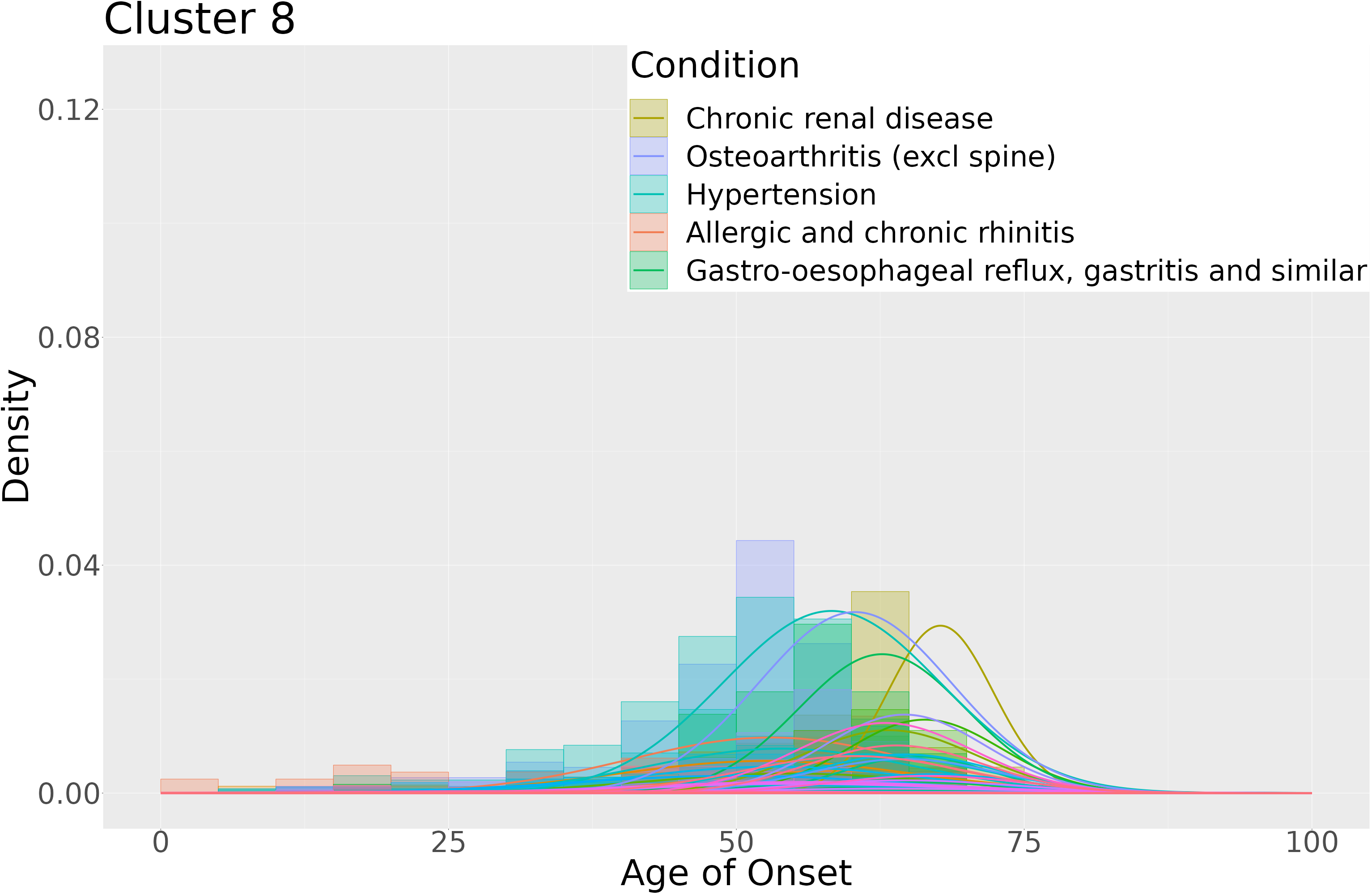}}
\end{figure}

\begin{figure}[ht]
\centering
\includegraphics[width=0.45\textwidth]{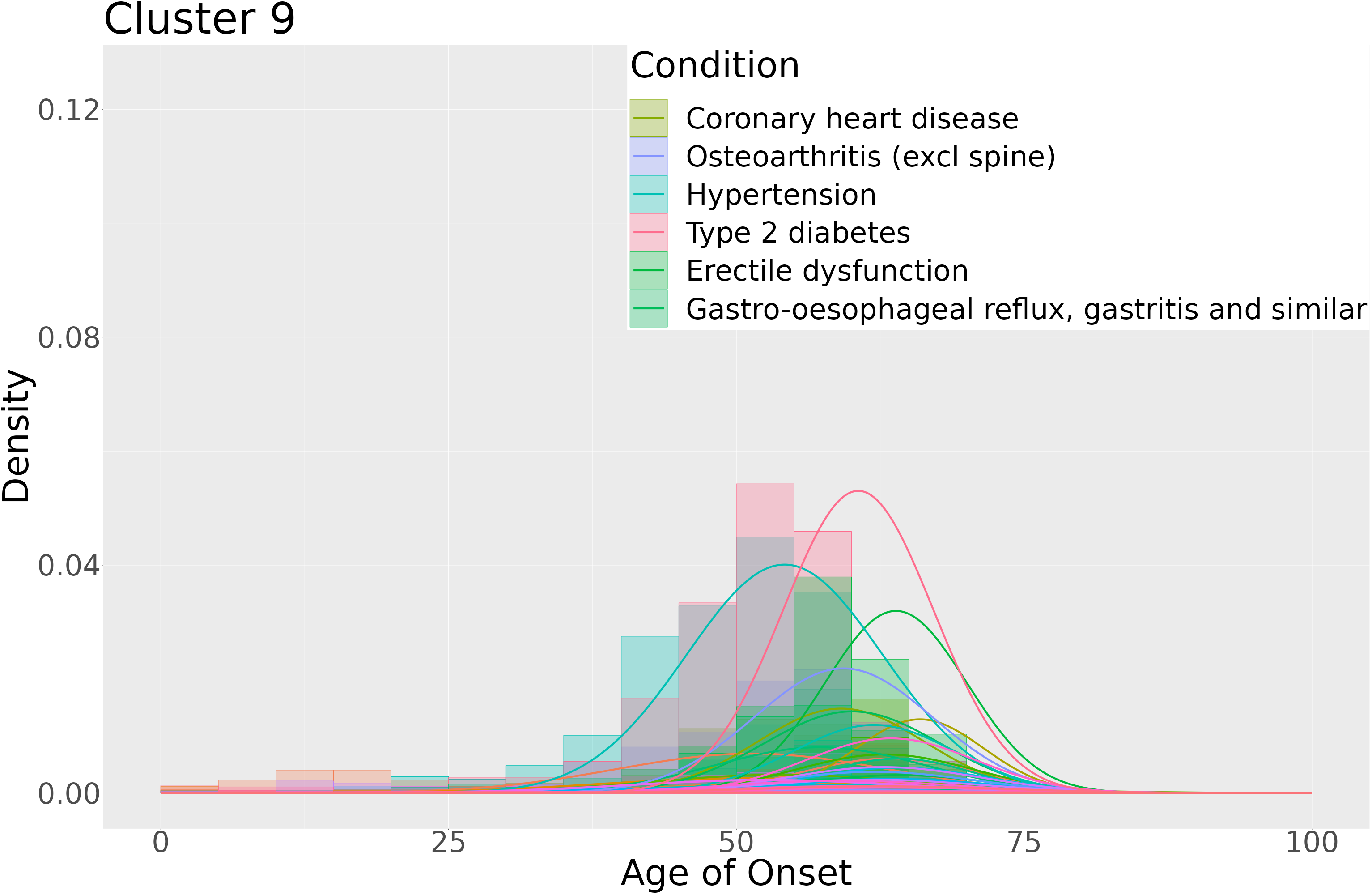}\qquad
\subfigure{\includegraphics[width=0.45\textwidth]{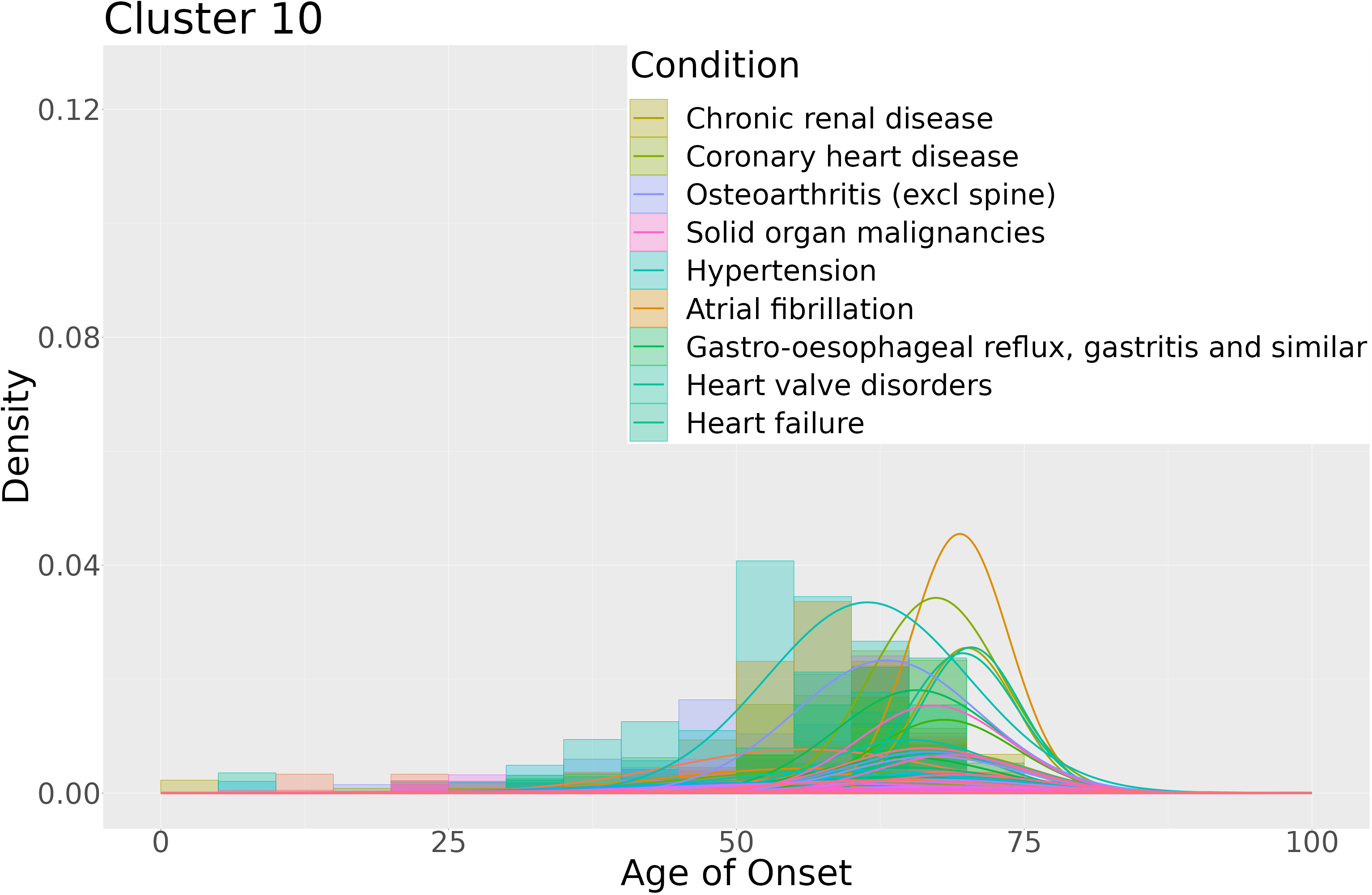}}
\end{figure}

\begin{figure}[ht]
\centering
\includegraphics[width=0.45\textwidth]{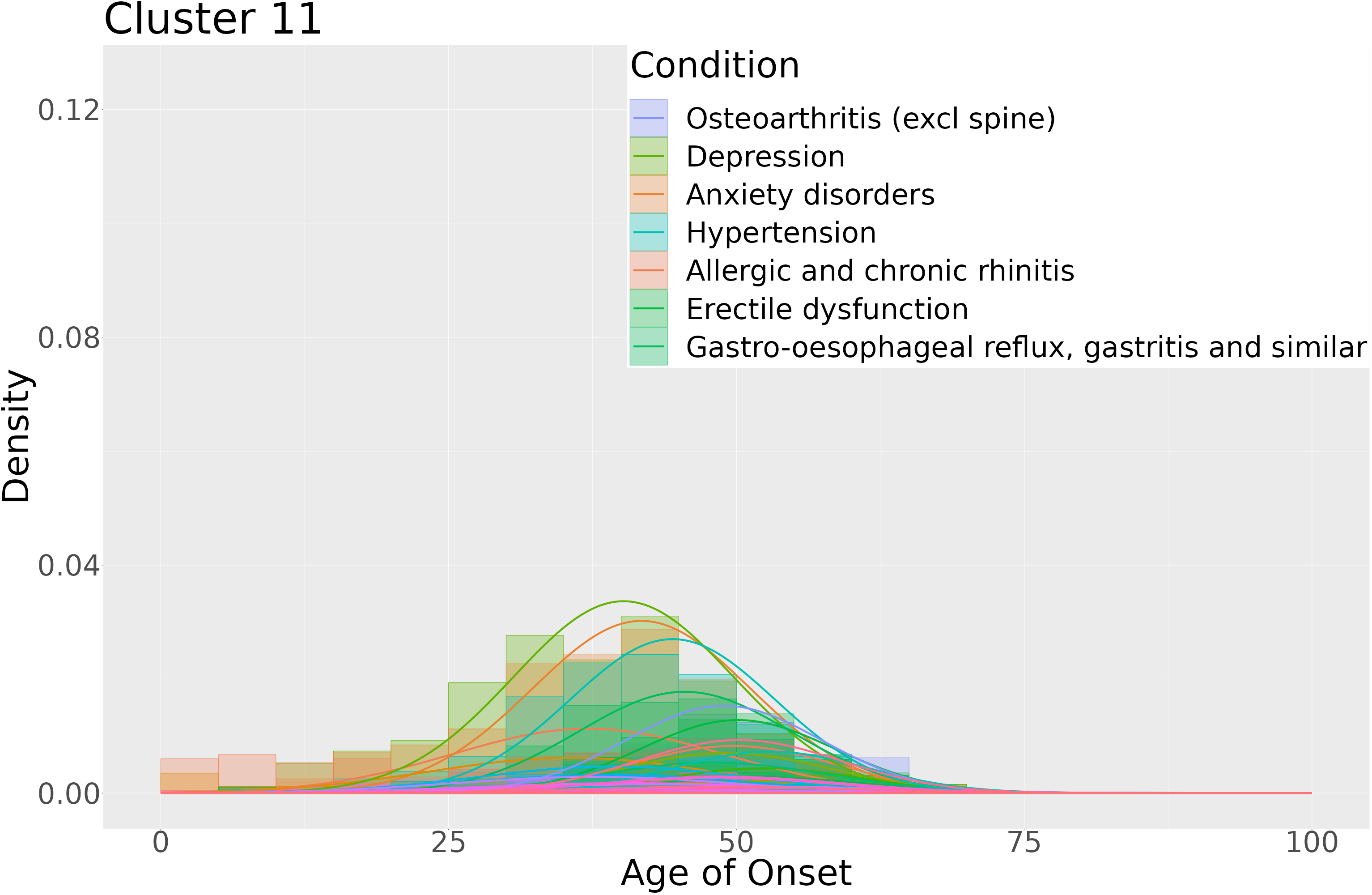}\qquad
\subfigure{\includegraphics[width=0.45\textwidth]{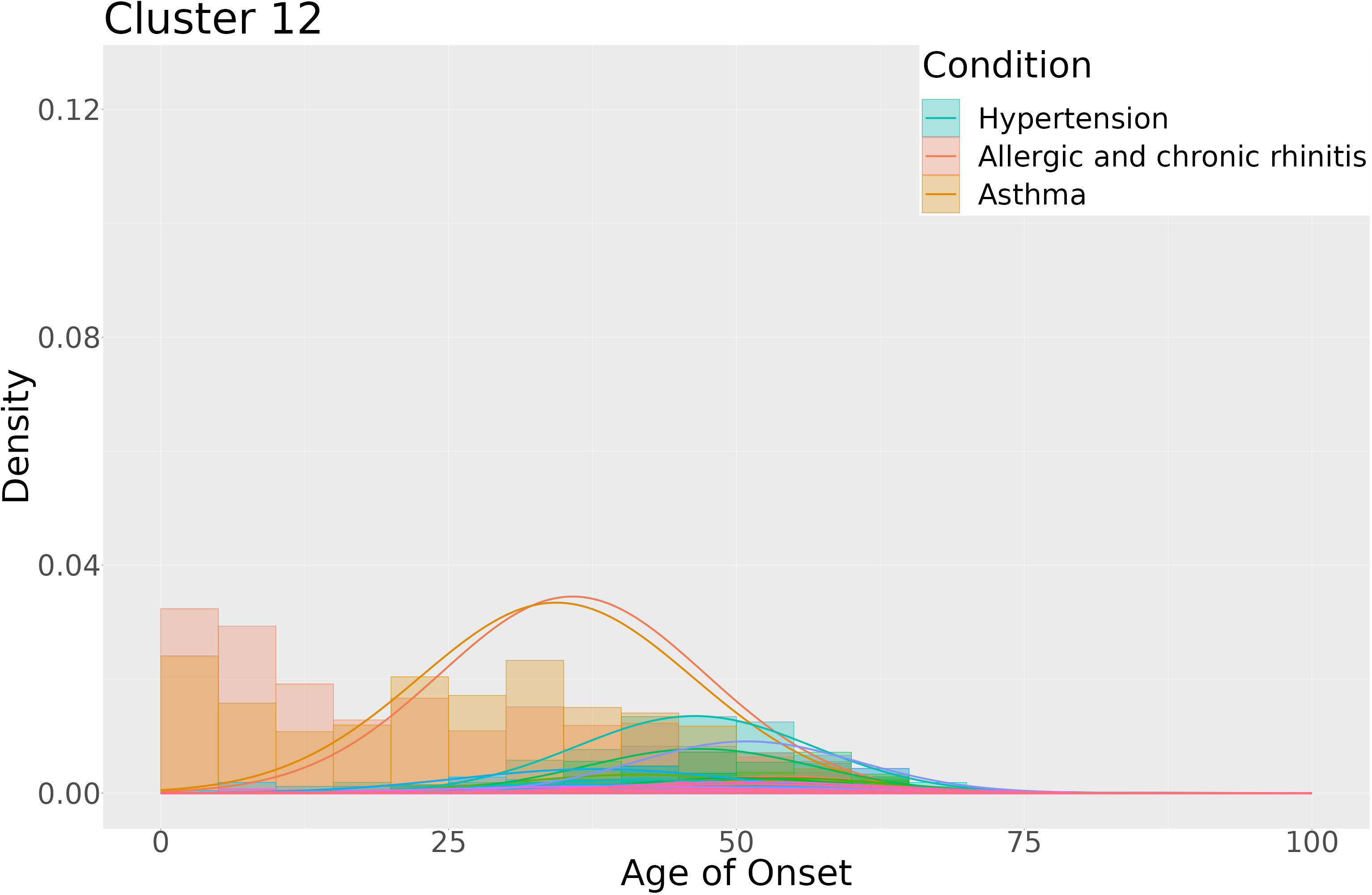}}
\end{figure}

\begin{figure}[ht]
\centering
\includegraphics[width=0.45\textwidth]{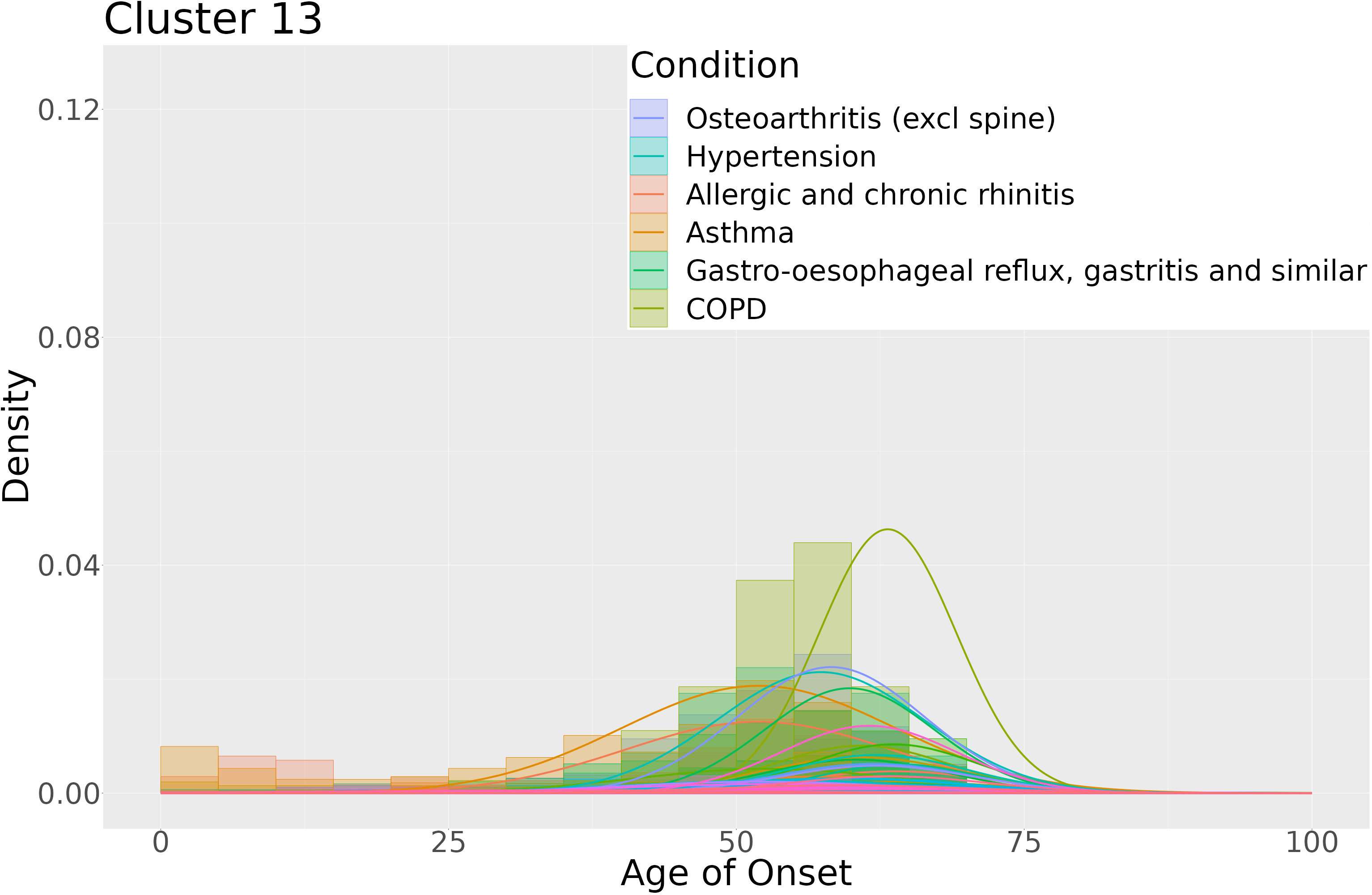}\qquad
\subfigure{\includegraphics[width=0.45\textwidth]{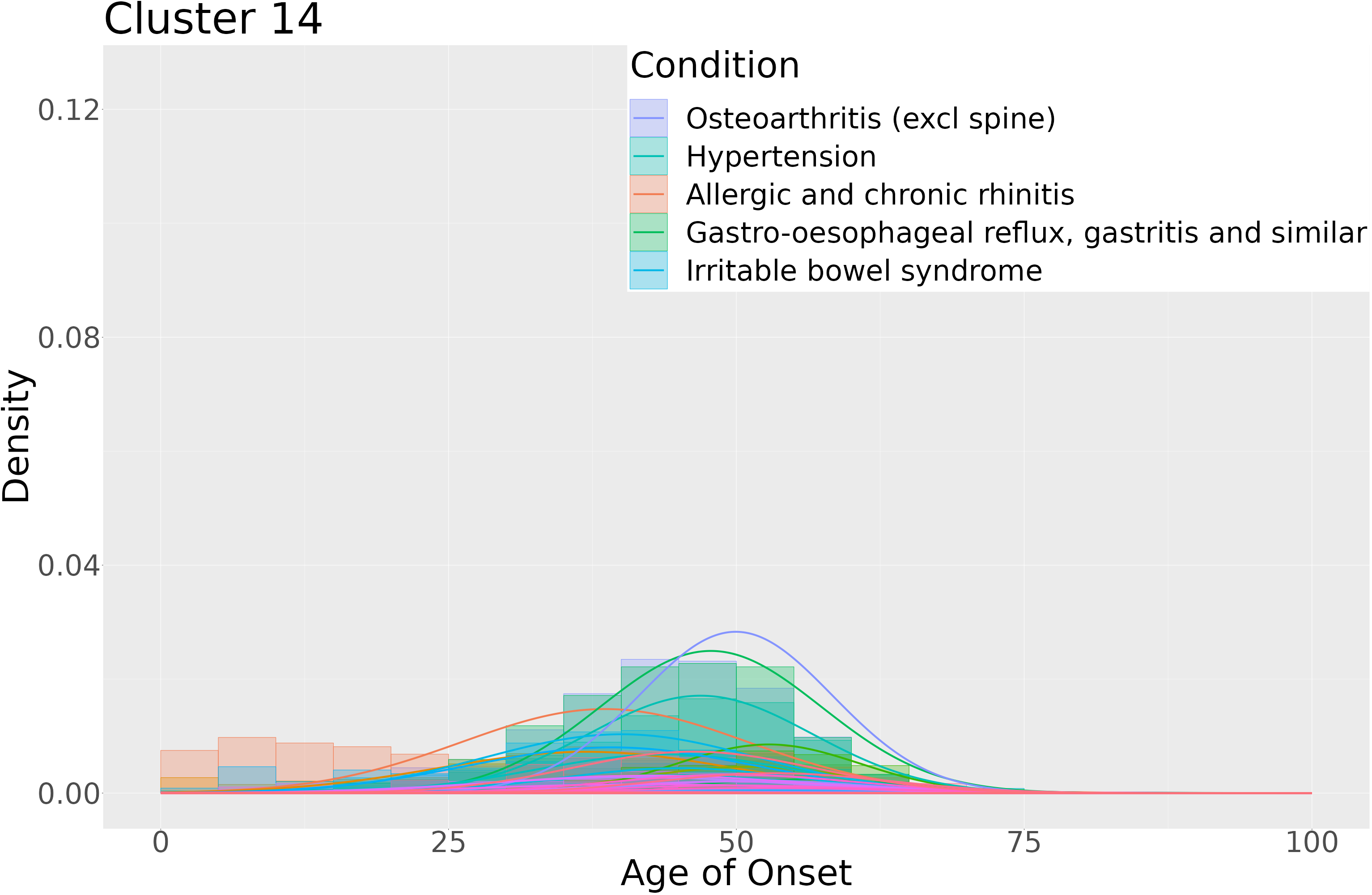}}
\end{figure}

\begin{figure}[ht]
\centering
\includegraphics[width=0.45\textwidth]{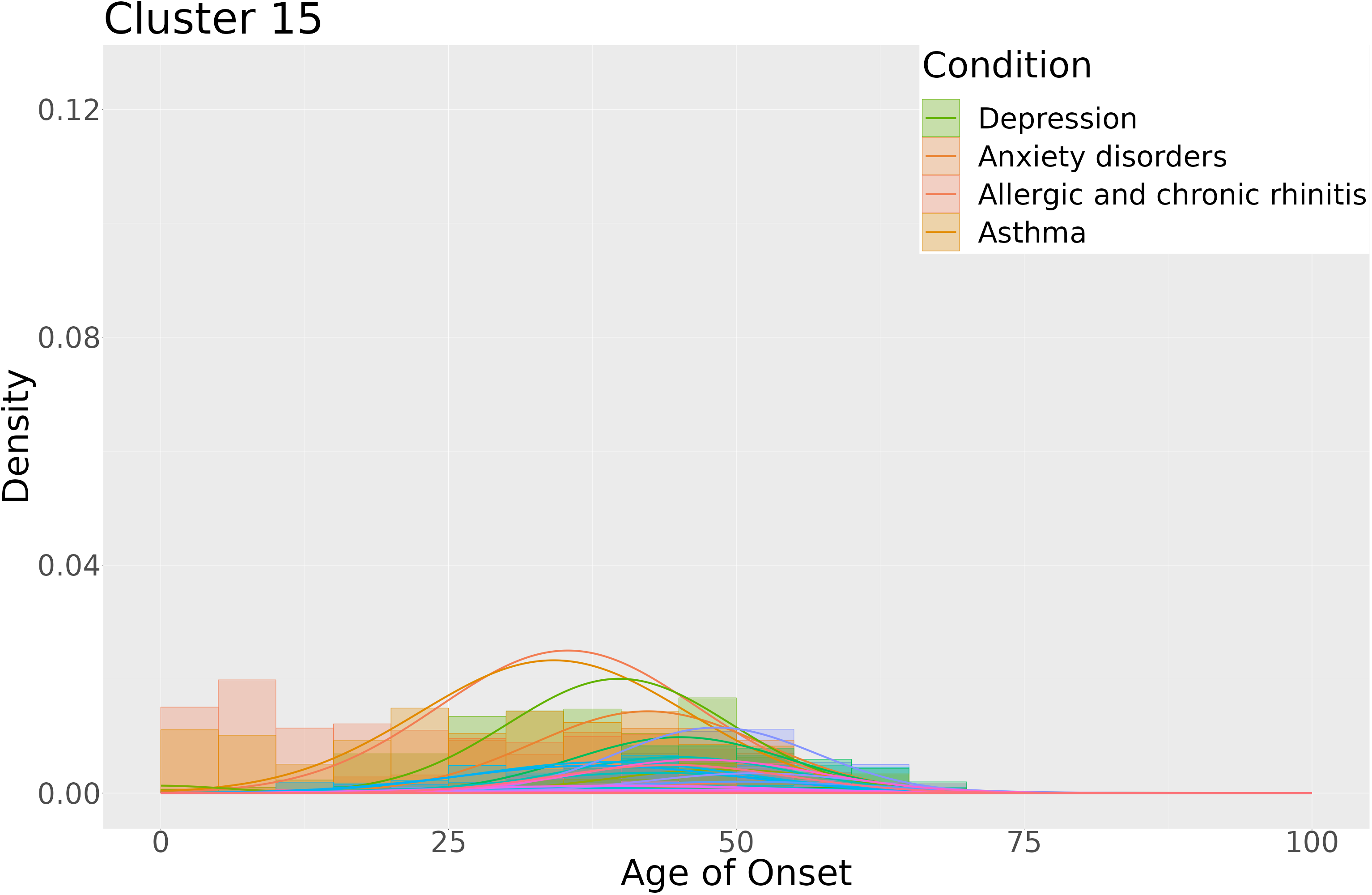}\qquad
\subfigure{\includegraphics[width=0.45\textwidth]{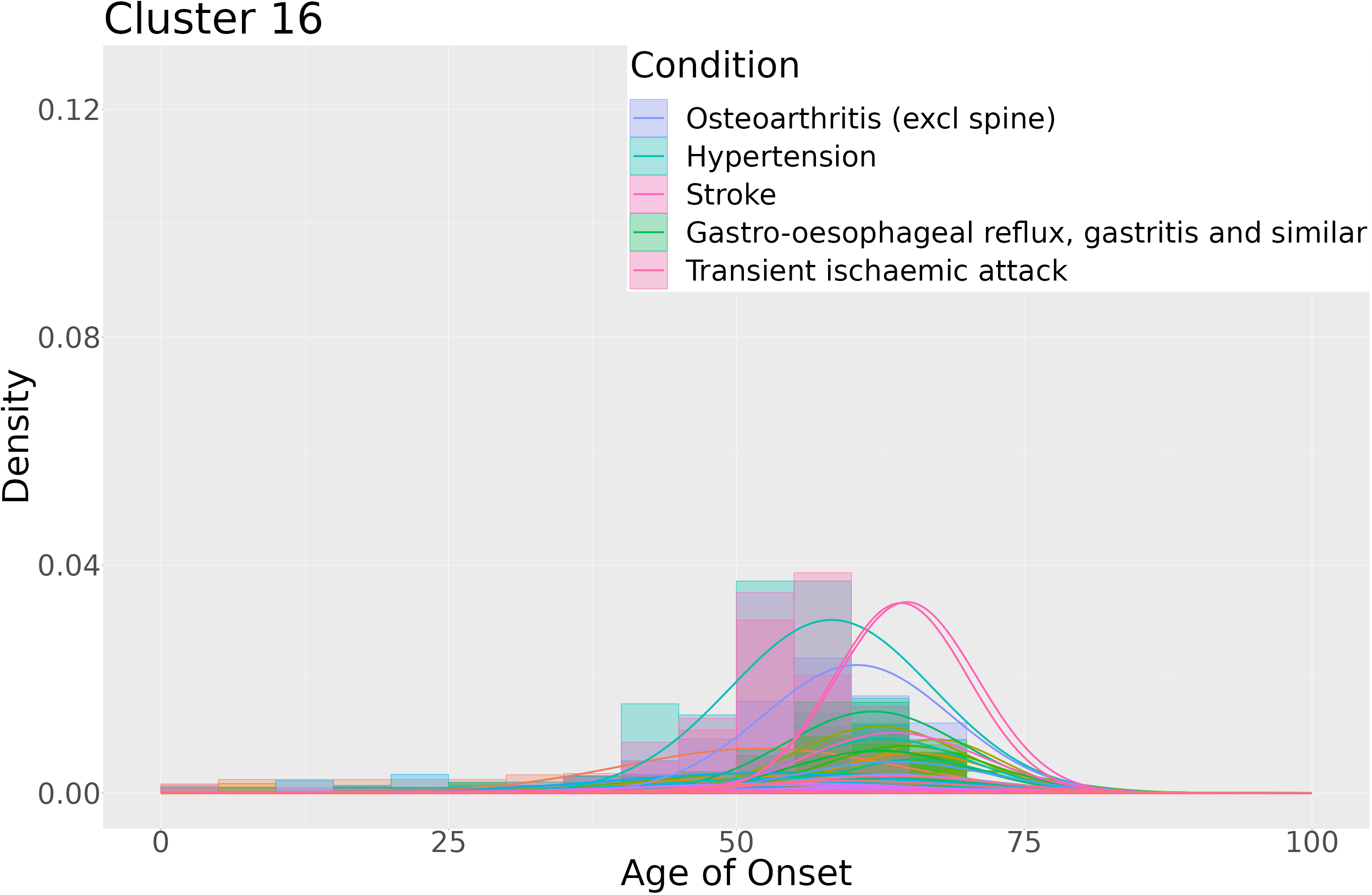}}
\end{figure}

\begin{figure}[ht]
\centering
\includegraphics[width=0.45\textwidth]{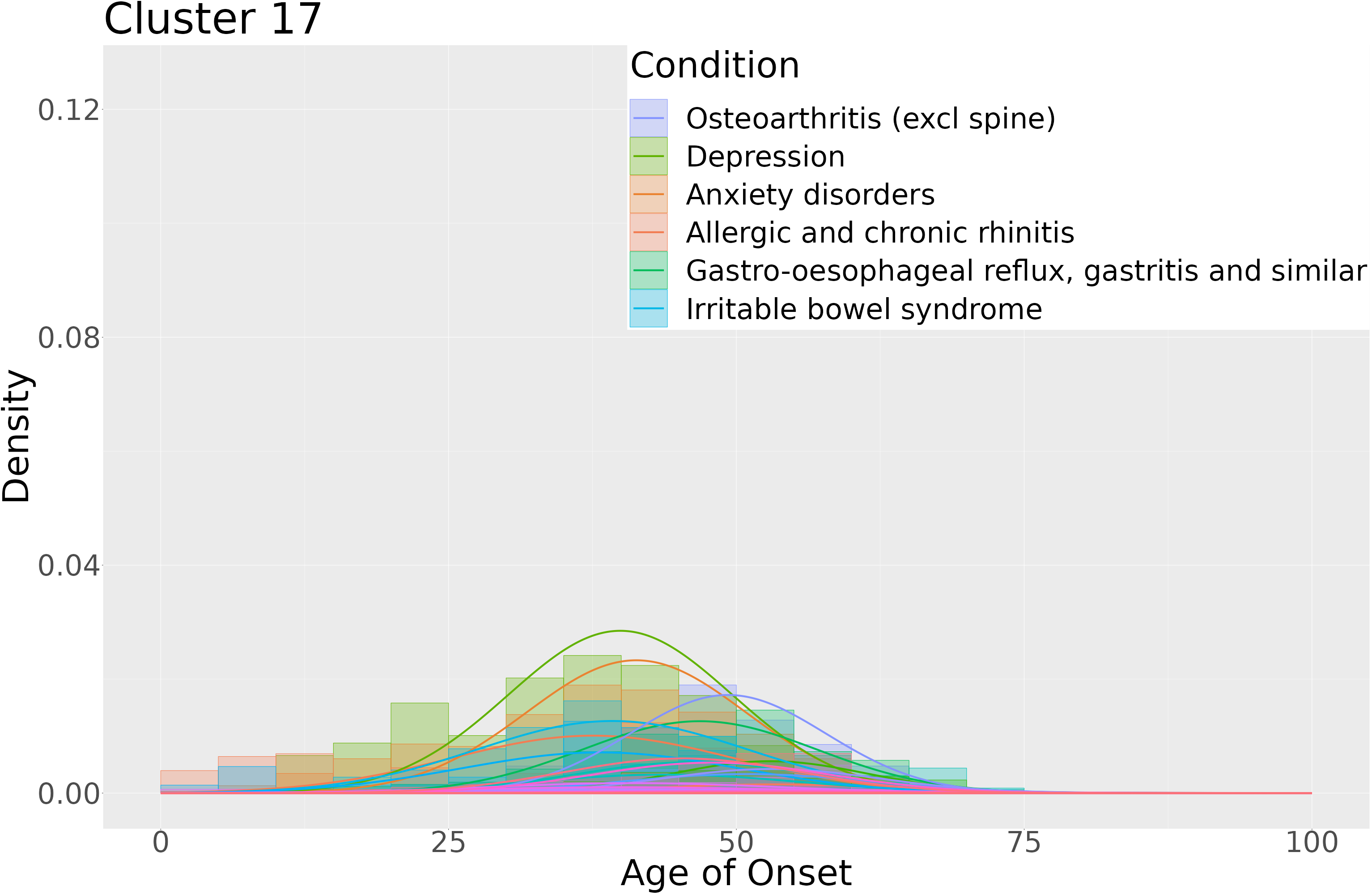}\qquad
\subfigure{\includegraphics[width=0.45\textwidth]{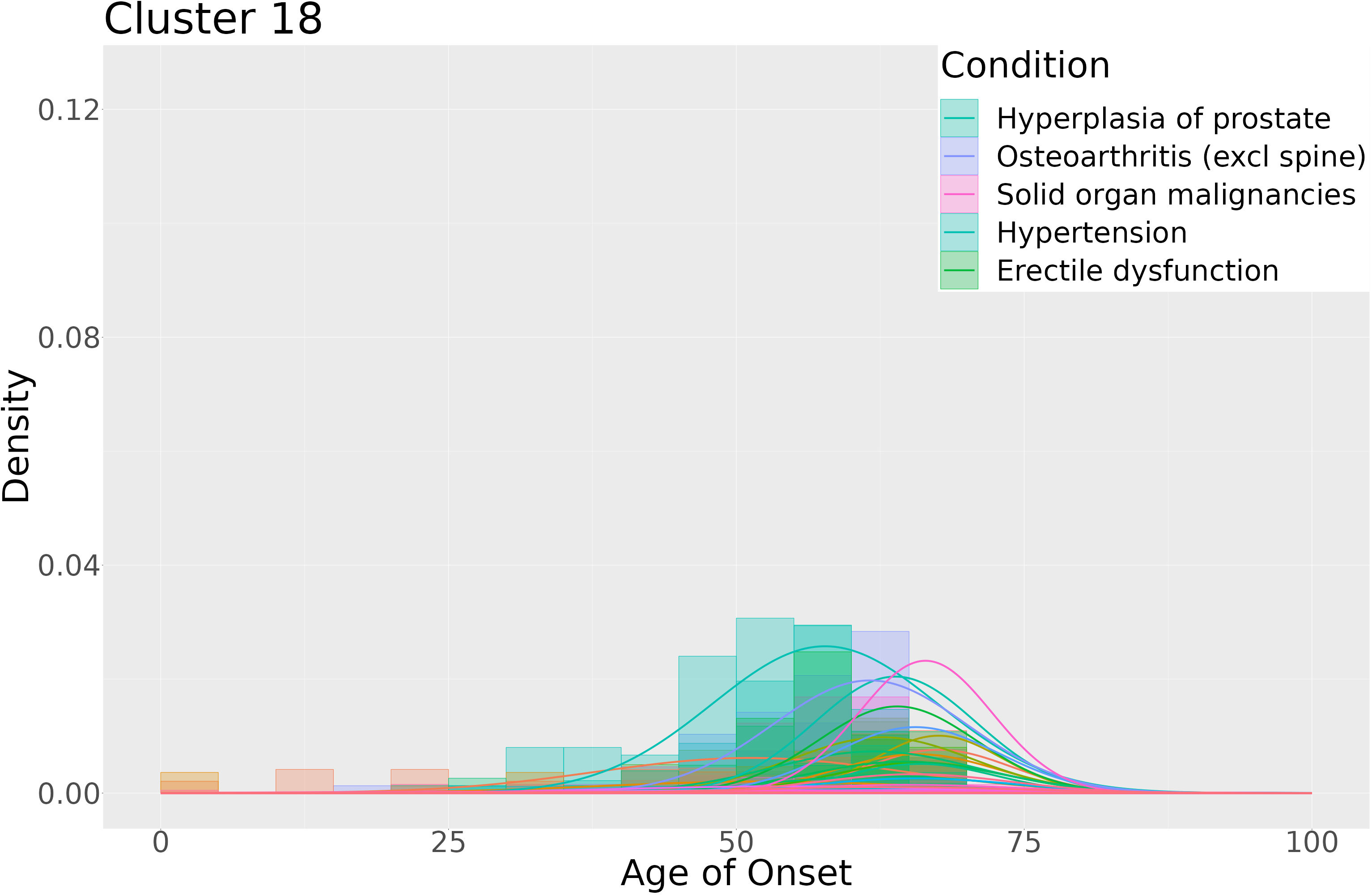}}
\end{figure}

\begin{figure}[ht]
\centering
\includegraphics[width=0.45\textwidth]{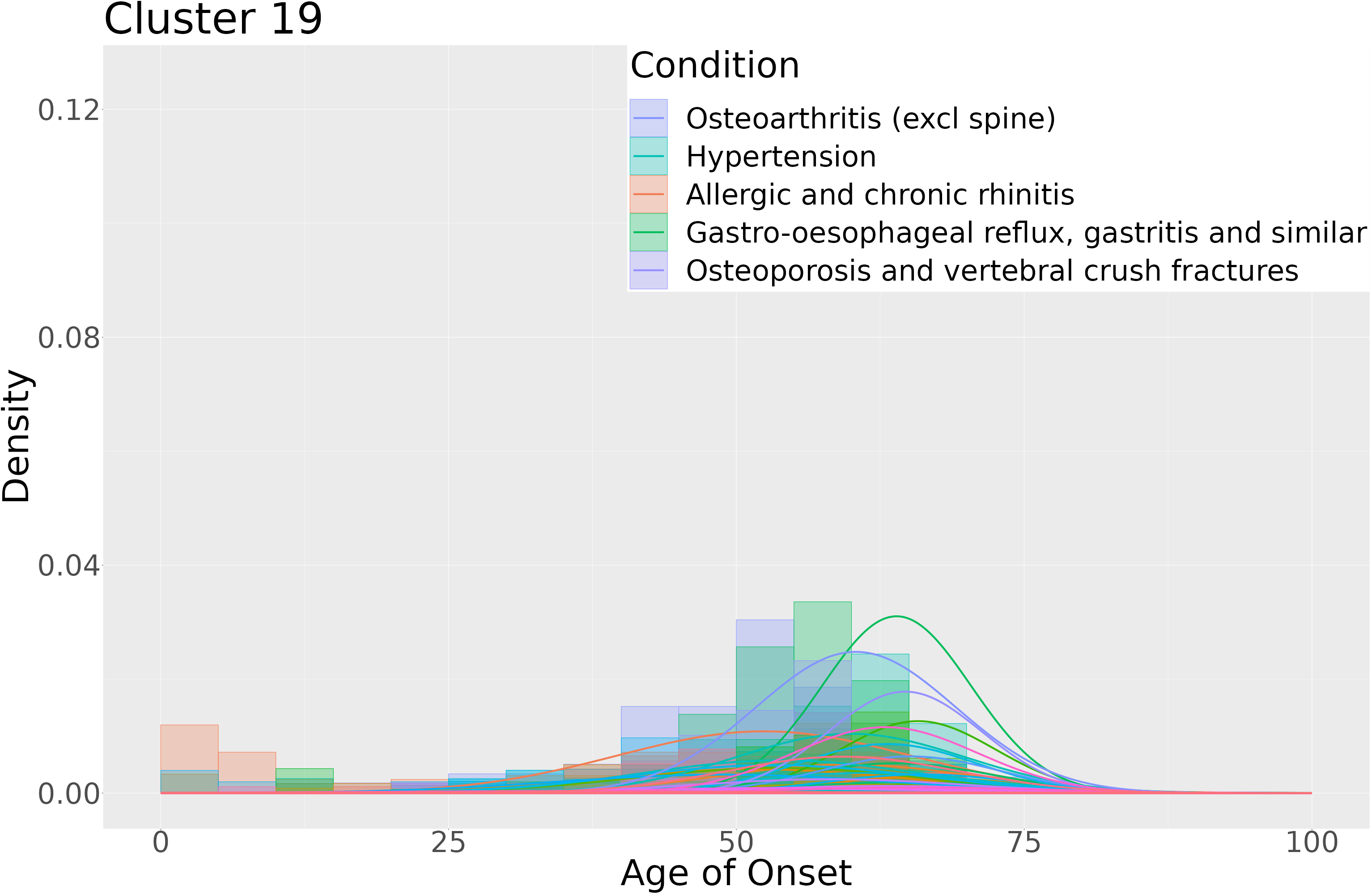}\qquad
\subfigure{\includegraphics[width=0.45\textwidth]{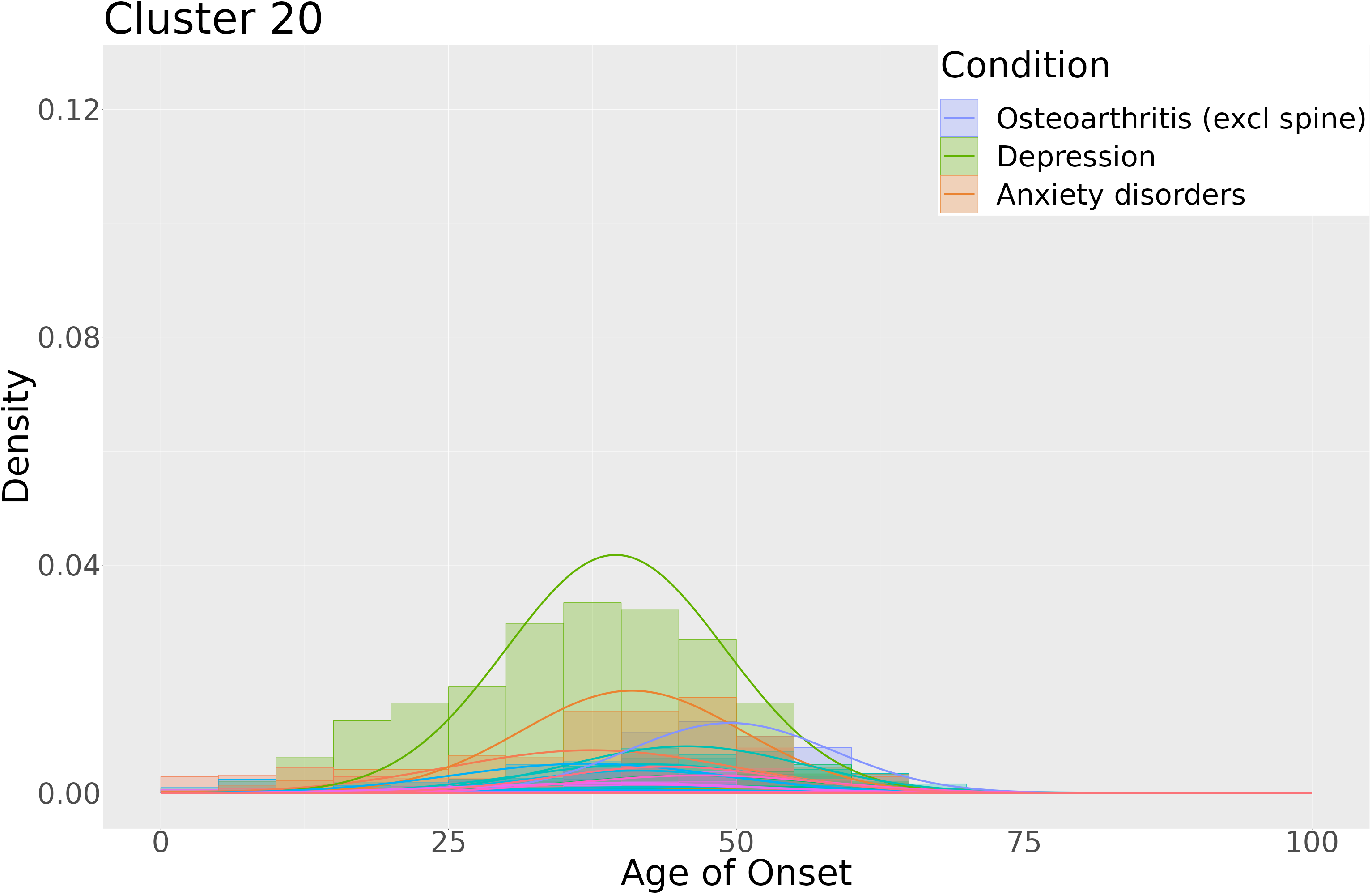}}
\end{figure}

\begin{figure}[ht]
\centering
\includegraphics[width=0.45\textwidth]{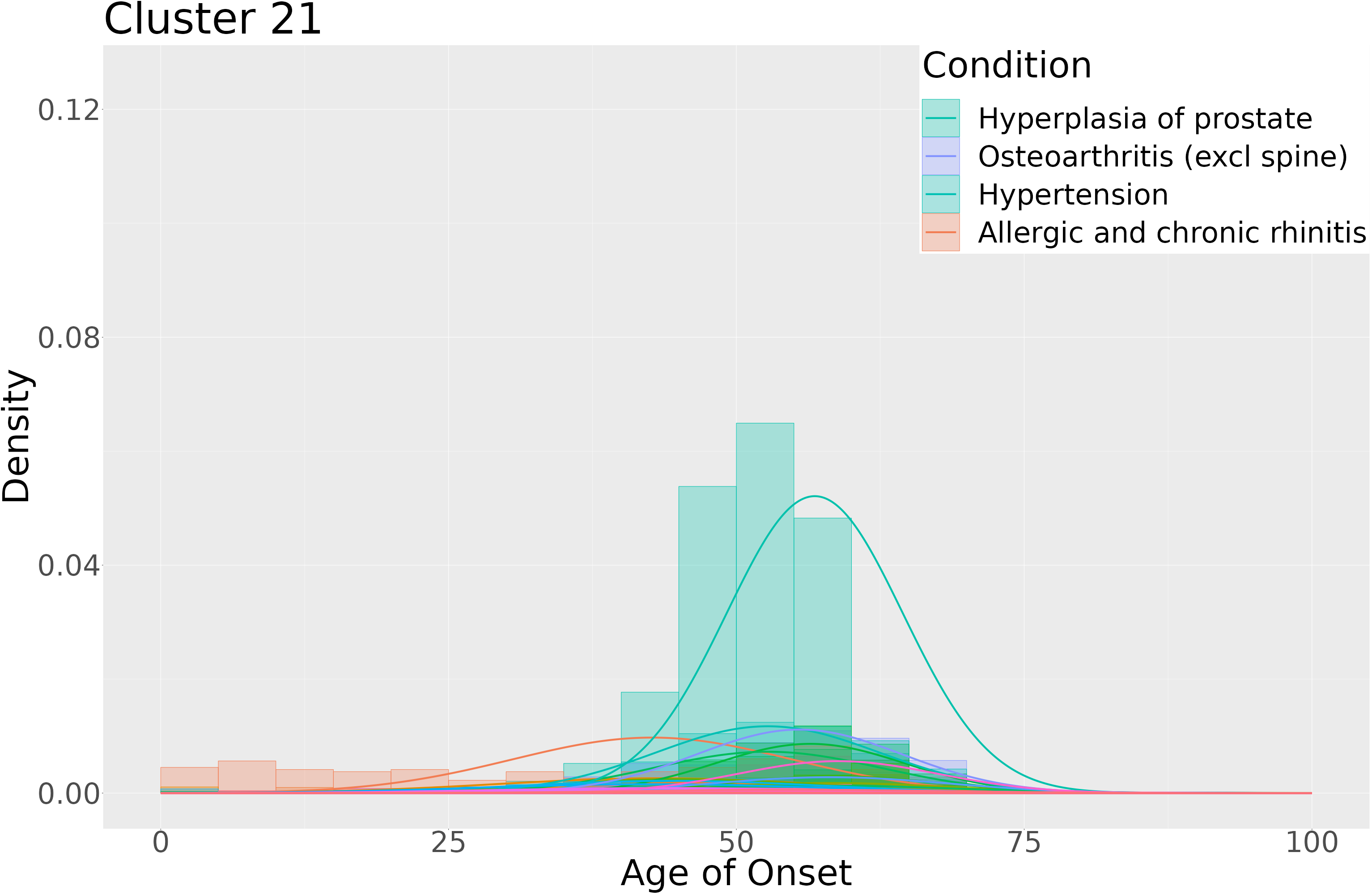}\qquad
\subfigure{\includegraphics[width=0.45\textwidth]{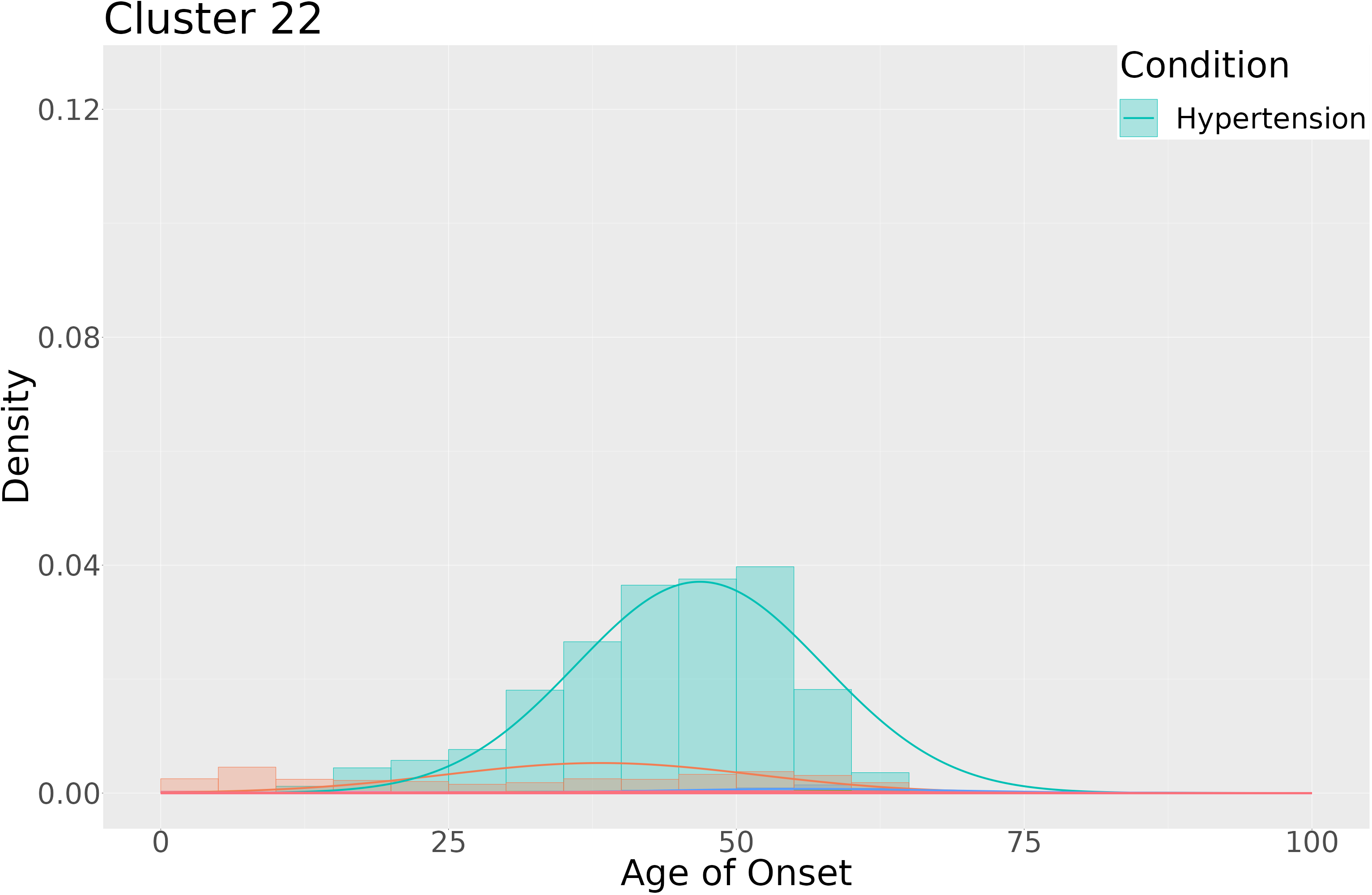}}
\end{figure}

\begin{figure}[ht]
\centering
\includegraphics[width=0.45\textwidth]{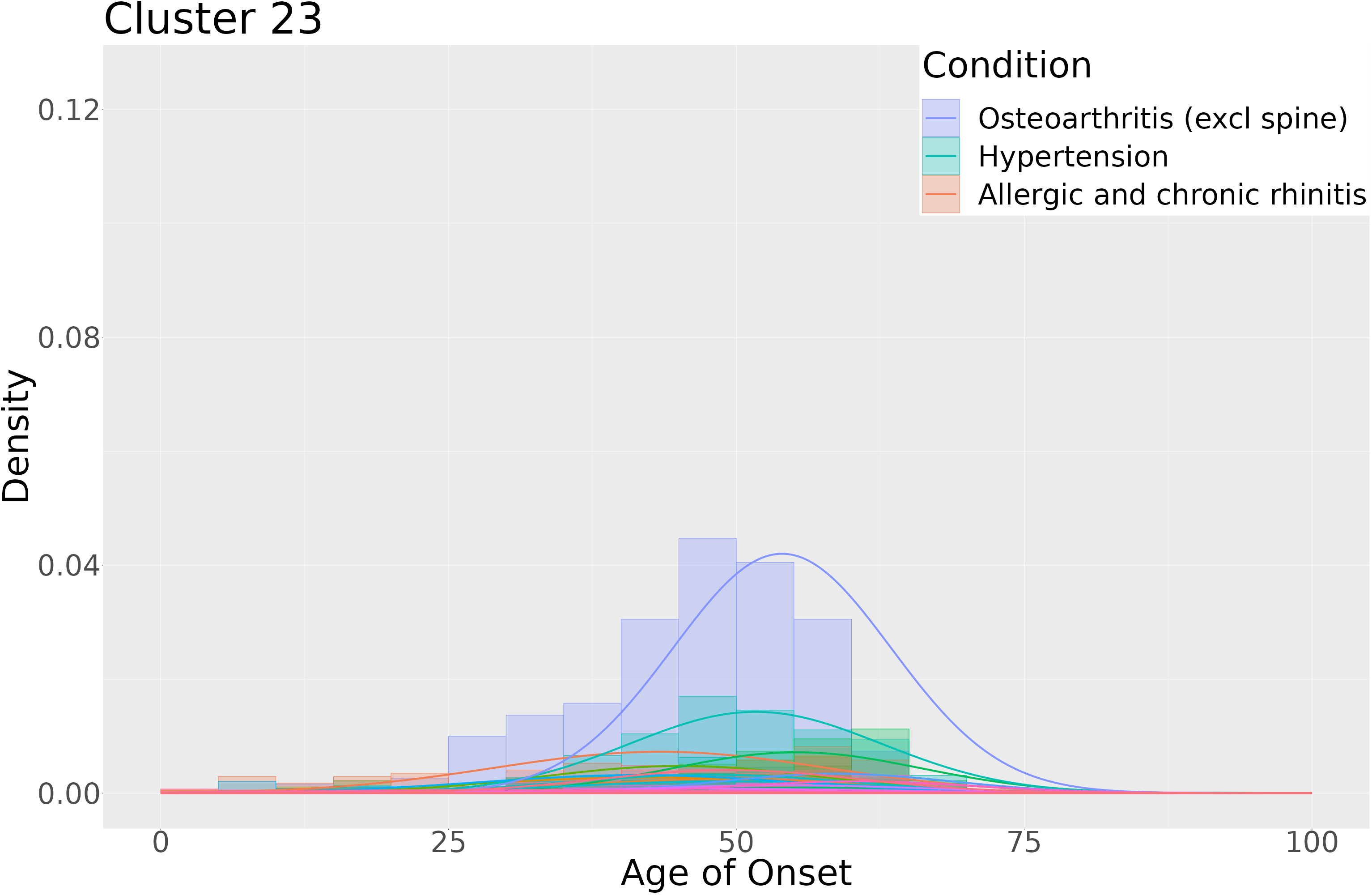}\qquad
\subfigure{\includegraphics[width=0.45\textwidth]{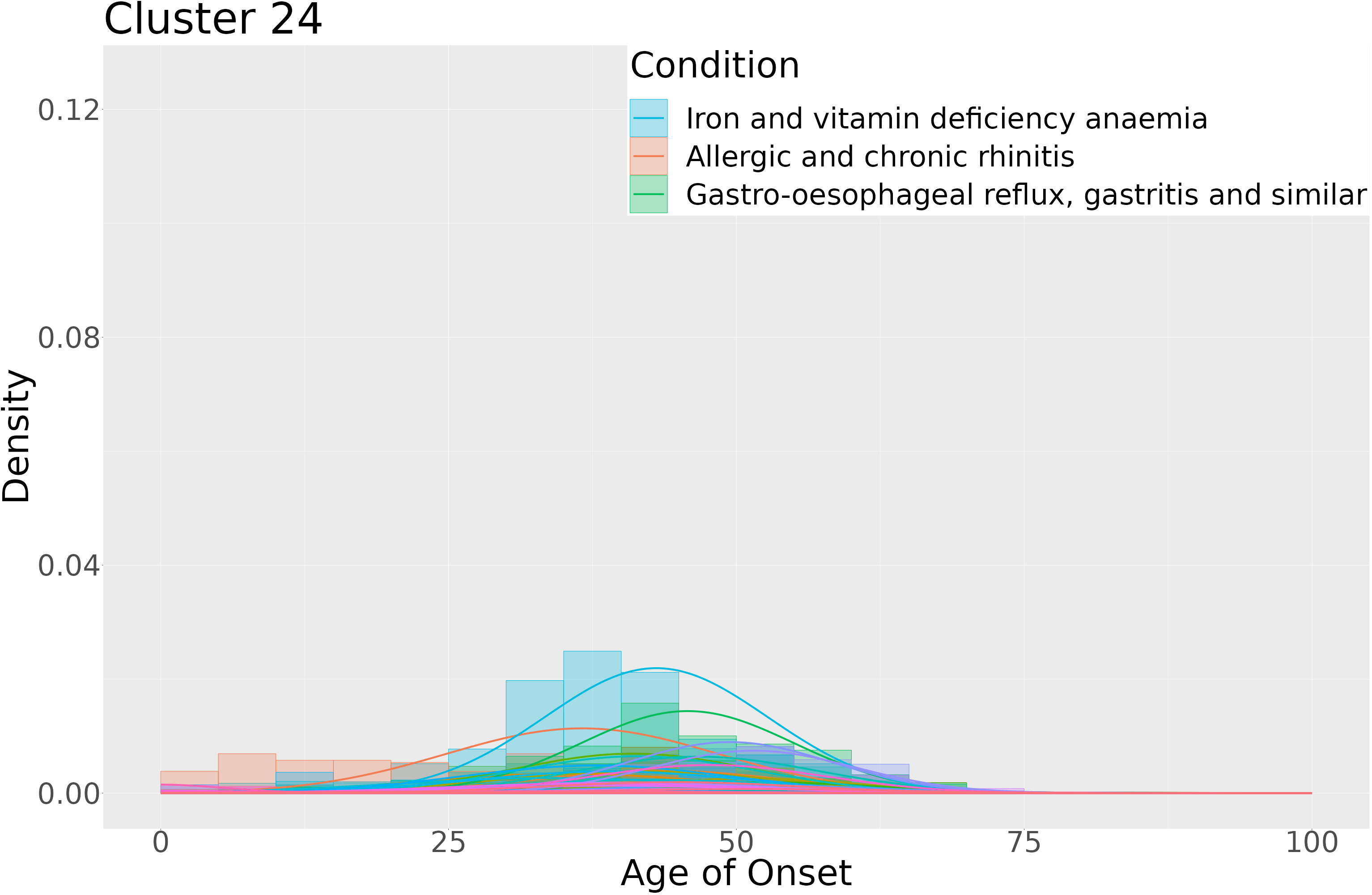}}
\end{figure}

\begin{figure}[ht]
\centering
\includegraphics[width=0.45\textwidth]{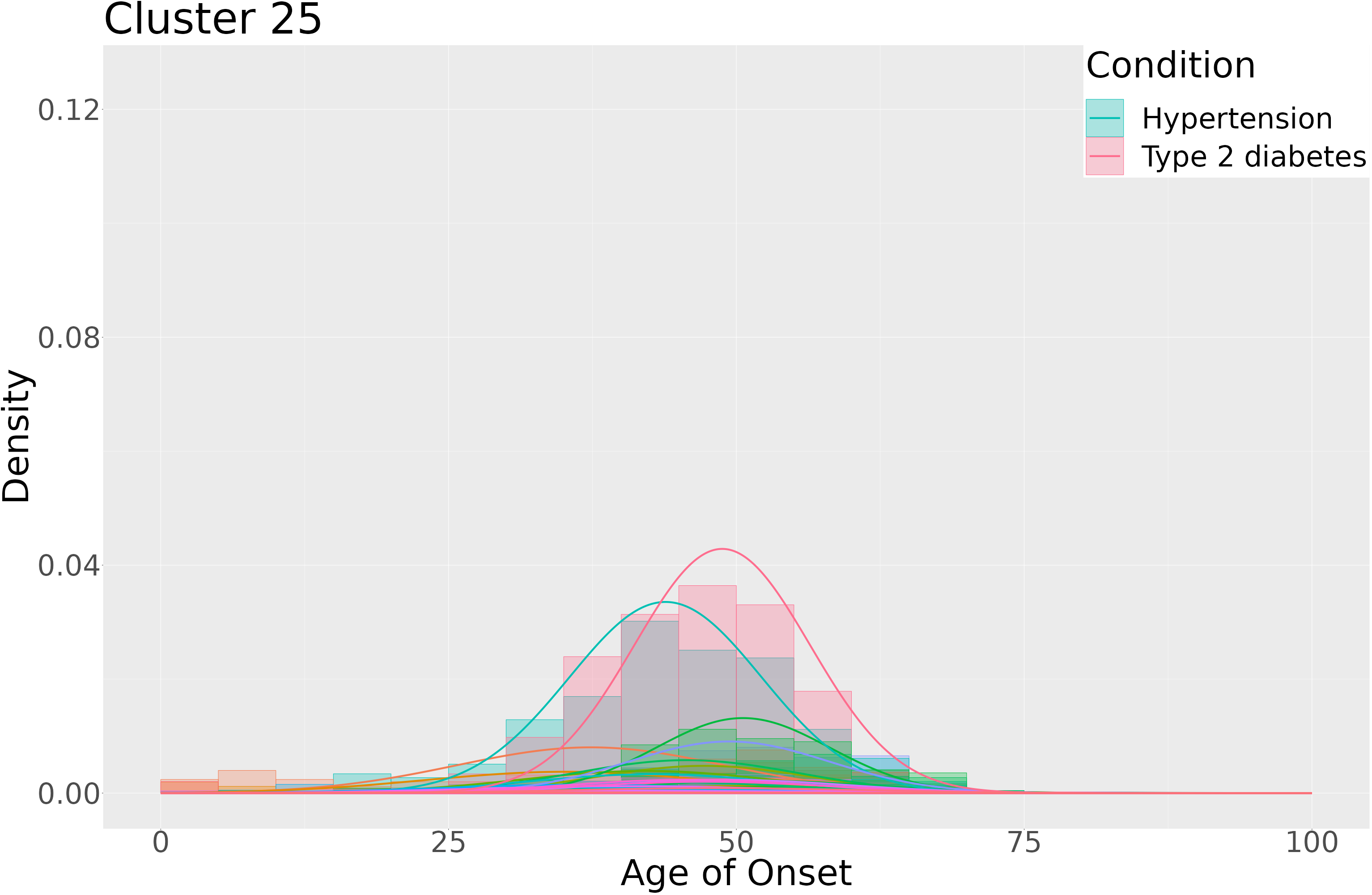}\qquad
\subfigure{\includegraphics[width=0.45\textwidth]{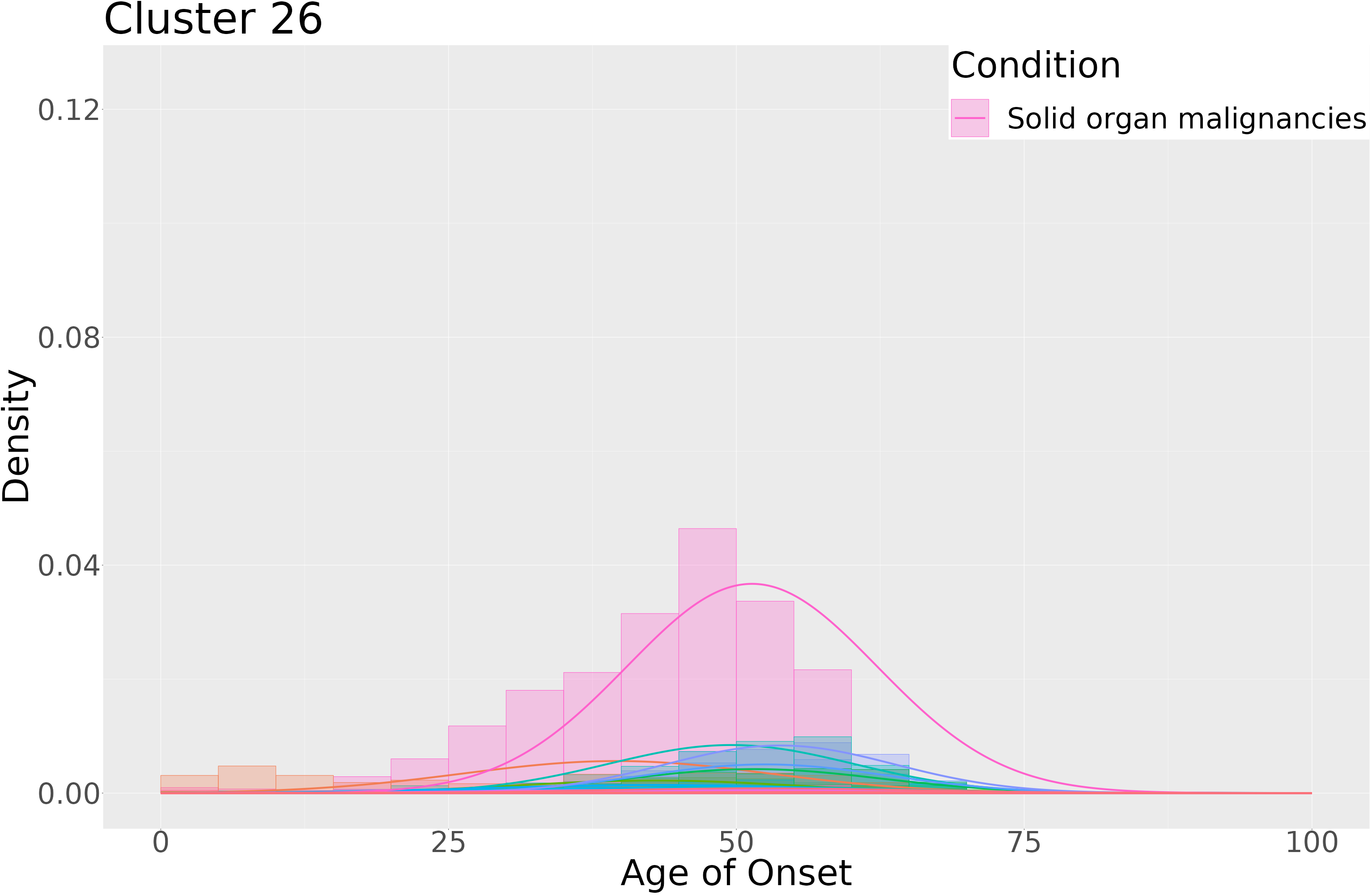}}
\end{figure}

\begin{figure}[ht]
\centering
\includegraphics[width=0.45\textwidth]{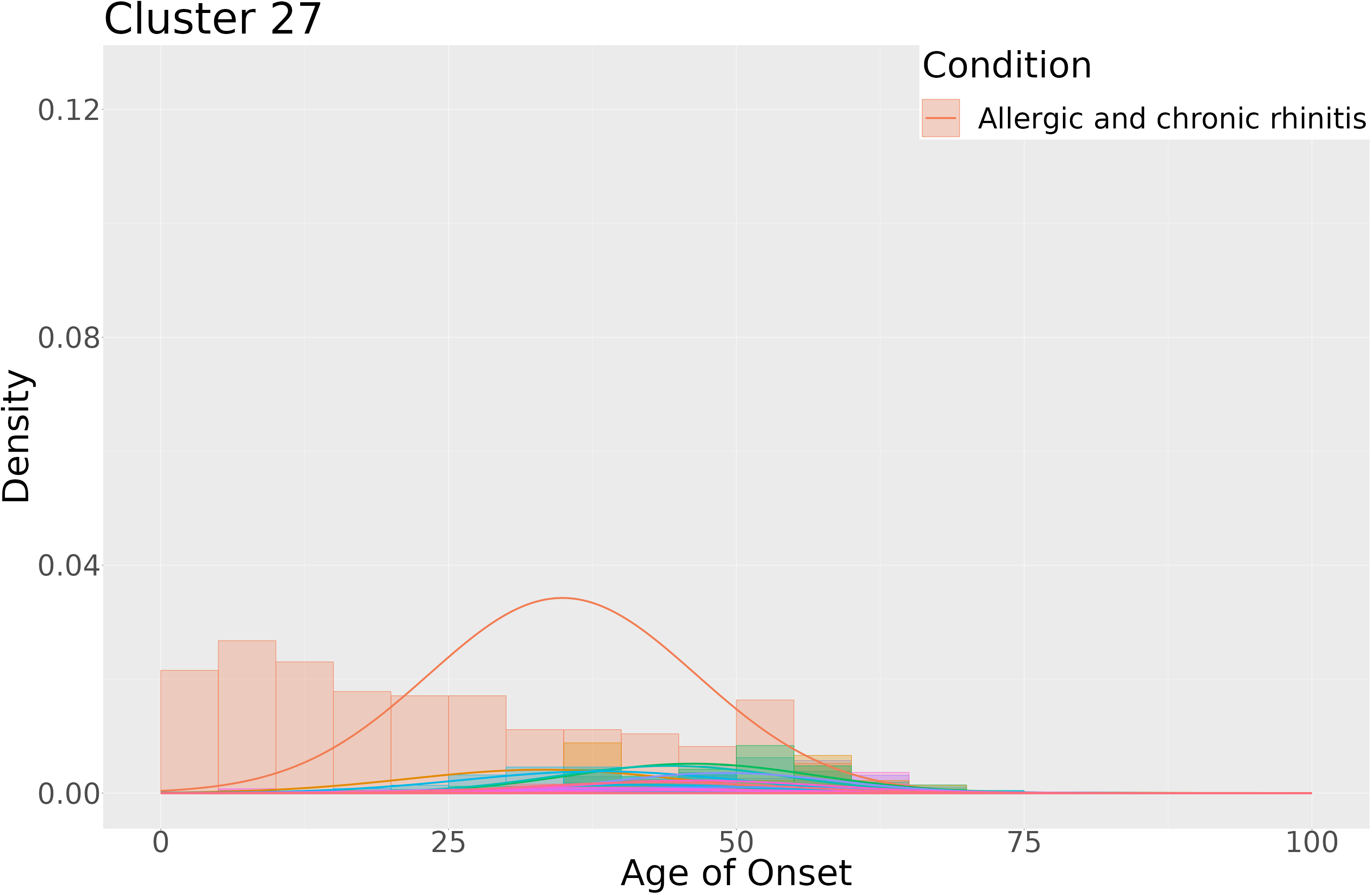}\qquad
\subfigure{\includegraphics[width=0.45\textwidth]{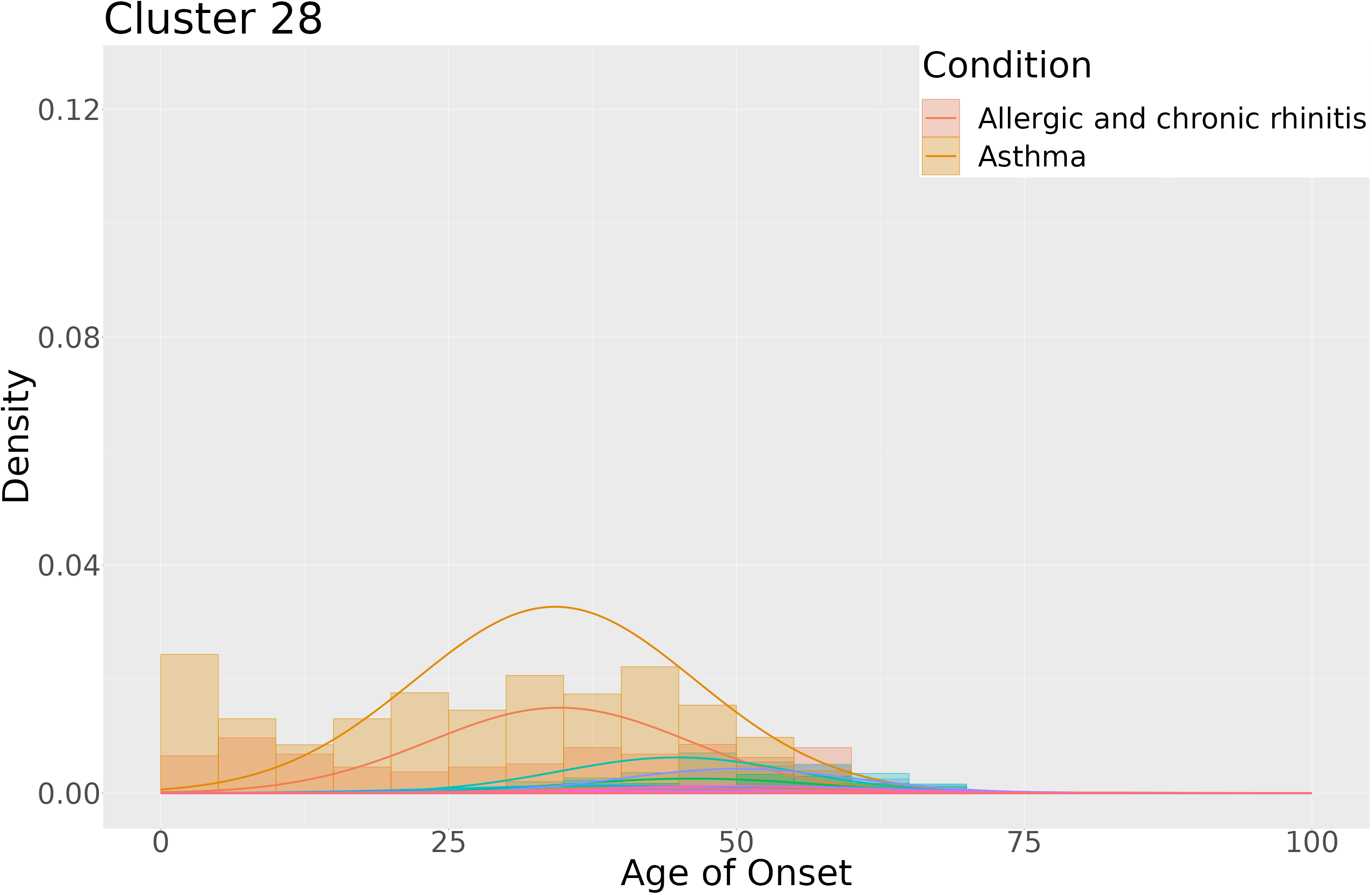}}
\end{figure}

\begin{figure}[ht]
\centering
\includegraphics[width=0.45\textwidth]{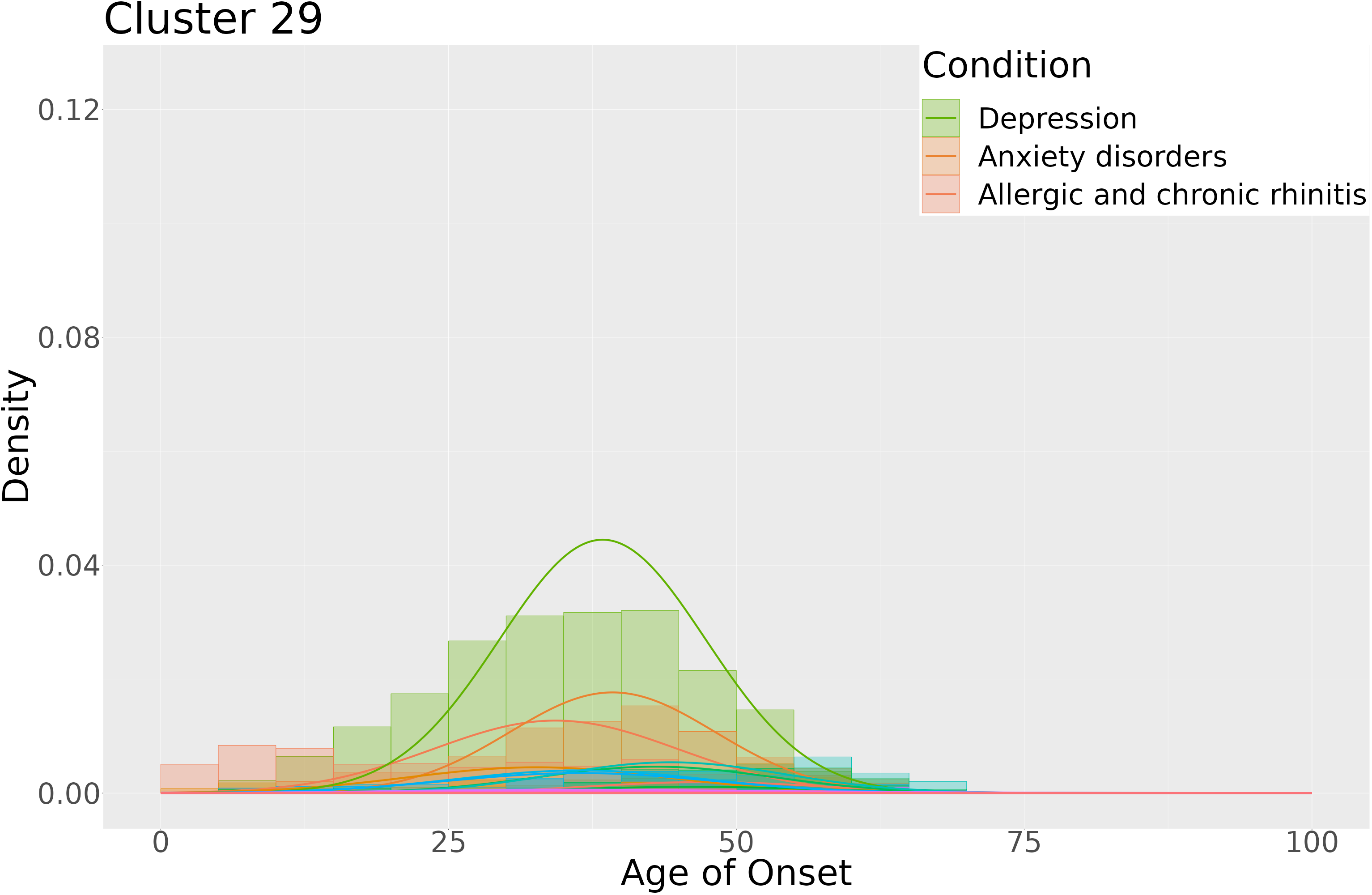}\qquad
\subfigure{\includegraphics[width=0.45\textwidth]{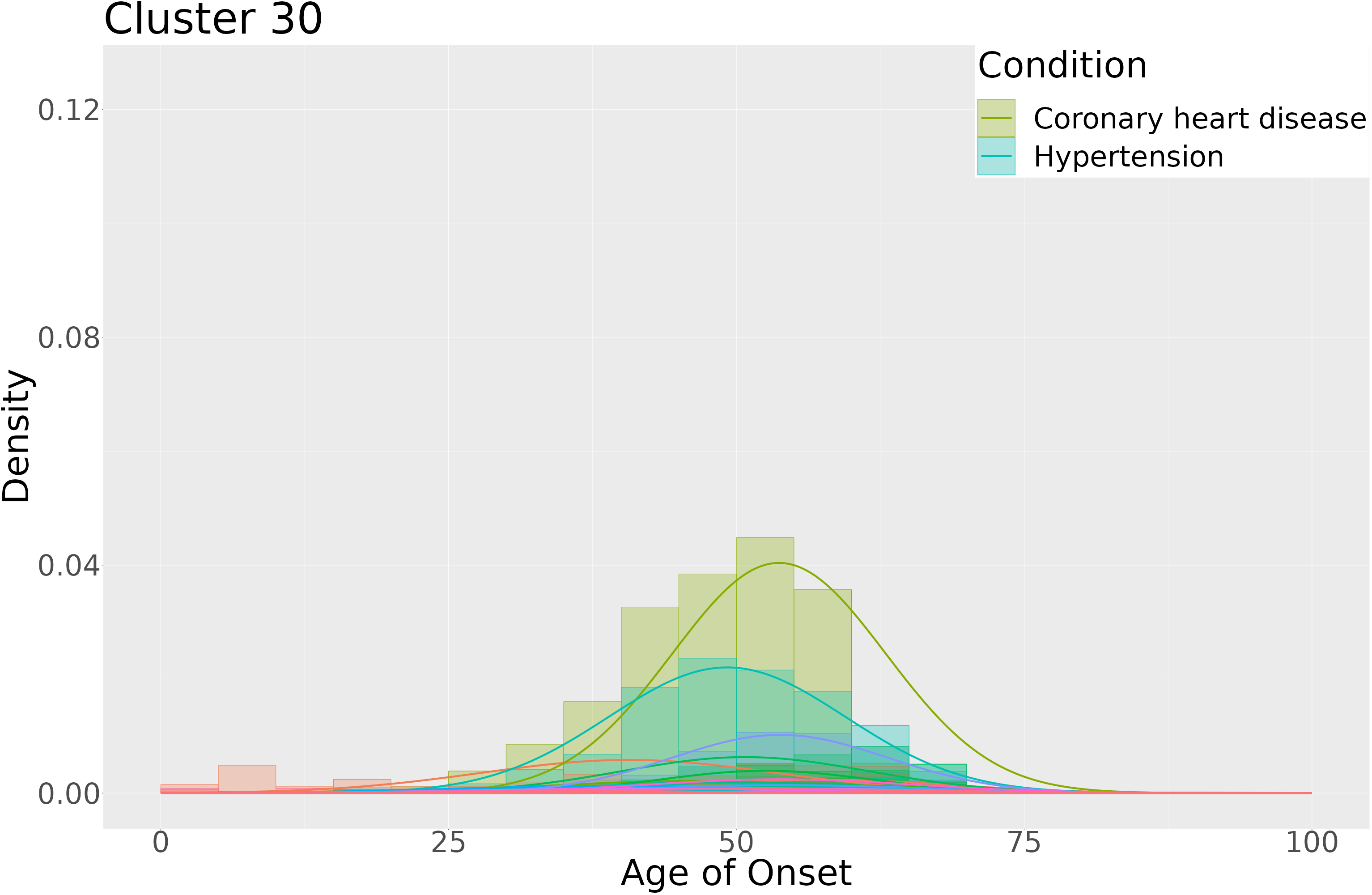}}
\end{figure}

\begin{figure}[ht]
\centering
\includegraphics[width=0.45\textwidth]{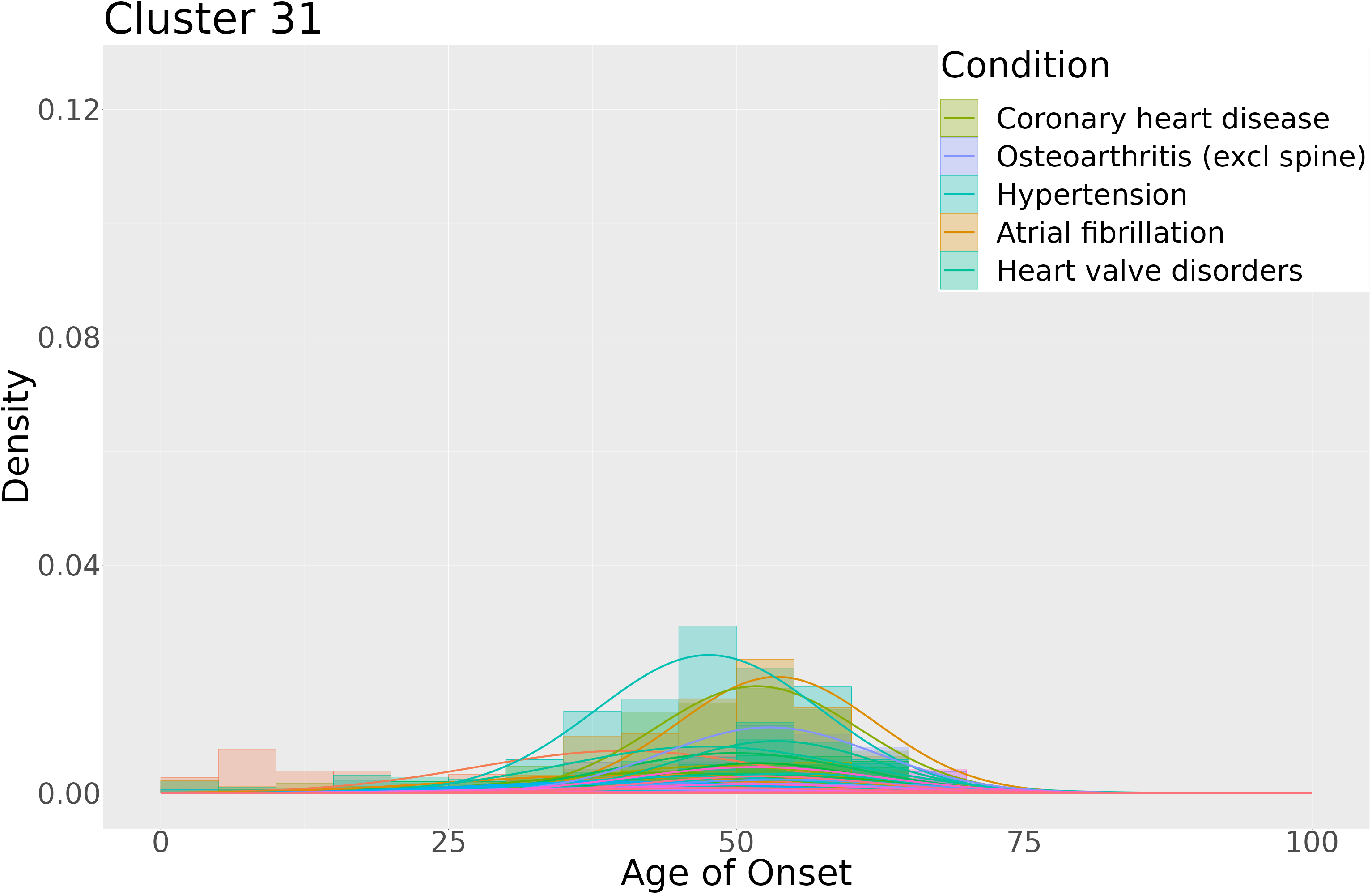}\qquad
\subfigure{\includegraphics[width=0.45\textwidth]{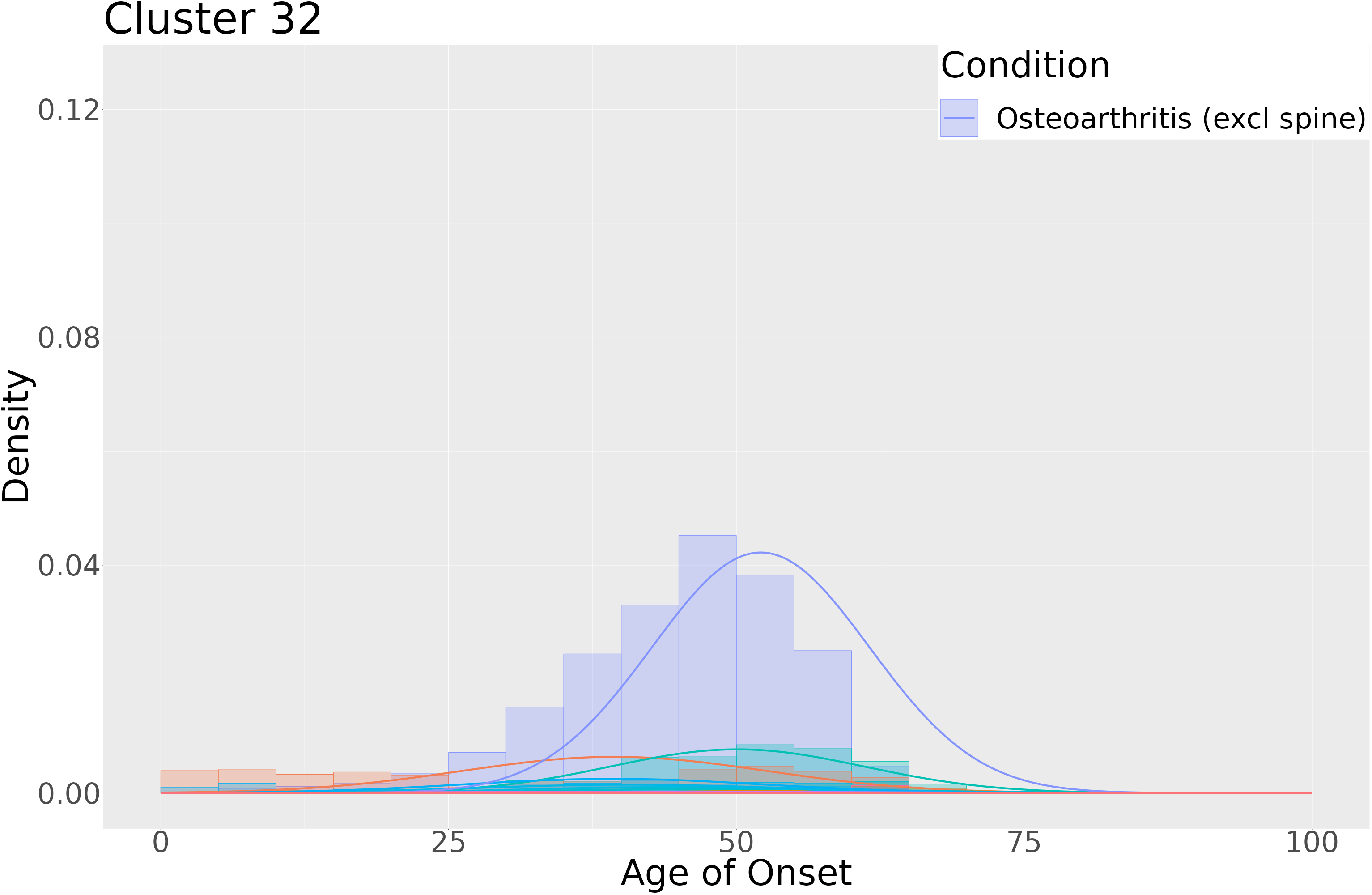}}
\end{figure}

\begin{figure}[ht]
\centering
\includegraphics[width=0.45\textwidth]{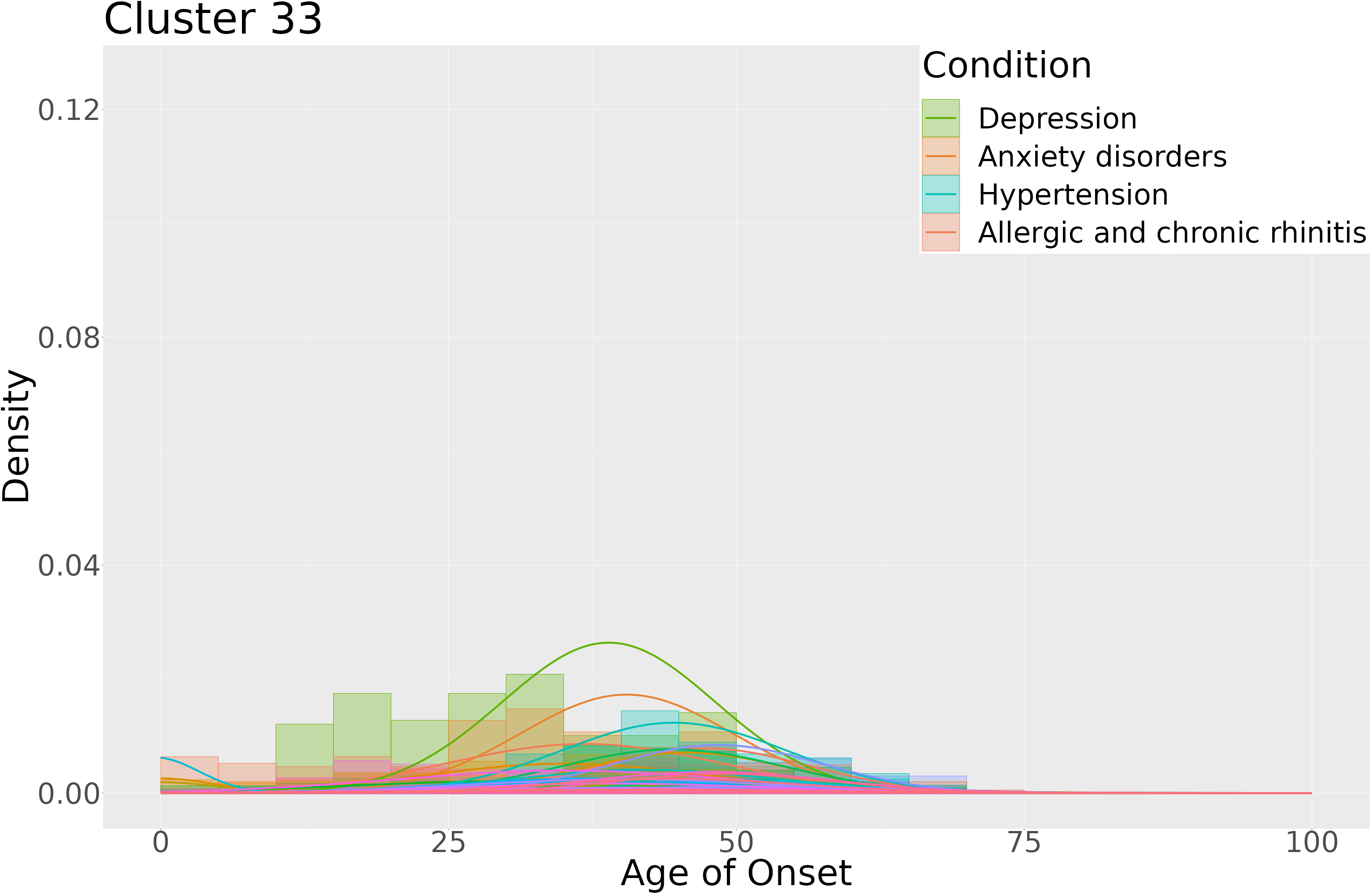}\qquad
\subfigure{\includegraphics[width=0.45\textwidth]{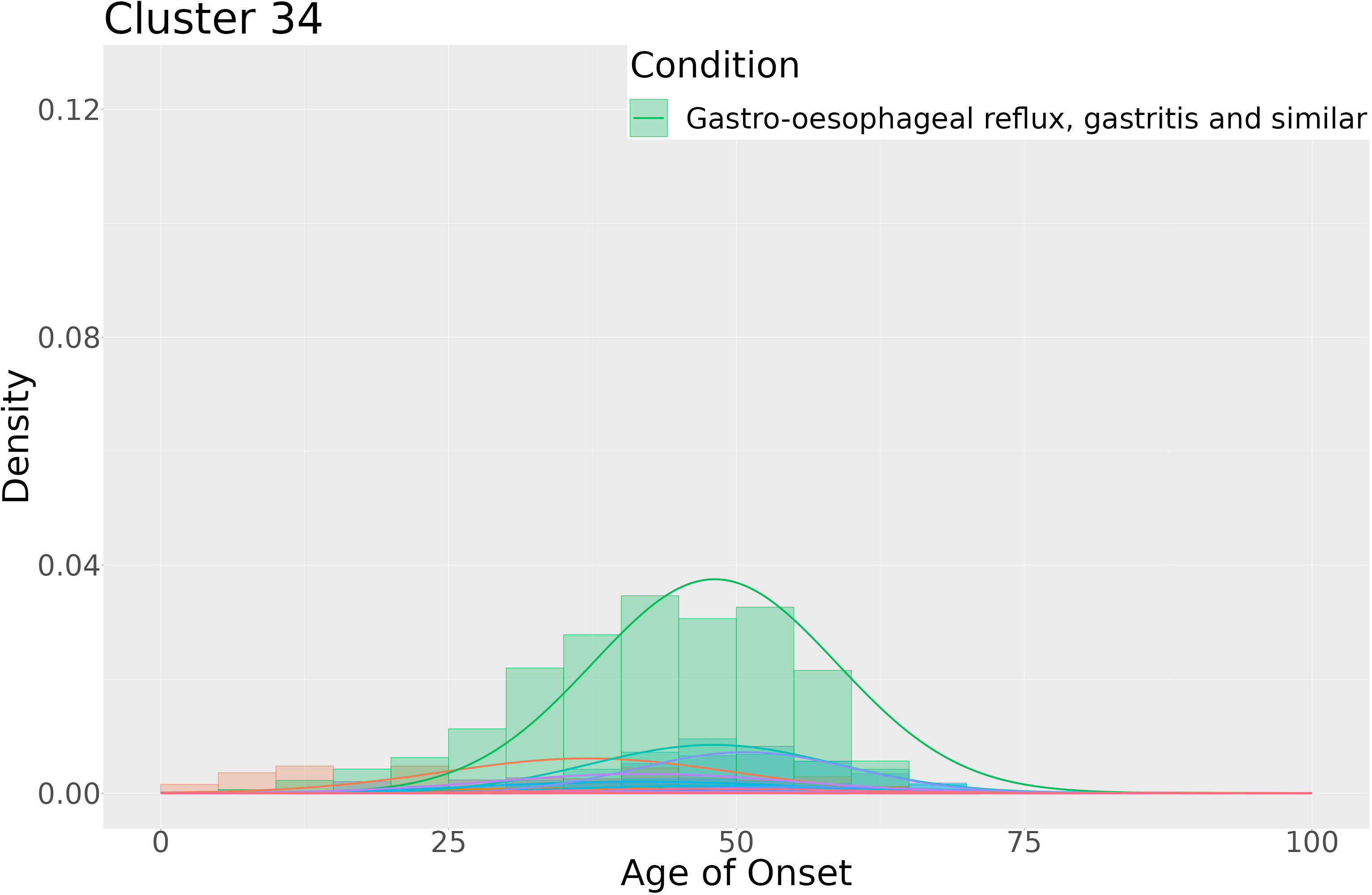}}
\end{figure}

\begin{figure}[ht]
\centering
\includegraphics[width=0.45\textwidth]{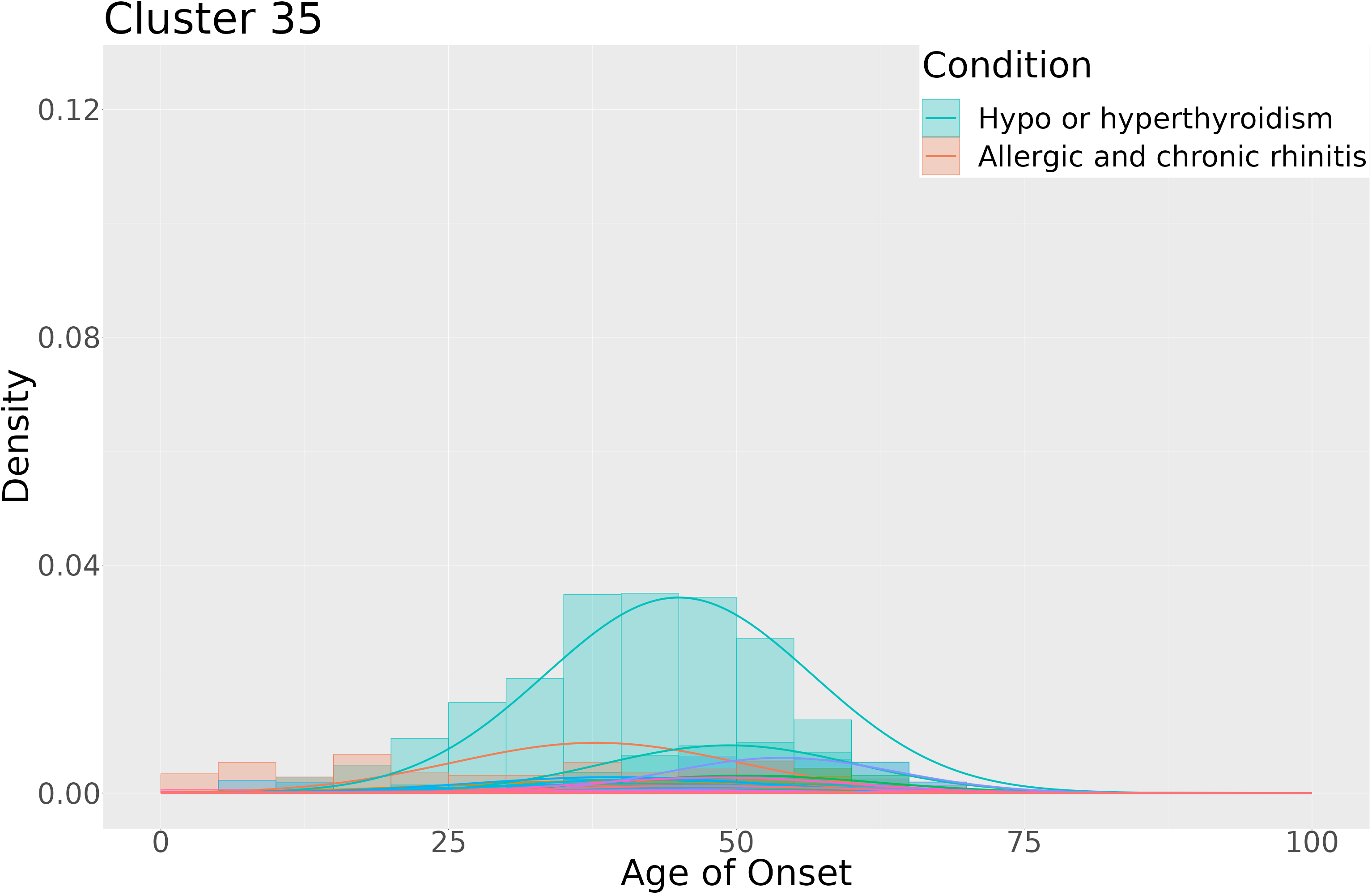}\qquad
\includegraphics[width=0.45\textwidth]{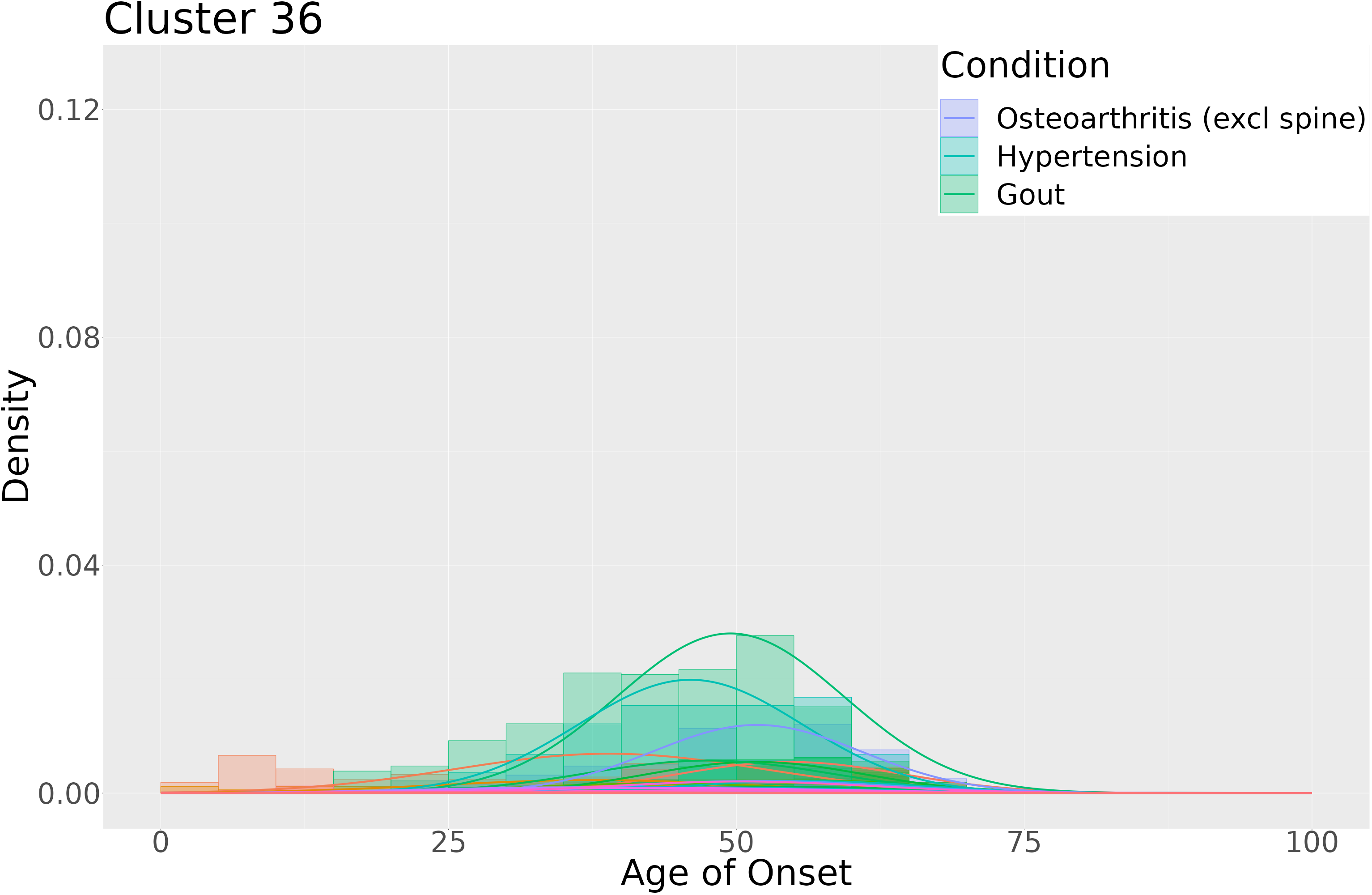}\qquad
\end{figure}

\begin{figure}[ht]
\centering
\includegraphics[width=0.45\textwidth]{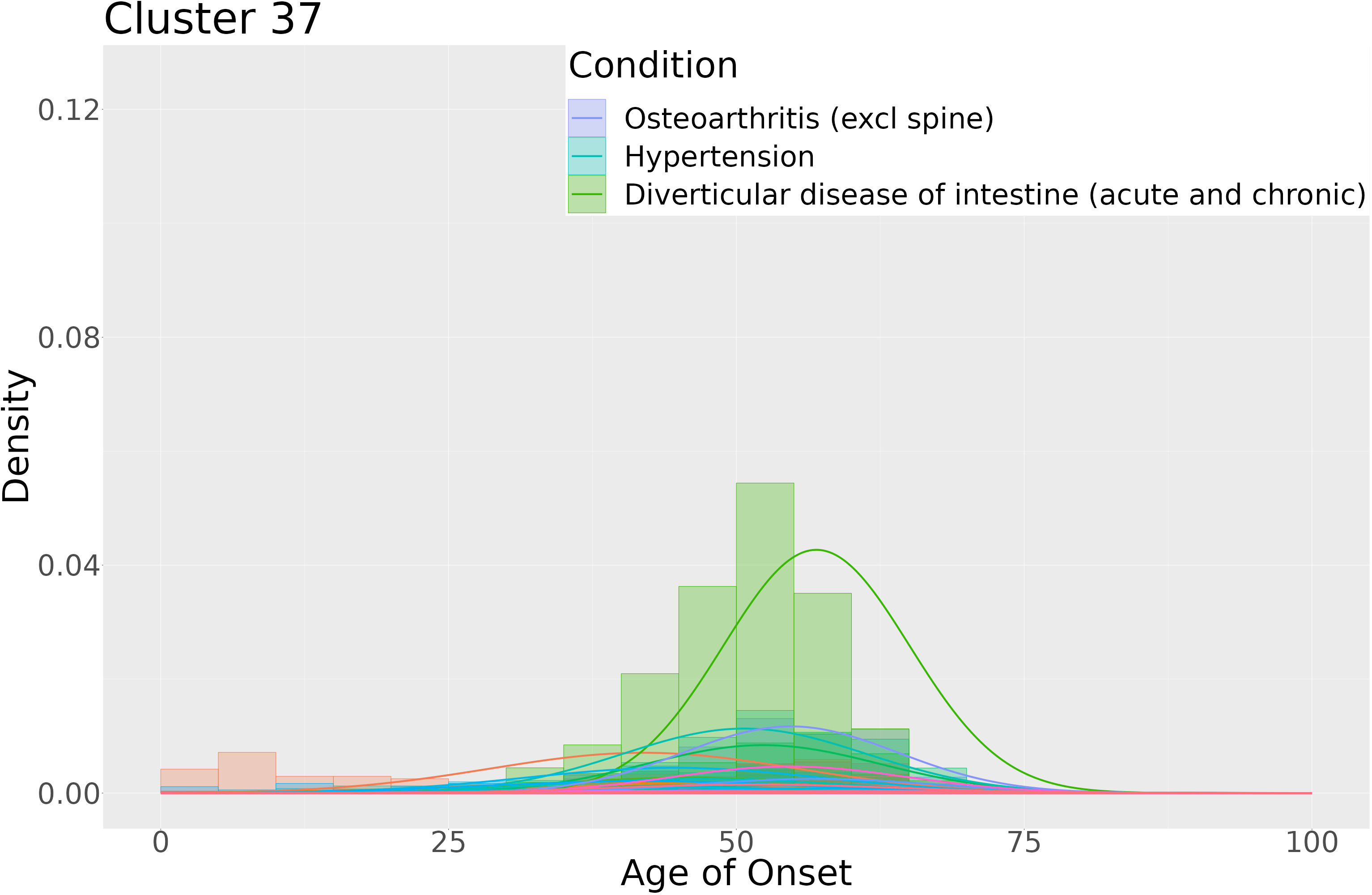}\qquad
\includegraphics[width=0.45\textwidth]{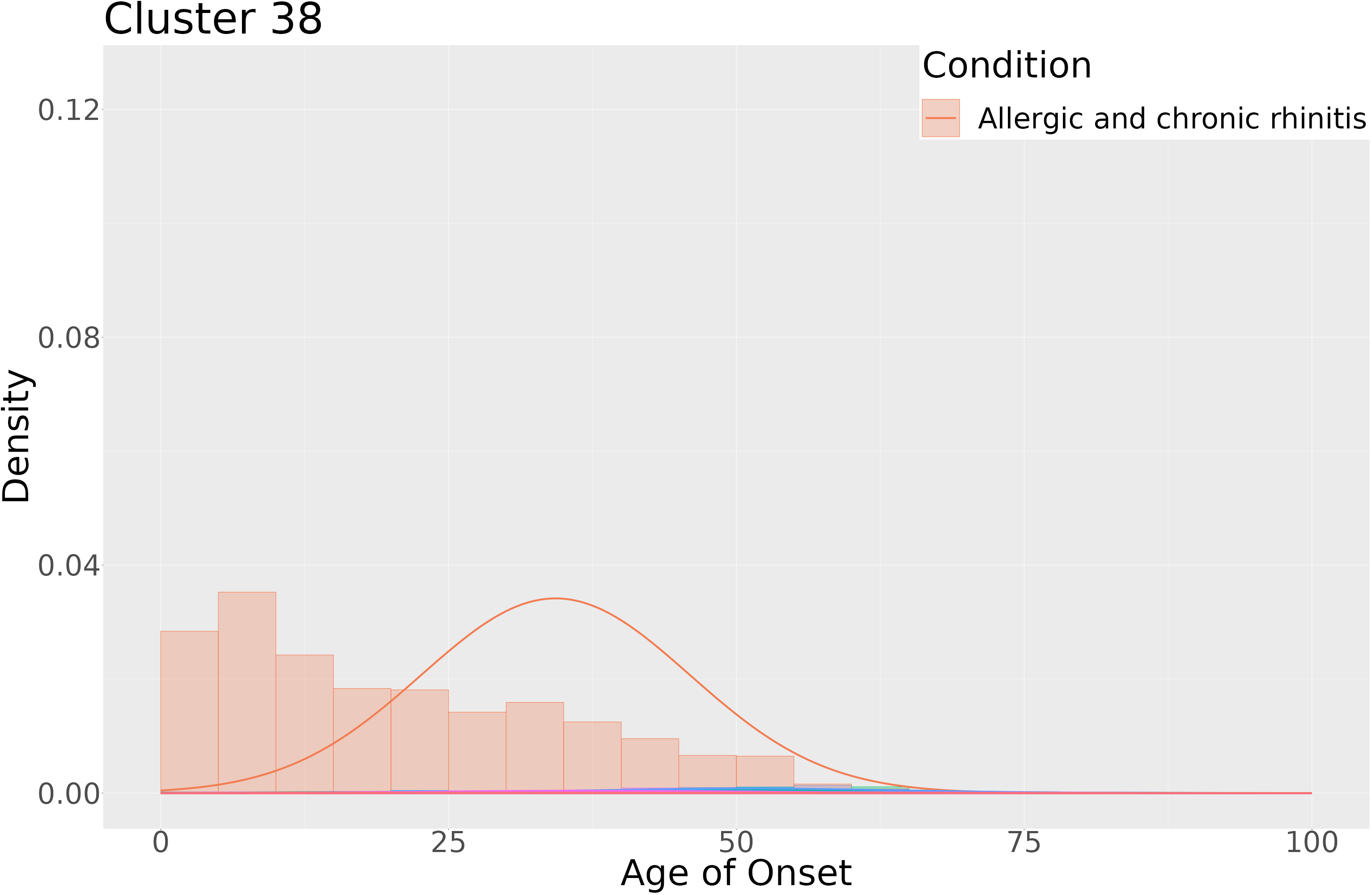}\qquad
\end{figure}

\begin{figure}[ht]
\centering
\includegraphics[width=0.45\textwidth]{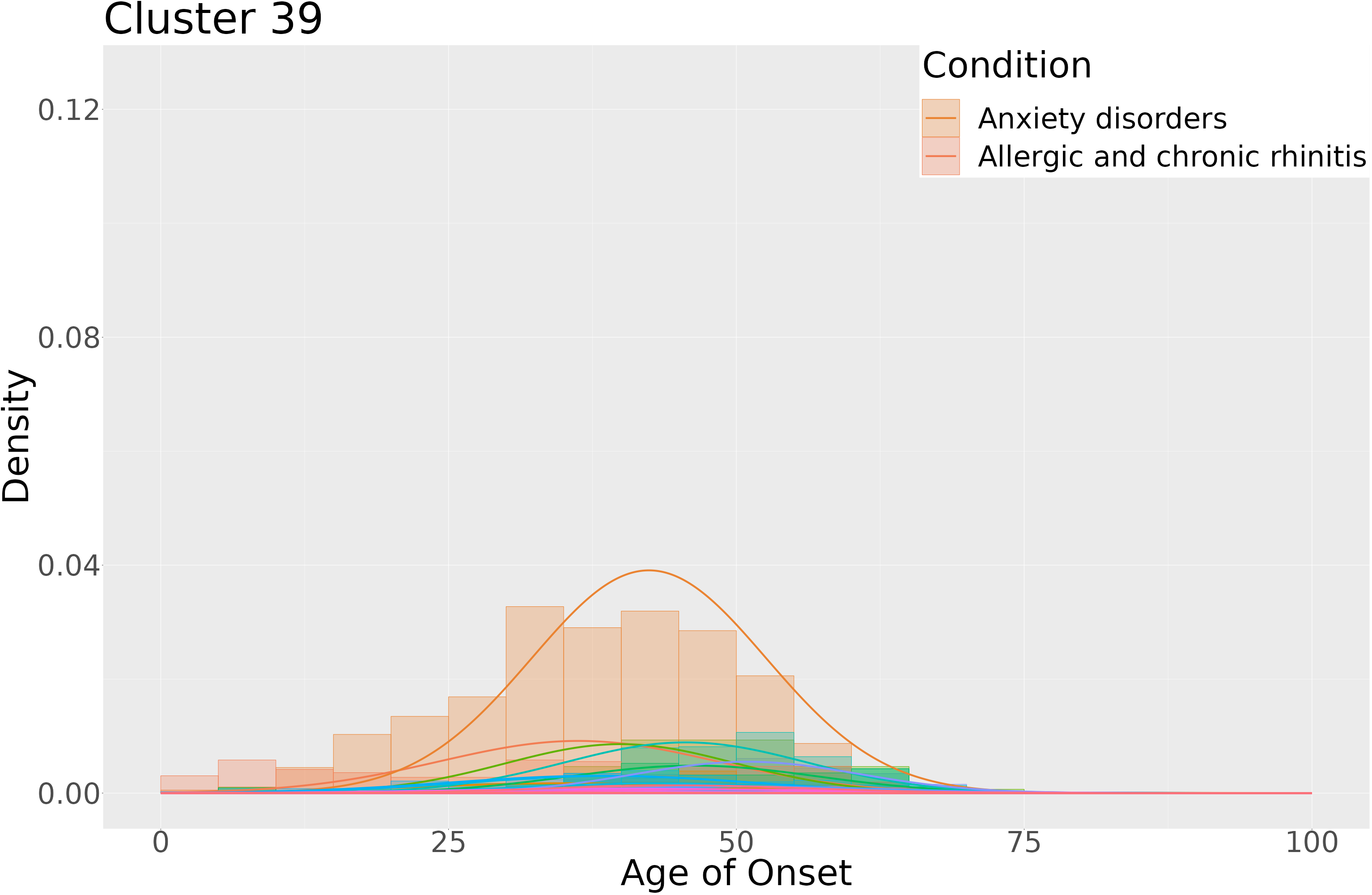}\qquad
\includegraphics[width=0.45\textwidth]{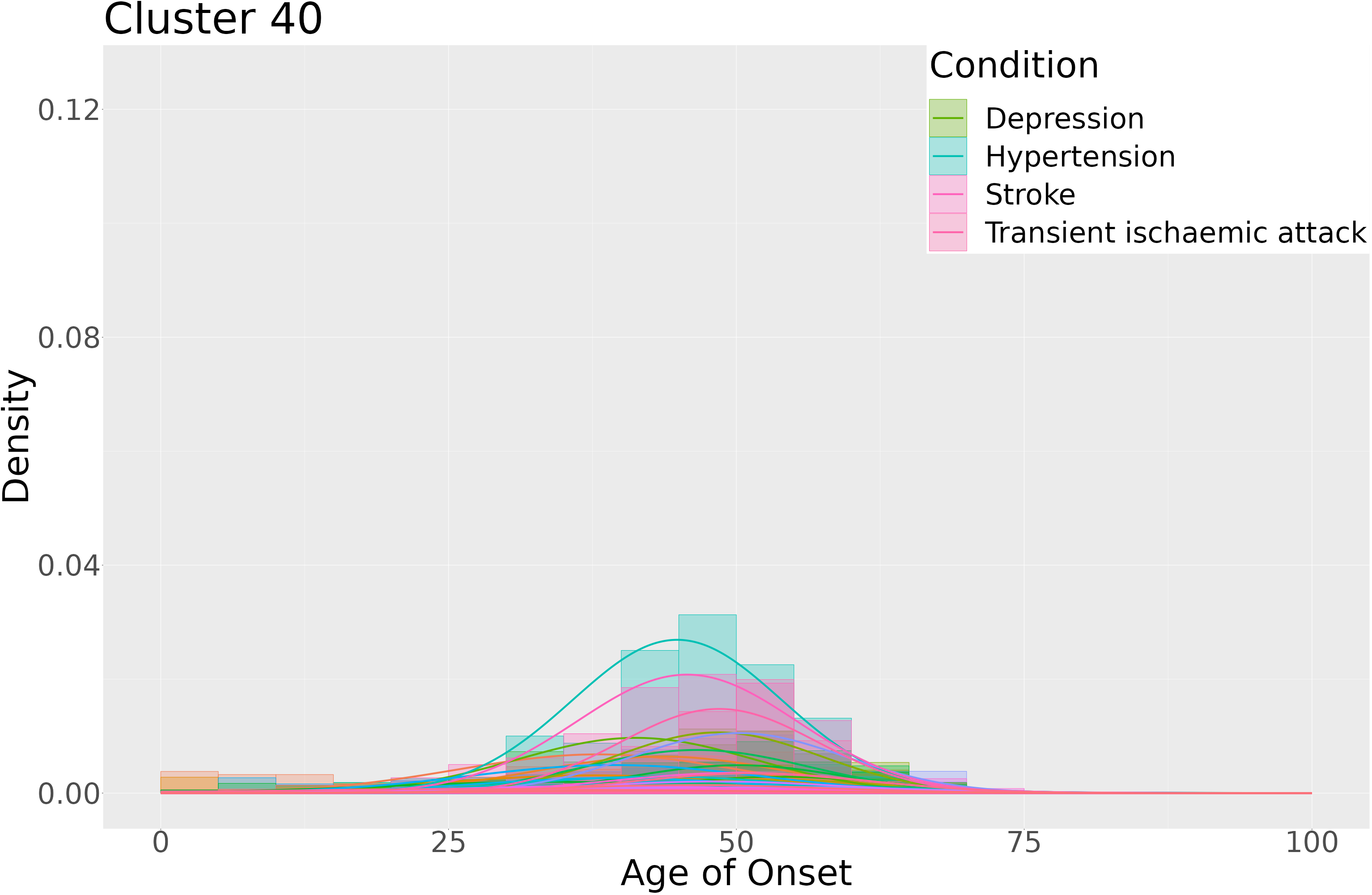}\qquad
\end{figure}

\begin{figure}[ht]
\centering
\includegraphics[width=0.45\textwidth]{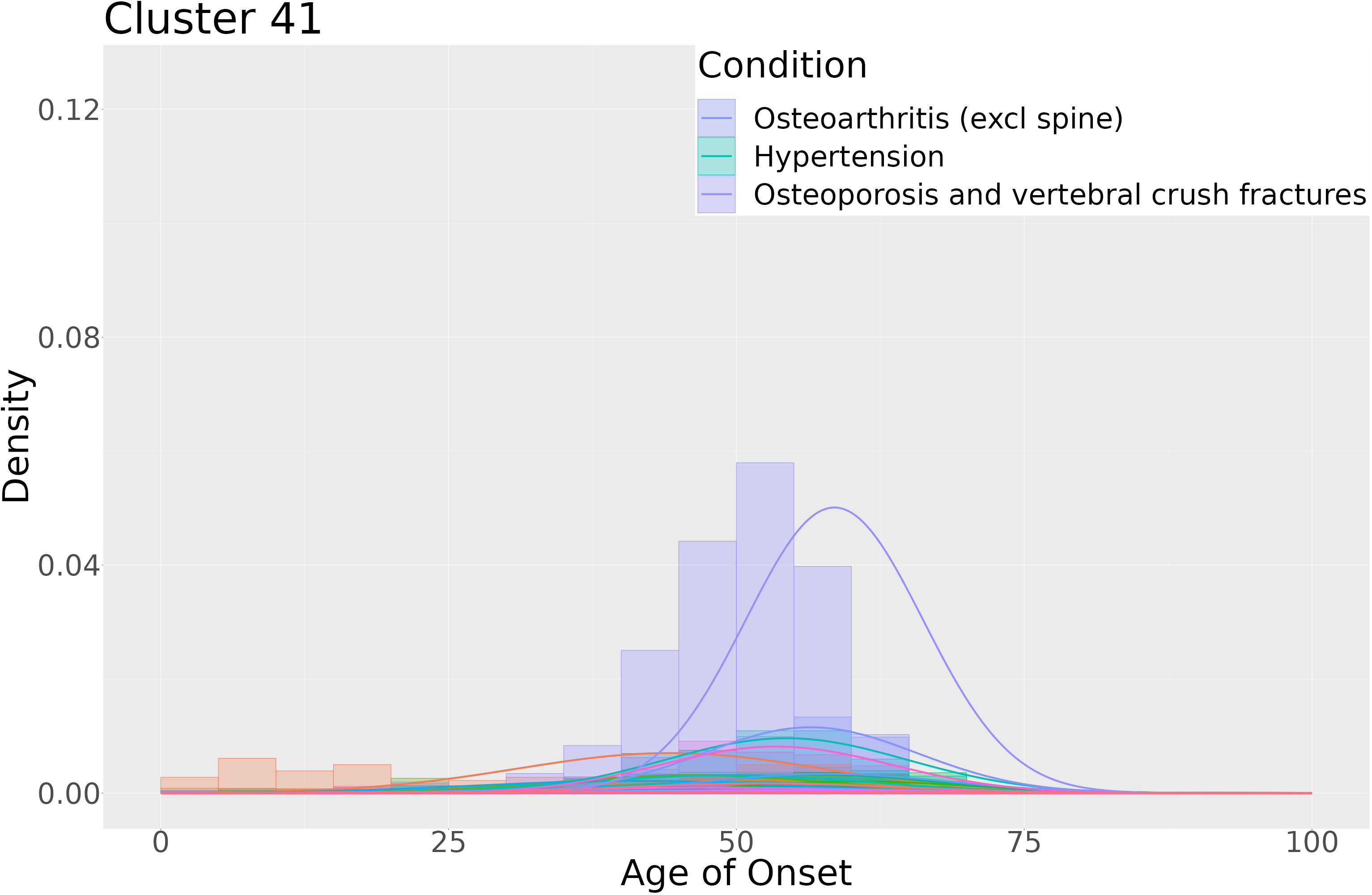}\qquad
\includegraphics[width=0.45\textwidth]{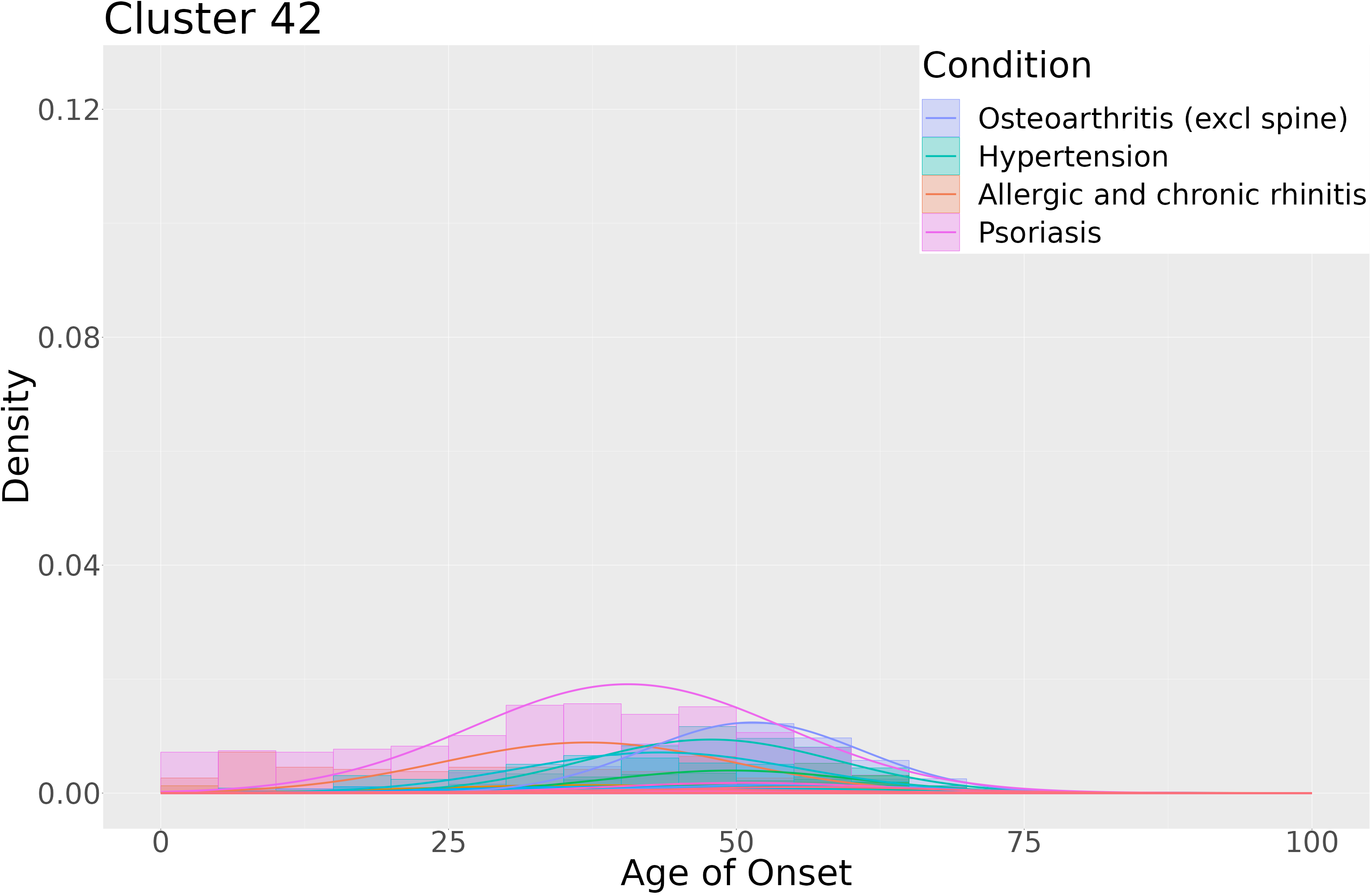}\qquad
\end{figure}

\begin{figure}[ht]
\centering
\includegraphics[width=0.45\textwidth]{50Cluster43.png}\qquad
\includegraphics[width=0.45\textwidth]{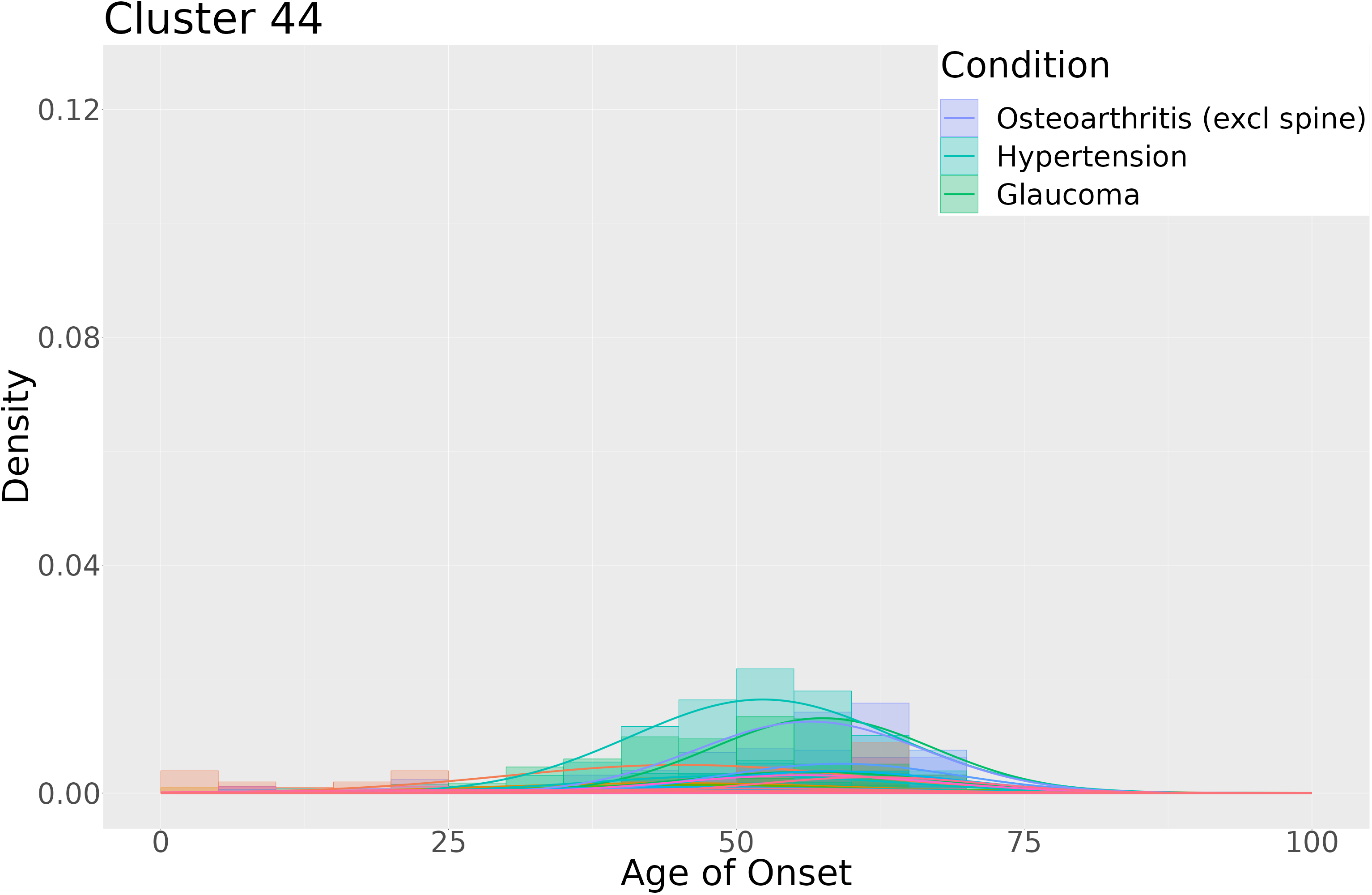}\qquad
\end{figure}

\begin{figure}[ht]
\centering
\includegraphics[width=0.45\textwidth]{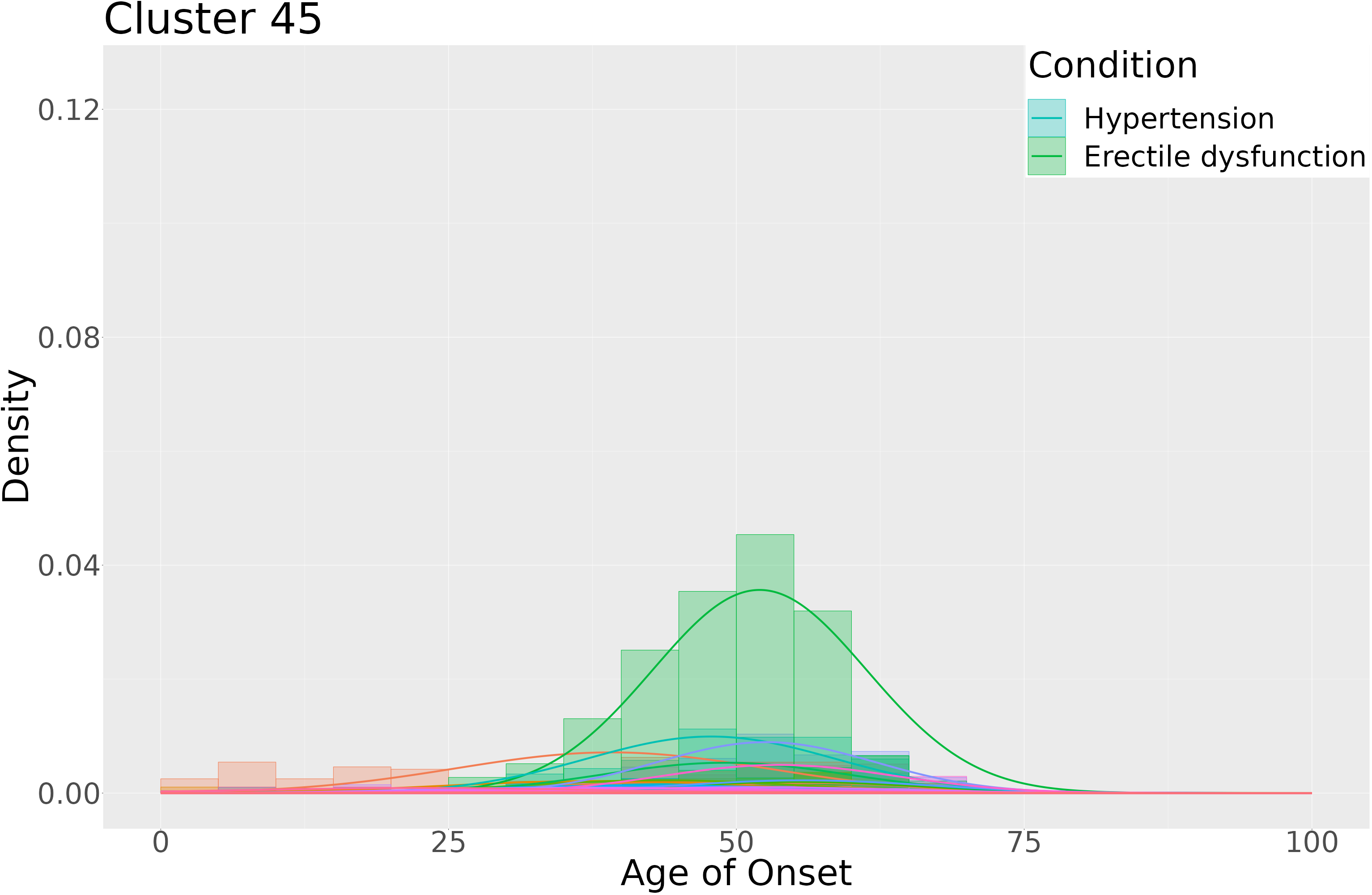}\qquad
\includegraphics[width=0.45\textwidth]{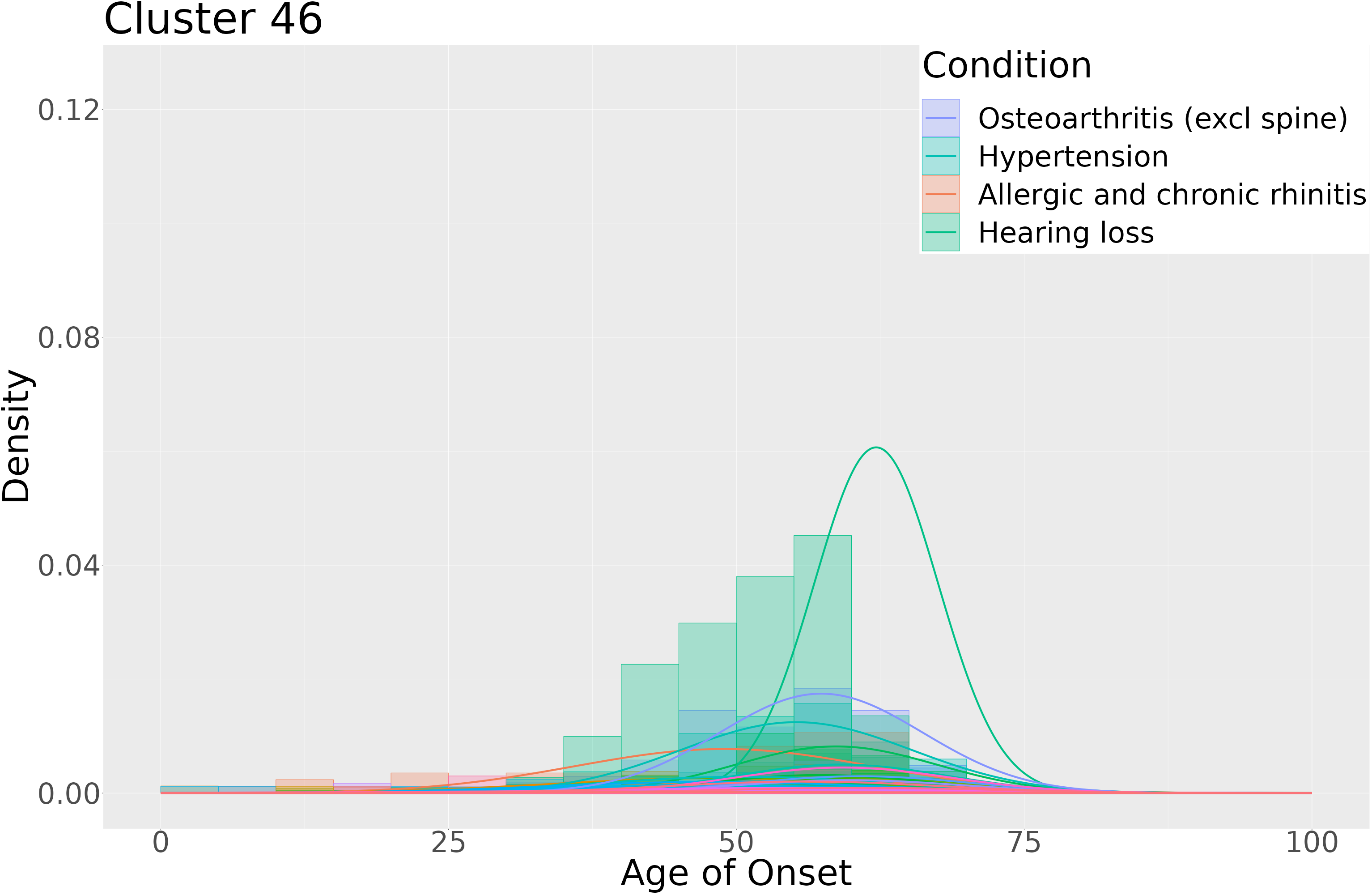}\qquad
\end{figure}

\begin{figure}[ht]
\centering
\includegraphics[width=0.45\textwidth]{50Cluster47.png}\qquad
\includegraphics[width=0.45\textwidth]{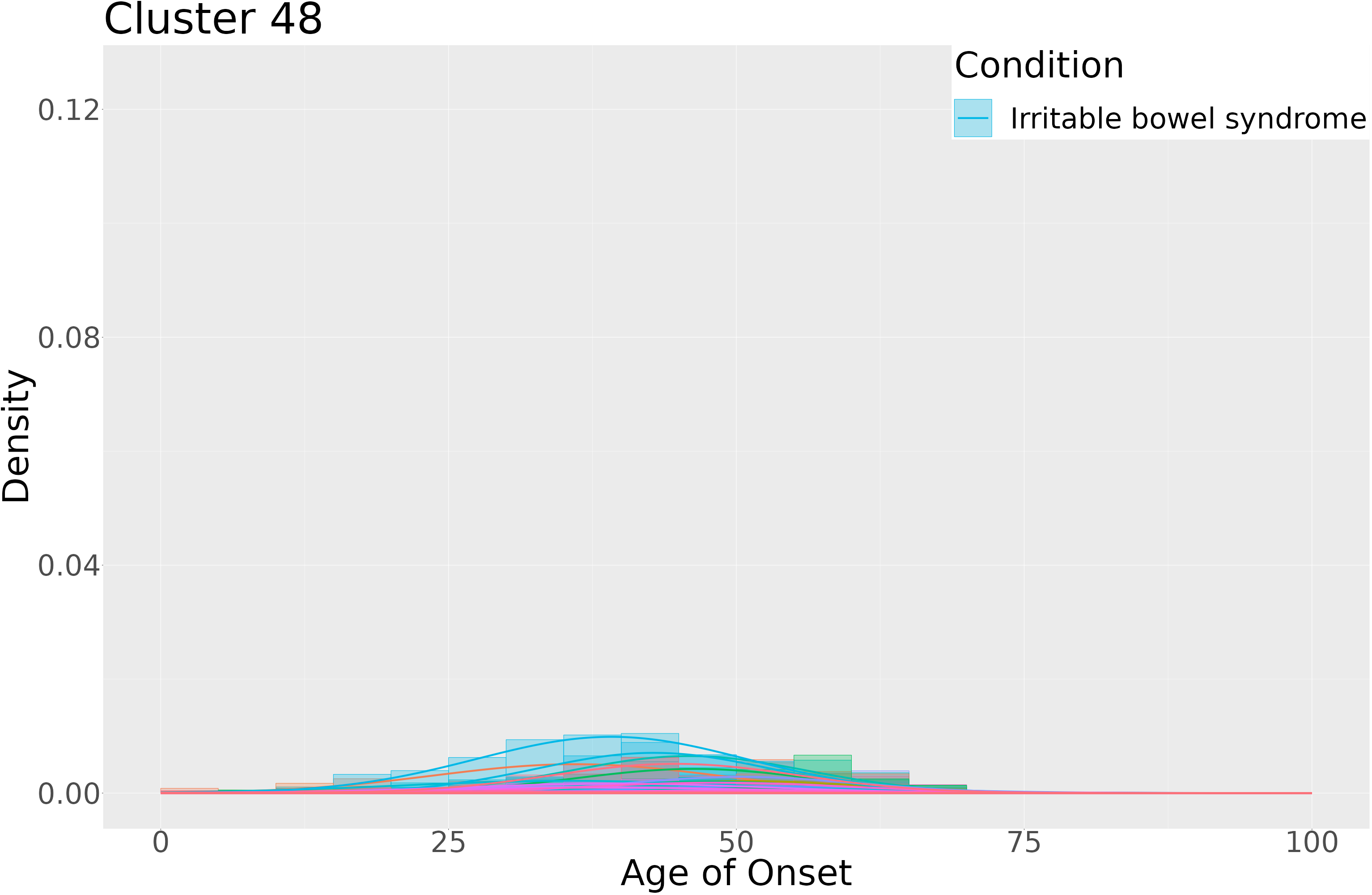}\qquad
\end{figure}

\begin{figure}[ht]
\centering
\includegraphics[width=0.45\textwidth]{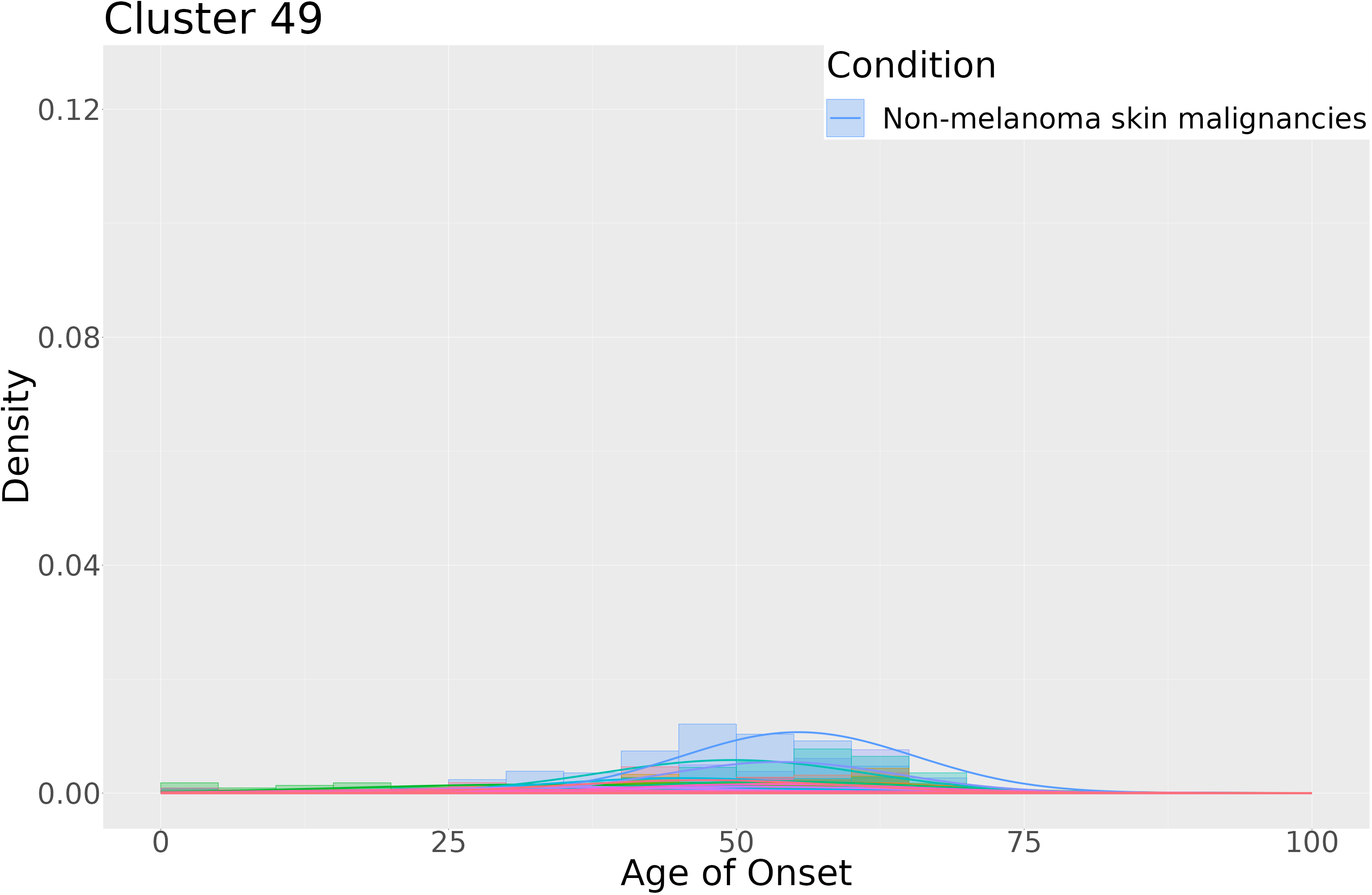}\qquad
\includegraphics[width=0.45\textwidth]{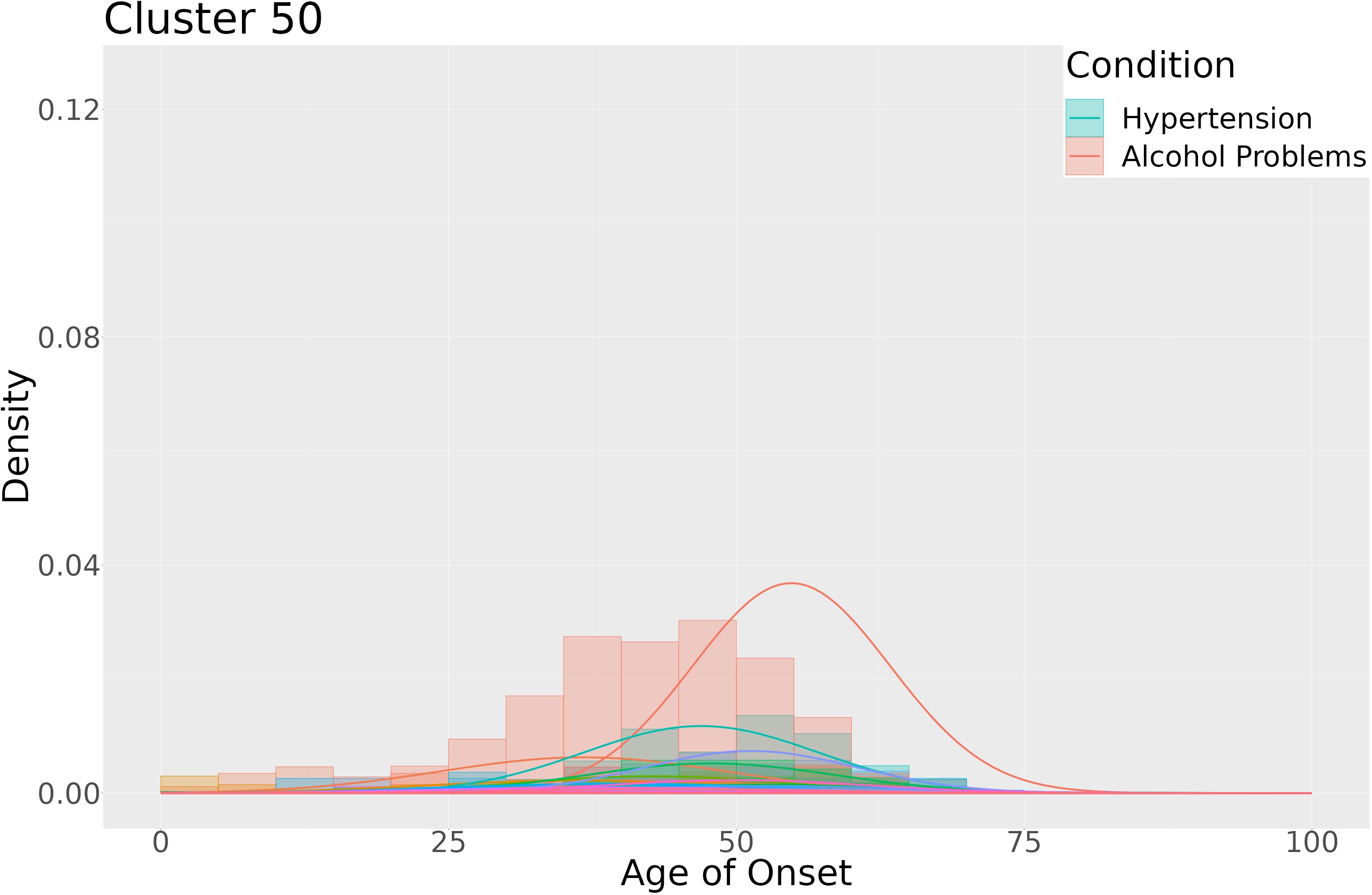}\qquad
\vspace*{6.5in}
\end{figure}

\fi

\end{document}